\documentclass[preprint,flushrt,a4]{aastex}

\usepackage[varg]{txfonts}
\usepackage{graphicx, natbib, amssymb}
\usepackage{enumerate}
\bibpunct{(}{)}{;}{a}{}{,}
\usepackage{longtable}
\usepackage{tabularx}
\usepackage[pagewise,displaymath, mathlines]{lineno}
\usepackage{pdflscape}

\newcommand{\kms}{km~s$^{-1}$}
\newcommand{\robs}{\vec{r}_{\mathrm{obs}}}
\newcommand{\vobs}{\vec{v}_{\mathrm{obs}}}
\newcommand{\vLIC}{\vec{v}_{\mathrm{LIC}}}

\shorttitle{The Warsaw Test Particle Model}
\shortauthors{Sok{\'o}{\l} et al. 2015}

\begin{document}

\title{Interstellar neutral helium in the heliosphere from \emph{IBEX} observations.\\ II. The Warsaw Test Particle Model (WTPM)}
    
\author{J.~M.~Sok{\'o}{\l}, M.~A.~Kubiak, M.~Bzowski, P.~Swaczyna}
\email{jsokol@cbk.waw.pl}

\affil{Space Research Centre of the Polish Academy of Sciences, 00-716 Warsaw, Poland}

\begin{abstract}
We have developed a refined and optimized version of the Warsaw Test Particle Model of interstellar neutral gas in the heliosphere, specially tailored for analysis of \emph{IBEX}-Lo observations. The former version of the model was used in the analysis of neutral He observed by \emph{IBEX} that resulted in an unexpected conclusion that the interstellar neutral He flow vector was different than previously thought and that a new population of neutral He, dubbed the Warm Breeze, exists in the heliosphere. It was also used in the reanalysis of \emph{Ulysses} observations that confirmed the original findings on the flow vector, but suggested a significantly higher temperature. The present version model has two strains targeted for different applications, based on an identical paradigm, but differing in the implementation and in the treatment of ionization losses. We present the model in detail and discuss numerous effects related to the measurement process that potentially modify the resulting flux of ISN~He observed by \emph{IBEX}, and identify those of them that should not be omitted in the simulations to avoid biasing the results. This paper is part of a coordinated series of papers presenting the current  state of analysis of \emph{IBEX}-Lo observations of ISN~He. Details of the analysis method are presented by \citet{swaczyna_etal:15a}, and results of the analysis are presented by \citet{bzowski_etal:15a}.
\end{abstract}


\section{Introduction}
\label{sec:Intro}
Our paper presents in detail the Warsaw Test Particle Model (WTPM), a previous version of which was used by \citet{bzowski_etal:12a} in their analysis of \emph{IBEX}-Lo data from 2009 and 2010 and by \citet{mccomas_etal:15a} in the preliminary analysis of \emph{IBEX} data from 2013 and 2014. It is an element of a coordinated series of papers presenting the current state of analysis of the interstellar neutral (ISN) He data using the methodology originally adopted by \citet{bzowski_etal:12a}, which belongs to a coordinated set of Special Issue papers on interstellar neutrals as measured by \emph{IBEX}, introduced and overviewed by \citet{mccomas_etal:15b}. In this series, the method of $\chi^2$-fitting of the data that feature various correlations, which is an extension and refinement of the method originally used, is presented by \citet{swaczyna_etal:15a}. That paper also discusses some observational aspects of the analysis, including the compensation of on board data throughput reduction and refinement of the spin axis determination. \citet{sokol_etal:15a} and \citet{galli_etal:15a} present an estimate for the energy threshold of the \emph{IBEX}-Lo sensitivity to ISN~He. \citet{bzowski_etal:15a} presents the results of the $\chi^2$ analysis and their interpretation. This coordinated analysis uses the WTPM model of ISN~He gas observations presented in this paper. 

WTPM has a long history of development and successful applications, going back to mid-1990s. The first version \citep{rucinski_bzowski:95b, bzowski_etal:97} addressed the issue of the influence of the time dependence of radiation pressure and ionization rate on the density and velocity of ISN~H inside the heliosphere. It was based on a simplified, idealized solar cycle variation of these quantities. Adaptation of this simplified model to ISN~He was presented by \citet{rucinski_etal:03}. Subsequently, the model was extended to accommodate the ionization rate dependence on the heliolatitude \citep{bzowski:03} and applied to infer the evolution of the latitudinal structure of the solar wind based on observations of the Ly$\alpha$ backscatter glow from SWAN on \emph{SOHO} \citep{bzowski_etal:03a}. The next phase of model development was introducing the dependence of radiation pressure on the radial velocity of atoms with respect to the Sun \citep{tarnopolski_bzowski:09} and a realistic, measurement-based ionization rate. It was applied to theoretical studies of the ISN~D distribution in the heliosphere \citep{tarnopolski_bzowski:08a} and to the determination of the ISN~H density at the termination shock and in the Local Interstellar Cloud (LIC) based on \emph{Ulysses} observations of H$^{+}$ pickup ions \citep{bzowski_etal:08a, bzowski_etal:09a}. Subsequently, the model was tailored to accommodate ISN~He observed by \emph{IBEX} \citep{bzowski_etal:12a}. It was also used by \citet{bzowski_etal:14a} to re-analyze observations from GAS/\emph{Ulysses}, including the first analysis of the previously not analyzed data from the last \emph{Ulysses} orbit in 2007, which had previously not been analyzed. This analysis brought a flow vector similar to the original analysis by \citet{witte:04}, but a temperature higher by at least $\sim1000$~K. It was also used by \citet{bzowski_etal:13b} and \citet{park_etal:14a} to analyze the abundance of Ne/O ratio in the LIC based on \emph{IBEX}-Lo measurements, and by \citet{kubiak_etal:14a} to discover the additional ISN~He population detected by \emph{IBEX}-Lo dubbed the Warm Breeze, which is very likely the secondary heliospheric population of ISN~He. The WTPM model was also used by \citet{kubiak_etal:13a} to predict possibilities of detection of the ISN~D flux by \emph{IBEX}-Lo, subsequently found in the \emph{IBEX}-Lo signal by \citet{rodriguez_etal:13a, rodriguez_etal:14a}.

For this round of analysis, the model was revised and optimized. For test and validation purposes, we developed its new version, the so-called analytic WTPM (aWTPM), which is effectively the classical hot model, first formulated by \citet{thomas:78}, adapted to the task of simulating the ISN~He flux observed by \emph{IBEX}. This model assumes that the ionization rate is constant over time and decreases with the square of heliocentric distance. Under these assumptions, the ionization losses can be calculated using an analytic formula: hence the name of the model. The new version of the original WTPM now becomes the numerical WTPM (nWTPM). Revisions and optimizations include adopting improved, more accurate algorithms for atom tracking and integration over spin-angle bins and observation time, which results in overall reduction of the computational load needed to compute a full simulation for one set of ISN~He parameters. aWTPM and nWTPM are independent codes based on an identical theoretical framework except for the treatment of ionization losses. nWTPM is coded in \texttt{Fortran} and \texttt{C}, and aWTPM is implemented in \texttt{Wolfram Research Mathematica}. A detailed comparison of aWTPM and nWTPM is provided in Table~\ref{tab:compWTPM} at the end of Section~\ref{sec:Outlook}.

The two versions of WTPM were thoroughly cross-validated with the goal of achieving an agreement no worse than $1\%$ when run under identical assumptions. This goal was successfully achieved, as we demonstrate in this paper. In the following, we present the foundations of WTPM and discuss various observational aspects that need to be addressed by a model intended for use in an analysis of \emph{IBEX}-Lo data as presented by \citet{swaczyna_etal:15a}, i.e., $\chi^2$-fitting of the observed count rate. Clearly, the accuracy of a model used to fit the data must be better than the uncertainties in the data, which are on the order of $1-2\%$ in the data points with the best statistics. Therefore one needs to consider all known observation effects that potentially affect the observed flux, even if by intuition they may seem subtle and not worth bothering with. We identify those that indeed may be neglected and those that should be taken into account in the analysis. Hence the description of the model is more detailed than usually provided in the science literature.

This paper has two main sections. In the first of them, Section~\ref{sec:modelDescr}, we present the baseline model and discuss differences between aWTPM and nWTPM which are summarized in Table~\ref{tab:compWTPM}. Cross validation of the two versions is presented in Section~\ref{sec:crossVal}. The second major section is Section~\ref{Sec:effects}, which presents --- to our knowledge, for the first time in the literature --- observation effects influencing the ISN~He flux measured by \emph{IBEX}-Lo, including, among others, the variation of the measured flux during an orbit due to the Earth's motion around the Sun and the satellite's motion around the Earth, effects of the tilt of the spin axis to the ecliptic, as well as effects of ionization losses and its uncertainty. The paper ends with a general summary and conclusions. 

\section{Model description}
\label{sec:modelDescr}
The WTPM is based on the concept of the hot model of neutral interstellar gas \citep{fahr:78, thomas:78, wu_judge:79a}. In this model, the local distribution function of neutral interstellar gas inside the heliosphere is calculated starting from an assumed homogeneous distribution function $f_{\mathrm{LIC}}\left(\vLIC; \vec{\pi}\right)$ of this gas in the so-called source region outside the heliosphere, where $\vLIC$ is the velocity vector of an individual atom and $\vec{\pi}$ a set of physical parameters of the assumed distribution function, including the mean velocity vector of the gas relative to the Sun $\vec{v_{\mathrm{B}}}$. The model bears an important assumption that the gas inside the heliosphere is collisionless, so the atoms can be treated as individual, non-interacting point-like objects and that far away from the heliosphere the gas is spatially homogeneous (i.e., the parameters $\vec{\pi}$ of the distribution function $f_{\mathrm{LIC}}$ do not depend on the location in space). The local distribution function of the gas $f\left(\vec{r}_{\mathrm{obs}}, \vec{v}_{\mathrm{obs}}, t_{\mathrm{obs}}; \vec{\pi} \right)$ for a time $t_{\mathrm{obs}}$, a heliocentric velocity vector $\vec{v}_{\mathrm{obs}}$, and a location in space given by a heliocentric radius vector $\vec{r}_{\mathrm{obs}}$ is given by the product:
	\begin{equation}
	f\left( \vec{r}_{\mathrm{obs}},\vec{v}_{\mathrm{obs}},t_{\mathrm{obs}};\vec{\pi} \right)=f_{\mathrm{LIC}}\left( \vLIC \left(\vec{r}_{\mathrm{obs}},\vec{v}_{\mathrm{obs}}\right);\vec{\pi}\ \right) w\left(\vec{r}_{\mathrm{obs}},\vec{v}_{\mathrm{obs}},t_{\mathrm{obs}},\beta \right)
	\label{eqLocDiFun}
	\end{equation}
where $\vLIC$ is a function of the local heliocentric velocity $\vec{v}_{\mathrm{obs}}$ of an atom at the heliocentric location $\vec{r}_{\mathrm{obs}}$ and $w$ is the probability of survival of the atom of the travel from the source region in front of the heliosphere to the local point $\vec{r}_{\mathrm{obs}}$. $\vLIC\left(\vec{v}_{\mathrm{obs}}, \vec{r}_{\mathrm{obs}}\right)$ is a relation that connects the velocity vector of the atom at $\vec{r}_{\mathrm{obs}}$ with the velocity $\vLIC$ of the atom in the source region of interstellar gas. $\beta$~is a function that describes all details of the ionization rate inside the heliosphere, including its dependence on heliolatitude, time, and solar distance. 

The survival probability $w$ and related ionization processes were extensively discussed by \citet{bzowski_etal:13b} and this discussion will not be repeated here. In short, the survival probability is calculated as an exponent of the exposure $\epsilon$ of the atom to ionization: 
	\begin{equation}
	w = \exp\left( \epsilon \right)= \exp \left( -\int \limits_{t_{\mathrm{LIC}}}^{t_{\mathrm{obs}}} \beta\left( \vec{r}\left(t\right), t \right) \mathrm{d}t \right)\
	\label{eqWdef}
	\end{equation}
where $\beta\left( \vec{r}\left(t\right), t \right)$ is the ionization rate at a time $t$ at a location inside the heliosphere defined by the radius vector $\vec{r}\left(t\right)$, which traces the trajectory of the atom. Thus, in a general case of ionization rates changing  with time, varying with heliolatitude, and falling off with solar distance different from $1/r^2$, one needs to calculate the survival probability by integrating the exposure in the exponent in Equation~\ref{eqWdef} numerically. Only for an ionization rate invariable with time and heliolatitude and falling off with the square of solar distance is it possible to calculate $w$ analytically using a formula shown later in the paper.	

Calculating the local distribution function for a local velocity $\vobs$ at a location $\robs$ requires finding the relation between the state vector of the atom $\left( \vobs, \robs \right)$ and the velocity vector of the atom $\vLIC$ in the source region. This relation is a function of the forces acting on the atom. In the case of hydrogen atoms, the forces include solar gravity and solar radiation pressure, which varies with solar activity and depends on the radial velocity of the atom \citep{tarnopolski_bzowski:09}, and thus is hard to take into account analytically. In the case of helium atoms (as well as oxygen and neon) the radiation pressure is negligible, the force is just due to solar gravity, and the relation $\vLIC \left( \vobs,\robs \right)$ can be given analytically. This will be presented later in the paper. 

With the local distribution function established it is easy to calculate its moments $m^{(n)}$, like density (zeroth moment), vector flux (first moment), etc. They are obtained by numerically calculating appropriate integrals \citep[see, e.g.,][]{bzowski_etal:97, rucinski_etal:03, tarnopolski_bzowski:09}:
	\begin{equation}
	m^{(n)}=\int v^n f\left( \vec{r}_{\mathrm{obs}},\vec{v}_{\mathrm{obs}},t_{\mathrm{obs}}; \vec{\pi} \right)\mathrm{d}^3 v.
	\label{eqDFMoments}
	\end{equation}
The integration is done in the solar inertial frame, but in principle can be performed in any inertial frame. 	
	
The version of the WTPM discussed in this paper has a different objective: instead of calculating moments of the local distribution function of interstellar gas in the solar inertial frame, it simulates results of observations obtained from the neutral atom detector \emph{IBEX}-Lo \citep{fuselier_etal:09b}. To that end, it must calculate the flux of atoms impinging on the detector and going through its collimator in the spacecraft inertial frame. The \emph{IBEX} spacecraft is spin-stabilized, with the spin-axis being changed periodically to approximately follow the Sun. The observed region is a strip on the sky perpendicular to the spin-axis and the instantaneous field of view (FOV) of the instrument, defined by the collimator aperture. The collimator makes the FOV hexagonal in shape, with transmission decreasing from a maximum value at the boresight to zero at the perimeter. 
	
The signal is sampled while the spacecraft is spinning at $\sim4.2$~rpm. The observations are accumulated in 60 identical time slots per spin, which is equivalent to registering them in $\Delta\psi = 6\degr$ spin-angle bins. While the spin axis is not varying during an orbit, the actual observation time is split into alternating sub-intervals corresponding to eight different energy settings of the instrument, the so-called energy steps. The observation interval adopted for analysis is a sum of sub-intervals of good times $\Delta t_{i,j}$, i.e., the time intervals $j$ for orbit $i$ with the data considered to be adequate for analysis \citep{mobius_etal:12a, leonard_etal:15a}. 
	
Consequently, the simulation software must be able to calculate the flux corresponding to a given line of sight of the detector, defined by the pointing of the spin-axis $\left(\lambda_{\mathrm{P}}, \phi_{\mathrm{P}} \right)$ and the spin-angle $\psi$ at a given time moment $t$, taking into account the collimator transmission function $T$. Denoting the observed flux for the $k$th spin-angle bin and time $t$ as $F\left(\lambda_{\mathrm{P}}, \phi_{\mathrm{P}}, \psi_{k}, t; \vec{\pi}\right)$, the program subsequently calculates average values of the flux over spin-angle bins, centered at $\psi_{k}$ and having a width $\Delta\psi = 6\degr$ and over good time intervals $\Delta t_{ij}$, which yields the value of the average flux $\langle F_{\mathrm{orb}}\left(\lambda_{\mathrm{P}}, \phi_{\mathrm{P}},\psi_k;\vec{\pi}\right)\rangle_{\Delta\psi,\mathrm{GT}}$ for a given orbit and spin-angle bin $\psi_k$: 
	\begin{equation}
	{\langle F_{\mathrm{orb}} \left( \lambda_{\mathrm{P}}, \phi_{\mathrm{P}}, \psi_{k};\vec{\pi} \right) \rangle } _{\Delta \psi, \mathrm{GT}} = \sum_{j=1}^{N_j}{\frac{\int\limits_{t_{ij}}^{t_{ij}+\Delta t_{ij}}{\left[ \int\limits_{\psi_{k}-\Delta\psi/2}^{\psi_{k}+\Delta\psi/2}{F\left( \lambda_{\mathrm{P}}, \phi_{\mathrm{P}}, \psi, t;\vec{\pi} \right)} \mathrm{d}\psi \right]}\mathrm{d}t}{\Delta \psi \sum_{j=1}^{N_j}{\Delta t_{ij}}}}
	\label{eqPhiOrb}
	\end{equation}	
The summation goes over all $N_j$ intervals of good times on orbit $i$. Details of the calculations are presented in the following sections.	

\subsection{Calculation of the distribution function in the LIC}
\label{sec:ballistics}
To calculate the local distribution function, defined in Equation~\ref{eqLocDiFun}, first one needs to calculate $f_{\mathrm{LIC}}\left(\vLIC\left( \vobs, \robs \right); \vec{\pi} \right)$, and to that end, one needs to find the relation $\vLIC \left( \vobs, \robs \right)$ between the state vector of an atom $\left(\vec{v}_{\mathrm{obs}}, \vec{r}_{\mathrm{obs}}\right)$ and the velocity of the atom $\vLIC$ in the source region of neutral interstellar atoms, assumed to be at a distance $r_{\mathrm{fin}}$ from the Sun (for the rationale, see Section~\ref{sec:rFin}). This relation can be found either by solving the equation of motion of the atom with the starting conditions $\left(\vobs, \robs \right)$, or --- in the case of the purely Keplerian motion of ISN~He atoms in the field of solar gravity --- analytically. The first solution was presented, e.g., by \citet{rucinski_bzowski:95a} and \citet{tarnopolski_bzowski:09} and will not be repeated here. The analytic solution is well known and has been widely used, recently, e.g., \citet{mueller_cohen:12a} and \citet{mueller_etal:13a}. The implementation used in the WTPM is shown here for the completeness of model presentation.

The atom is moving on a hyperbolic Keplerian orbit with the Sun in the focus and we know the velocity $\vobs$ and position $\robs$ of the atom in a given time moment. The speed of the atom is $v_{\mathrm{obs}} = \left(\vobs \cdot \vobs\right)^{1/2}$ and the distance from the Sun $r_{\mathrm{obs}} = \left(\robs \cdot \robs\right)^{1/2}$. Thus we can immediately calculate the total mechanical energy $E$ and angular momentum $\vec{L}$ per unit mass: 
	\begin{equation}
	E=\frac{v_{\mathrm{obs}}^{2}}{2}-\frac{GM}{r_{\mathrm{obs}}} > 0; \quad \vec{L} = \vec{r}_{\mathrm{obs}} \times \vec{v}_{\mathrm{obs}},
	\label{eqEnerAngMom}
	\end{equation}
with $GM$ being the product of the gravity constant and solar mass, best implemented as the Gauss solar gravity constant due to its high accuracy. The motion is planar and the angular momentum vector determines the direction perpendicular to the orbital plane. We also calculate the local radial speed $v_{\mathrm{r,obs}}$:	
	\begin{equation}
	v_{\mathrm{r,obs}} = \left( \vec{r}_{\mathrm{obs}}/r_{\mathrm{obs}} \right) \cdot \vec{v}_{\mathrm{obs}}.
	\label{eqVrad}
	\end{equation}	

With this definition, a negative value of $v_{\mathrm{r,obs}}$ implies the atom is approaching the Sun. The initial velocity vector $\vobs$ is a sum of two vectors in the orbital plane: the radial ($\vec{v}_{\mathrm{r,obs}}$) and transversal ($\vec{v}_{\mathrm{t,obs}}$) velocity vectors. We point out that the radial velocity unit vector is of course parallel to the radial direction, but its direction depends on the sign of the radial speed. The transversal velocity vector is obtained from vector subtraction of the radial velocity vector from the full velocity vector:	
	\begin{equation}
	\vec{v}_{\mathrm{t,obs}}=\vec{v}_{\mathrm{obs}} - \vec{v}_{\mathrm{r,obs}}.
	\label{eqVtrans}
	\end{equation}
The unit vectors $\hat{v}_{\mathrm{r,obs}}$, $\hat{v}_{\mathrm{t,obs}}$ of the radial and transversal velocity vectors can be used to form the basis of the reference system with the $x--y$ plane corresponding to the orbital plane, which will be specified further in the text.	

The heliocentric distance $r$ of the atom at an arbitrary point on its trajectory is defined by:
	\begin{equation}
	r = \frac{p}{1+e \cos \theta},
	\label{eqHyperTraj}
	\end{equation}
where $\theta$ is a true anomaly that measures the angular distance between the direction to the perihelion and the actual location of the atom at $r$ and $p$ is the orbital parameter defined by:	
	\begin{equation}
	p=\frac{L^2}{GM},
	\label{eqOrbPar}
	\end{equation}	
$e > 1$ is the eccentricity of the orbit, equal to:		
	\begin{equation}
	e=p/r_{\mathrm{peri}},
	\label{eqEccentricity}
	\end{equation}
with $r_{\mathrm{peri}}$ being the perihelion distance, obtained from:
	\begin{equation}
	r_{\mathrm{peri}} = \frac{\left( \left( GM \right)^2 + 2EL^2 \right)^{1/2} - GM}{2E}.
	\label{eqRperi}
	\end{equation}
To calculate the velocity vector of the atom in the source region $\vLIC$ at a distance $r_{\mathrm{LIC}}$ from the Sun, we must calculate its true anomaly $\theta_{\mathrm{LIC}}$ for this distance. In addition, we will need the angle swept by the atom on its way from the source region to the local position $\robs$, for a purpose that will be explained in the next section. The true anomaly $\theta_{\mathrm{obs}}$ of the atom at $\robs$ is obtained from its sine and cosine functions, calculated as follows:
	\begin{equation}
	\cos\theta_{\mathrm{obs}}=p/r_{\mathrm{obs}}-1; \quad \sin\theta_{\mathrm{obs}} = \frac{v_{\mathrm{r,obs}}}{\left| v_{\mathrm{r,obs}} \right|} \sin\left( \arccos \left(\cos\theta_{\mathrm{obs}}\right) \right).
	\label{eqThetaObs}
	\end{equation}
The true anomaly of the atom in the source region $\theta_{\mathrm{LIC}}$ is obtained from the solution of Equation~\ref{eqHyperTraj} for the hyperbolic orbit for $r = r_{\mathrm{LIC}}$ with the prerequisite that the atom is moving toward the Sun, i.e., its radial velocity at $r_{\mathrm{LIC}}$ is negative. Thus,
	\begin{equation}
	\theta_{\mathrm{LIC}} = - \arccos\left[ \left( p/r_{\mathrm{LIC}} - 1 \right)/e \right]
	\label{eqThetaLIC}
	\end{equation}
and we can calculate the velocity vector of the atom in the LIC in the orbital reference frame: its $z$-component is 0, the transversal coordinate from the conservation of angular momentum is 	
	\begin{equation}
	v_{\mathrm{t,LIC}} = L/r_{\mathrm{LIC}},
	\label{eqVtLIC}
	\end{equation}
and the radial component from the conservation of energy and the prerequisite that the radial velocity is negative	
	\begin{equation}
	v_{\mathrm{r,LIC}} = -\left[ 2\left( E+GM/r_{\mathrm{LIC}} \right) - v_{\mathrm{t,LIC}}^2 \right]^{1/2}.
	\label{eqVrLIC}
	\end{equation}	

Defining the basis unit vectors $\left\{ \hat{x}, \hat{y}, \hat{z} \right\}$  for the reference system with the $x--y$ plane coplanar with the orbital frame, 
	\begin{eqnarray}
	\hat{x} & = & \hat{r}_{\mathrm{obs}} \cos\theta_{\mathrm{obs}} - \hat{v}_{\mathrm{t,obs}} \sin\theta_{\mathrm{obs}} \nonumber \\
	\hat{y} & = & \hat{r}_{\mathrm{obs}} \sin\theta_{\mathrm{obs}} + \hat{v}_{\mathrm{t,obs}} \cos\theta_{\mathrm{obs}} \nonumber \\
	\hat{z} & = & \vec{L}/L 
	\label{eqOrbRef}
	\end{eqnarray}	
we calculate the components of $\vLIC$ in the reference system in which vectors $\robs$, $\vobs$ are defined:	
	\begin{eqnarray}
	\vec{v}_{\mathrm{orbit}} & = & \left\{ v_{\mathrm{r,LIC}} \cos\theta_{\mathrm{LIC}} - v_{\mathrm{t,LIC}}\sin\theta_{\mathrm{LIC}}, v_{\mathrm{r,LIC}} \sin\theta_{\mathrm{LIC}} + v_{\mathrm{t,LIC}}\cos\theta_{\mathrm{LIC}}, 0 \right\} \nonumber \nonumber \\
	v_{\mathrm{x,LIC}} & = & \vec{v}_{\mathrm{orbit}} \cdot \hat{x} \nonumber \\
	v_{\mathrm{y,LIC}} & = & \vec{v}_{\mathrm{orbit}} \cdot \hat{y} \nonumber \\
	v_{\mathrm{z,LIC}} & = & \vec{v}_{\mathrm{orbit}} \cdot \hat{z}.
	\label{eqVLICVec}
	\end{eqnarray}	
		
The velocity vector of the atom in the source region $\vLIC$ should be inserted into Equation~\ref{eqLocDiFun}. The analytical version of WTPM works in the ecliptic reference system, and in this case, with $\vec{v}_{\mathrm{B}}$, $\robs$, $\vobs$ defined in this system, no further transformations are needed. In the numerical version of WTPM, with a fully time- and location-dependent ionization rate, for which the natural reference plane is the solar equatorial plane, it is convenient to carry out the calculations in heliographic coordinates. Here, the initial vectors as well as the bulk velocity vector of interstellar gas relative to the Sun must first be transformed into heliographic coordinates (the non-rotating reference system based on the solar rotation axis as the $z$-axis is the heliocentric inertial reference system; \citet{burlaga:84a}). 

In the derivation above as well as in both versions of WTPM, we adopted a finite distance to the source region. In the classical hot model, this distance is set to infinity. If one wants to use this assumption, the only modification needed in the above formulae is to make a transition with $r_{\mathrm{LIC}} \rightarrow \infty$. Discussion of this assumption is presented in Section~\ref{sec:rFin}.	

In the current version of WTPM (both analytical and numerical) we use the analytic formulae presented in this section to calculate the velocity vector of the atom in the source region. In the previous versions, we tracked the atoms numerically. Numerical experiments showed, however, that using the analytic formulae gives more accurate results and with radiation pressure ineffective for helium, we do not have to address the complexities related to radiation pressure being variable with time and depending on radial velocity of the atom. In the fully numerical version of WTPM we still track the atoms numerically (i.e., we seek the full solution for the trajectory of the atom) to precisely take into account the time, latitude, and solar distance dependence of the ionization rate, as will be discussed in the next section. The numerical tracking results are used solely for this latter purpose of calculating the survival probabilities. Experience showed that because most of the losses occur relatively close to the Sun, the slow decay in precision of the numerical solution of the equation of motion does not severely degrade the accuracy of the ionization losses and the precision-setting parameters in the trajectory integration routine can be less stringent, thus enabling the program to run faster. 

\subsection{Calculation of survival probability}
\label{sec:calcSurPro}
Calculation of survival probability is one of the main differences between the two strains of WTPM. In the newly developed analytic version we strictly adhere to the assumptions of the classical hot model: we assume that the ionization rate is spherically symmetric and falls off with the square of the solar distance. As shown very early in the heliospheric studies \citep[e.g.][]{fahr:68a, axford:72}, the survival probability $w$ under these assumptions can be calculated from a simple formula
	\begin{equation}
	w = \exp\left[ -\beta_{0} r_{\mathrm{E}}^2 \Delta \theta / L \right],
	\label{eqW}
	\end{equation}
where $\beta_0$ is the ionization rate at $r_{\mathrm{E}}= 1$~AU from the Sun, $L$ is the angular momentum defined in Equation~\ref{eqEnerAngMom}, and $\Delta\theta$ is the angle swept by the atom on its way from $\vec{r}_{\mathrm{LIC}}$ to $\robs$. The latter can be calculated as
	\begin{equation}
	\Delta \theta = | \theta_{\mathrm{obs}} - \theta_{\mathrm{LIC}} |, 
	\label{eqDeltaTheta}
	\end{equation}
where $\theta_{\mathrm{obs}}$ is given by Equation~\ref{eqThetaObs} and $\theta_{\mathrm{LIC}}$ by Equation~\ref{eqThetaLIC}. 

In the full numerical version of WTPM, the survival probability is calculated numerically by solving the equation of motion supplemented with an additional term, which is equal to the time derivative of the exposure to ionization. The definition of exposure is given by \citet{bzowski_etal:13b} in Equation~3, and the formulation of the equation of motion with the additional term to calculate the survival probability by \citet{tarnopolski_bzowski:09} in Equation~3, where one must put the radiation pressure factor $ \mu = 0$. Details of the ionization rate used in the analytic version of WTPM are presented by \citet{bzowski_etal:13b} and for the current model of photoionization in \citet{sokol_bzowski:14a}; in brief, the local ionization rate is calculated for a given time moment and heliolatitude (i.e., the rate is assumed to be three-dimensional and time-dependent). More information is provided in Section~\ref{sec:ionEffects}.

The ionization rate model is organized on a 2D mesh in time and heliolatitude. The mesh pitch in time is the Carrington rotation period and in latitude $10\degr$. The total ionization rates (photo-, charge exchange, and electron rates, separately) are tabulated as a function of time and heliolatitude and bi-linearly interpolated for the required time and heliolatitude. To adjust the obtained rates for the solar distance, the dependence of individual rates on $r$ is subsequently folded in. In that way, an arbitrary evolution of the ionization rate with time, heliolatitude, and distance can be simulated. For validation and test purposes, the complex behavior of the ionization rate is simplified to conform to the assumptions of the classical hot model \citep{thomas:78}.

\subsection{Calculation of the differential flux on the sky}
\label{sec:diffFlux}
The calculation of the local distribution function, discussed in the preceding sections, is universal for many purposes, including the calculation of the moments (see Equation~\ref{eqDFMoments}) and the simulation of the flux observed by \emph{IBEX}-Lo. Calculation of the latter one, however, is specific because it must take the Galilean transformation between two reference systems. 

We have the \emph{IBEX} spacecraft located at the radius vector $\robs$, moving at a velocity $\vec{v}_{\mathrm{IBEX}}$ relative to the Sun. The latter velocity is, evidently, a sum of the Earth velocity relative to the Sun and the \emph{IBEX} velocity relative to the Earth. We want to calculate the differential flux of ISN~He atoms $\Phi\left(\psi,\alpha\right)$, which in the spacecraft-inertial reference system come from a direction determined in the spacecraft coordinate system by azimuth $\psi$ and elevation $\alpha$. This flux will be later used to calculate the flux transmitted through the collimator, i.e., integrated over a solid angle corresponding to the collimator FOV. Thus, the most convenient coordinates to express the differential flux are spherical. The velocity vector of the atom relative to the spacecraft is defined as 
	\begin{equation}
	\vec{u}_{\mathrm{rel}} = -u_{\mathrm{rel}} \left\{ \cos\psi \cos\alpha, \sin\psi \cos\alpha, \sin\alpha \right\}
	\label{eqUrel}
	\end{equation}
where $u_{\mathrm{rel}} > 0$ is the speed of the atom relative to the spacecraft. This vector must be rotated into the reference frame in which the atom tracking is performed, i.e., to the ecliptic reference frame. This is done by the transformation:	
	\begin{equation}
  \vec{u}_{\mathrm{rel}}^{\mathrm{ecl}} = \mathbf{M}_{\mathrm{IBEX} \to \mathrm{ecl}} \cdot \vec{u}_{\mathrm{rel}}
	\label{eqIBEX2ecl}
	\end{equation}
where $\mathbf{M}_{\mathrm{IBEX} \to \mathrm{ecl}}$ is the matrix of transformation from the \emph{IBEX} coordinates to ecliptic coordinates. The \emph{IBEX} coordinates are defined by the direction of the \emph{IBEX} spin-axis $\left( \lambda_{\mathrm{P}}, \phi_{\mathrm{P}} \right)$, which determines the $+z$-axis of the spacecraft coordinate system, and the spin-angle 0 point. The transformation matrix $\mathbf{M}_{\mathrm{IBEX} \to \mathrm{ecl}}$ is defined as follows: 
	\begin{equation}
  \mathbf{M}_{\mathrm{IBEX} \to \mathrm{ecl}} =  \left( 
  	\begin{array}{ccc}
  	-\cos\lambda_{\mathrm{P}}\sin\phi_{\mathrm{P}} & \sin\lambda_{\mathrm{P}} & \cos\lambda_{\mathrm{P}}\cos\phi_{\mathrm{P}} \\
  	-\sin\lambda_{\mathrm{P}}\sin\phi_{\mathrm{P}} & -\cos\lambda_{\mathrm{P}} & \cos\phi_{\mathrm{P}}\sin\lambda_{\mathrm{P}} \\
  	\cos\phi_{\mathrm{P}} & 0 & \sin\phi_{\mathrm{P}}
  	\end{array}
   \right) .
	\label{eqIBEX2eclTrans}
	\end{equation}		
The velocity of this atom relative the Sun $\vobs$ is given by the formula:		
	\begin{equation}
	\vec{v}_{\mathrm{obs}} = \vec{u}_{\mathrm{rel}}^{\mathrm{ecl}} + \vec{v}_{\mathrm{IBEX}}.
	\label{eqVobs}
	\end{equation}
To calculate the differential flux $\Phi \left( \psi, \alpha, t; \vec{\pi} \right)$ in the spherical coordinates we must calculate the integral:	
	\begin{equation}
	\Phi \left( \psi, \alpha, t; \vec{\pi} \right) = \int\limits_{u_{\mathrm{min}}}^{u_{\mathrm{max}}}{u_{\mathrm{rel}}f\left( \vec{r}_{\mathrm{obs}}, \vec{v}_{\mathrm{obs}} \left( \vec{u}_{\mathrm{rel}} \right), t; \vec{\pi} \right) u_{\mathrm{rel}}^{2} \mathrm{d}u_{\mathrm{rel}}}.
	\label{eqDiffFlux}
	\end{equation}		
In this equation we integrate over the relative speed of the atom and the spacecraft, but the distribution function is calculated for velocity $\vobs$ calculated from Equation~\ref{eqVobs} for a given spin-axis direction and $u_{\mathrm{rel}}$, $\psi$, and $\alpha$. The local distribution function is expressed in the solar inertial frame and defined in Equation~\ref{eqLocDiFun}. The integration is effectively along a curved path through velocity space in the solar-inertial reference frame. This path is defined by the fixed viewing direction $\psi$ and $\alpha$ and speed $u_{\mathrm{rel}}$, varying from $u_{\mathrm{min}}$ to $u_{\mathrm{max}}$ in the spacecraft inertial frame. The transformation from the spacecraft-inertial frame to the solar inertial frame is done analytically ``on the fly'' during the calculations, separately for each atom. This way, the effect of the velocity transformation on the differential flux is taken into account self-consistently and without any simplifications because we assume in the model that we know the source distribution function in front of the heliosphere accurately. 	

\subsubsection{Determination of the integration boundaries}
\label{sec:intBndr}
Specifying the integration boundaries $u_{\mathrm{min}}$ and $u_{\mathrm{max}}$ in Equation~\ref{eqDiffFlux} requires some attention. Formally, $u_{\mathrm{min}} = 0$ and $u_{\mathrm{max}} = \infty$. In practice, $u_{\mathrm{min}}$ represents the minimum velocity of an atom that is able to trigger the \emph{IBEX}-Lo instrument. In the modeling, we determine the integration boundaries individually for each simulation and each look direction $\left( \psi,\alpha \right)$ on the sky in a multi-tier refinement process.  

In the first step, the boundaries are determined approximately. The lower boundary is assessed starting from the realization that the slowest atom expected in the solar system at $\robs$ from the Sun follows a parabolic trajectory. Thus, its total energy in the solar-inertial frame is 0 and its speed relative to the Sun at $\robs$ is given by $\left( 2 GM/r_{\mathrm{obs}} \right)^{1/2}$. However, the direction of motion of this atom relative to the Sun is unknown; we only know its direction of motion relative to the moving \emph{IBEX} spacecraft. In practice, ISN~He atoms with the lowest possible energy are still well above the \emph{IBEX}-Lo energy threshold during the spring observations. However, during fall observations and for the wing of the Warm Breeze this threshold becomes important \citep{kubiak_etal:14a, galli_etal:15a, sokol_etal:15a}.     

To determine $u_{\mathrm{min}}$, we start by looking for the velocity vector of the atom in the spacecraft frame $\vec{V}_{\mathrm{a}}^{\mathrm{sc}} = V_{\mathrm{a}}^{\mathrm{sc}}\left\{ v_{\mathrm{a},x}^{\mathrm{sc}}, v_{\mathrm{a},y}^{\mathrm{sc}}, v_{\mathrm{a},z}^{\mathrm{sc}} \right\}$, where $V_{\mathrm{a}}^{\mathrm{sc}}$ is the speed for which we are searching, and $v_{\mathrm{a},i}^{\mathrm{sc}}$ are the directional coordinates of the atom velocity in the spacecraft frame that we know. We should solve the following equation:
	\begin{equation}
	\vec{V}_{\mathrm{a}}^{\mathrm{sc}} = \vec{V}_{\mathrm{sc}}^{\odot}-\vec{V}_{\mathrm{a}}^{\odot}.
	\label{eqUminShort}
	\end{equation}
$\vec{V}_{\mathrm{sc}}^{\odot} = \left\{ V_{\mathrm{sc},x}^{\odot}, V_{\mathrm{sc},y}^{\odot}, V_{\mathrm{sc},z}^{\odot} \right\}$ is the velocity vector of the spacecraft relative to the Sun (all quantities known), and $\vec{V}_{\mathrm{a}}^{\odot} = V_{\mathrm{a}}^{\odot}\left\{ v_{\mathrm{a},x}^{\odot}, v_{\mathrm{a},y}^{\odot}, v_{\mathrm{a},z}^{\odot} \right\}$ is the velocity vector of the atom relative to the Sun, for which we know only $V_{\mathrm{a}}^{\odot}$. It means that we should solve Equation~\ref{eqUminShort} in the following form
	\begin{equation}
	V_{\mathrm{a}}^{\mathrm{sc}}\left\{ v_{\mathrm{a},x}^{\mathrm{sc}}, v_{\mathrm{a},y}^{\mathrm{sc}}, v_{\mathrm{a},z}^{\mathrm{sc}} \right\} = \left\{ V_{\mathrm{sc},x}^{\odot},V_{\mathrm{sc},y}^{\odot},V_{\mathrm{sc},z}^{\odot} \right\} - V_{\mathrm{a}}^{\odot}\left\{ v_{\mathrm{a},x}^{\odot},v_{\mathrm{a},y}^{\odot},v_{\mathrm{a},z}^{\odot} \right\}
	\label{eqUminFull}
	\end{equation}
with an additional condition:
	\begin{equation}
	\sqrt{{v_{\mathrm{a},x}^{\odot}}^2+{v_{\mathrm{a},y}^{\odot}}^2+{v_{\mathrm{a},z}^{\odot}}^2}=1
	\label{eqUminAdd}
	\end{equation}
to get $V_{\mathrm{a}}^{\mathrm{sc}}$ (the speed of the atom with respect to the spacecraft). The formula resulting from Equation~\ref{eqUminFull} for the speed of the atom with respect to the spacecraft is the following:
	\begin{equation}
	V_{\mathrm{a}}^{\mathrm{sc}}=v_{\mathrm{a},x}^{\mathrm{sc}}v_{\mathrm{sc},x}^{\odot} + v_{\mathrm{a},y}^{\mathrm{sc}}v_{\mathrm{sc},y}^{\odot} + v_{\mathrm{a},z}^{\mathrm{sc}}v_{\mathrm{sc},z}^{\odot} \pm \sqrt{\left( v_{\mathrm{a},x}^{\mathrm{sc}}v_{\mathrm{sc},x}^{\odot} + v_{\mathrm{a},y}^{\mathrm{sc}}v_{\mathrm{sc},y}^{\odot} + v_{\mathrm{a},z}^{\mathrm{sc}}v_{\mathrm{sc},z}^{\odot} \right)^2 + {V_{\mathrm{a}}^{\odot}}^2 - \left( {V_{\mathrm{sc},x}^{\odot}}^2 + {V_{\mathrm{sc},y}^{\odot}}^2 +{V_{\mathrm{sc},z}^{\odot}}^2 \right)}
	\label{eqUminFin}
	\end{equation}
From Equation~\ref{eqUminFin} we obtain two solutions for $V_{\mathrm{a}}^{\mathrm{sc}}$ (positive and negative) and we take the positive one. We finish by taking the larger from the value thus obtained and the speed resulting from the pre-requisite energy sensitivity threshold.

To set the upper boundary $u_{\mathrm{max}}$, we require that the simulation does not miss more than $\Delta_{n}$ of the total population in front of the heliopause. In other words, we are potentially interested in atoms whose speed in the reference frame of the interstellar gas is inside a range $\left(0, U_{\mathrm{lim}}\right)$ obtained from the condition:
	\begin{equation}
	1-\Delta_{n} = \int\limits_{\mathrm{sphere}} {\mathrm{d}\Omega \int\limits_{0}^{U_{\mathrm{lim}}}{v^2 f_{\mathrm{LIC}}\left( v,\omega \right) \mathrm{d}v} },
	\label{eqUlimDef}
	\end{equation}	
where $v$ is the speed of the atom in the gas frame and $\omega$ is its direction of motion in this reference system. For interstellar gas moving at $v_\mathrm{B}$ relative to the Sun, the maximum allowable speed of an atom at infinity is $v_\mathrm{B} + U_{\mathrm{lim}}$, and at $r_{\mathrm{obs}}$ (from the conservation of energy): $u_{\mathrm{lim}} = \left( \left( v_\mathrm{B} +U_{\mathrm{lim}}\right)^2 + 2 GM/r_{\mathrm{obs}}\right)^{1/2}$. In practice, we require $\Delta_n = 10^{-5}$ for a Maxwellian distribution function, which results in a speed of the fastest atoms at $\sim1$~AU of $\sim62$~\kms~relative to the Sun. Since, similarly as for the lower boundary, only the speed relative to the Sun is known, and the direction is not, we repeat the procedure described for $u_{\mathrm{min}}$ to determine the maximum speed relative to \emph{IBEX} for a given direction $\left(\psi, \alpha\right)$.  

With the integration boundaries in the spacecraft frame tentatively determined, we refine them to reduce the calculation load. We profit from the fact that the integrand function in Equation~\ref{eqDiffFlux} features a single maximum in $u_{\mathrm{rel}}$ and is expected to asymptotically go to $0$ at least at the high end of its domain. Therefore we seek to further constrain the integration boundaries. We tabulate the integrand function from Equation~\ref{eqDiffFlux} between $u_{\mathrm{min}}$ and $u_{\mathrm{max}}$ in 34 equally spaced mesh points (with the step in relative speed equal to $\delta u$) and we calculate the first estimate of the integral defined in Equation~\ref{eqDiffFlux}. Subsequently, we test for the contributions of individual mesh points to the integral, going from the boundaries inward to the integration range and looking for the range for the mesh points inside which the relative contribution to the integral exceeds $1 - 0.001$. Having found these boundary points, we extend the range by $\delta u$ each way for safety (however, making sure we do not exceed the original boundaries $u_{\mathrm{min}}$, $u_{\mathrm{max}}$ determined above) and we end up with the refined integration boundaries $\left(u_{\mathrm{min,1}}, u_{\mathrm{max,1}}\right)$. 

Further integration from $u_{\mathrm{min,1}}$ to $u_{\mathrm{max,1}}$ is done using the trapezoidal rule, with the step $\delta u$ halved in each iteration until the integral varies by less than $0.001$ in aWTPM and $10^{-5}$ in nWTPM from one iteration to the following one. This procedure is repeated for each direction on the sky for which we wish to calculate the differential flux.

\begin{figure}
	\centering
	\begin{tabular}{c}
	\includegraphics[scale=0.5]{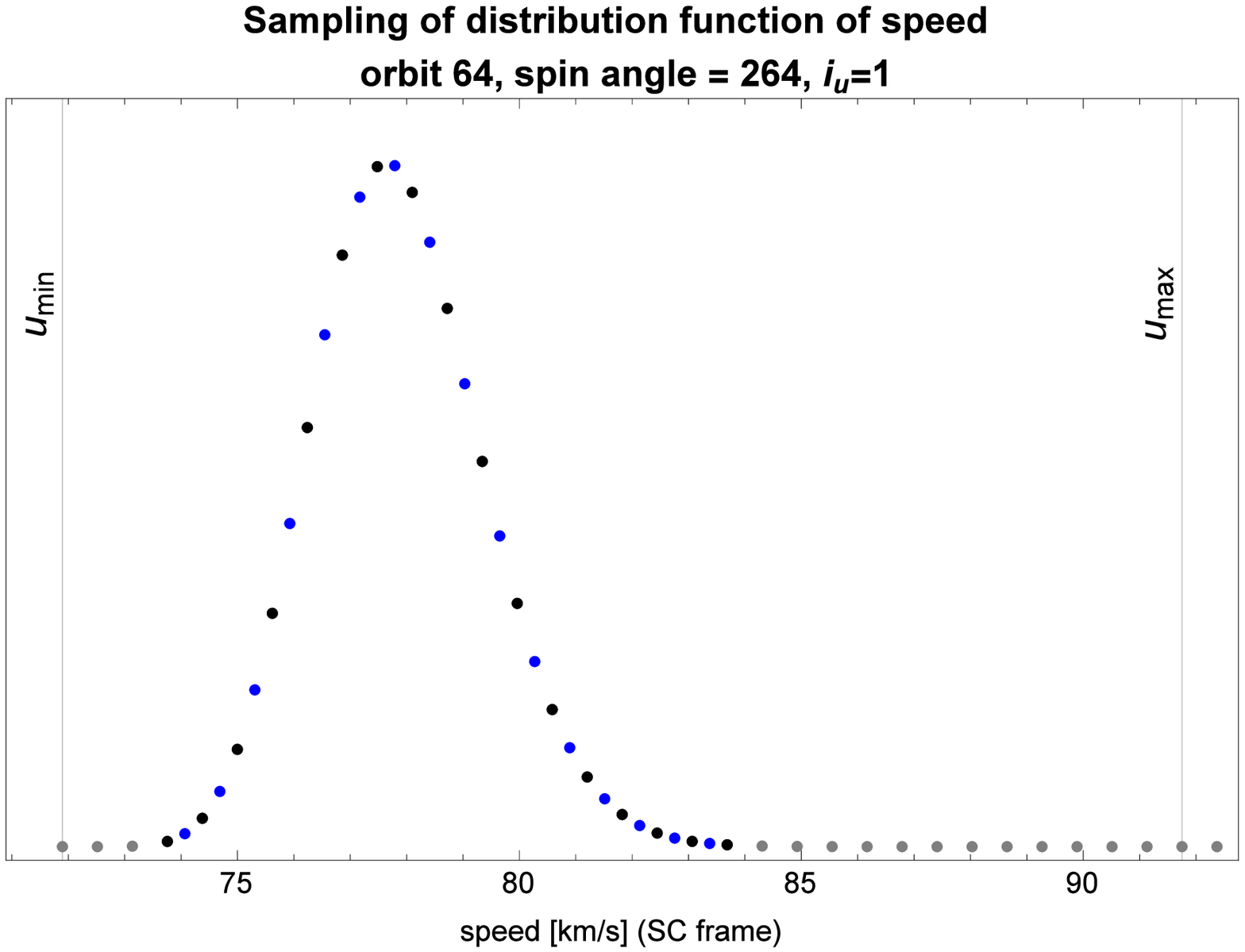}\\
	\includegraphics[scale=0.5]{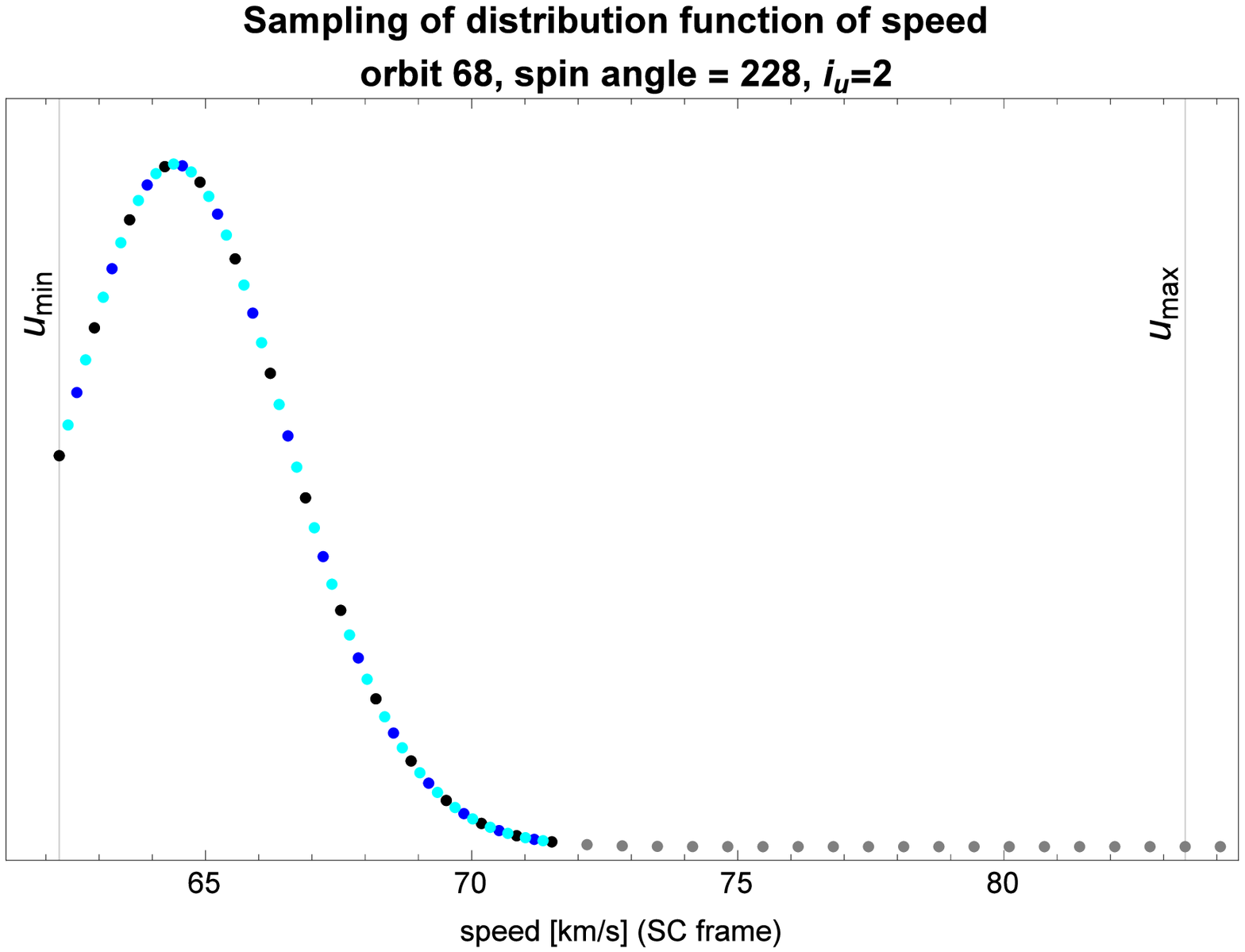}\\
	\end{tabular}
	\caption{Illustration of the integration boundary setting and integration step selection for two example cases of differential flux. Shown are the integrand functions in Equation~\ref{eqDiffFlux} for one selected look direction for orbits 64 (upper panel) and 68 (lower panel) as a function of atom speed in the spacecraft frame. The vertical bars represent the first guess for the integration boundaries, obtained from the application of Equation~\ref{eqUminFin} to calculate $u_{\mathrm{min}}$, $u_{\mathrm{max}}$. Gray dots represent the first division of the integration interval. The original integration region is subsequently narrowed to the region $\left(u_{\mathrm{min,1}},\, u_{\mathrm{max,1}}\right)$, occupied by the black dots. Blue dots represent a subdivision of one step further $\left(i_u = 1\right)$. This subdivision was sufficient to achieve the desired accuracy in the upper panel, but the lower panel required one more subdivision step, represented by cyan dots $\left(i_u = 2\right)$. The lower panel exemplifies a case where the integrand function is cut off at the lower boundary due to the parabolic speed limit, even though the function value at this boundary is not negligible. This is due to physical reasons, i.e., we reject atoms at elliptical orbits.}
	\label{figDistFunSpeedBoun}
	\end{figure}

In a typical case of parameters $\vec{\pi}$ of ISN~He gas, integration over the full speed range with a relative accuracy of $0.001$ requires just one subdivision of the original mesh in $u_{\mathrm{rel}}$. Thus, a typical step in the integration over speed is  $\delta u = \sim0.3$~\kms. In some cases, the number of subdivisions increases to 3 or 4. This happens mostly when the visible signal is close to the boundary of the FOV. An illustration of the integrand function for integration over speed and of the operation of the boundary and step selection logic is illustrated in Figure~\ref{figDistFunSpeedBoun}.

\subsection{Integration of the flux over the collimator}
\label{sec:collimTransmiss}
Integration of the differential flux over the collimator results in a flux  $F\left( \lambda_{\mathrm{P}},\phi_{\mathrm{P}},\psi,t;\vec{\pi} \right)$ (see Equation~\ref{eqPhiOrb}). The definition of the collimator-averaged flux is the following:
	\begin{equation}
	F\left( \lambda_{\mathrm{P}}, \phi_{\mathrm{P}}, \psi, t; \vec{\pi} \right) = \frac{\int\limits_{\mathrm{FOV}} {\Phi \left( \psi, \omega, t; \vec{\pi} \right) T\left( \omega \right) \mathrm{d}\Omega }}{ \int\limits_{\mathrm{FOV}}{T \left( \omega \right) \mathrm{d}\Omega}}
	\label{eqCollAvDef}
	\end{equation}
where $\psi$ is the spin-angle of the collimator axis, $\omega$ is the direction around the collimator axis, parameterized by the angle from the collimator axis $\rho$ and the anti-clockwise angle around the axis $\varphi$. $T(\omega)$ is the attenuation of the incoming atom flux as a function of the deviation of its direction from the boresight direction, and $\mathrm{d}\Omega$ is the solid angle differential.

Equation~\ref{eqCollAvDef} is a general formula. Its implementation in the code is different in the two versions of the program. It will be presented after the presentation of the adopted collimator transmission function, which follows. 

\subsubsection{Collimator transmission function}
The \emph{IBEX}-Lo collimator is composed of three quadrants: one high-resolution and three low-resolution (see Figure~3 in \citet{fuselier_etal:09b}). In the low-resolution observation mode, all four quadrants are active, while in the high-resolution mode only the high-resolution quadrant is active. The quadrants are built up as a hexagonal mesh so that the FOV of a given quadrant is hexagonal in shape. Linear dimensions of the low-resolution quadrants are identical and the orientation of all the hexagonal grids is the same. Thus the transmission functions of the three low-resolution quadrants are identical. 

Effectively, the transmission function is given by the formula
	\begin{equation}
	T\left( \rho, \varphi \right) = 3 S_{\mathrm{low}}T_{\mathrm{low}}\left( \rho, \varphi \right) + S_{\mathrm{high}}T_{\mathrm{high}}\left( \rho, \varphi \right),
	\label{eqCollTotal}
	\end{equation}
where $T_{\mathrm{low}}$ is the transmission function of the low-resolution quadrant, and $T_{\mathrm{high}}$ is the transmission function of the high-resolution quadrant. The coefficients $S_{\mathrm{low}}$ and $S_{\mathrm{high}}$ reflect the effective areas of the apertures of individual quadrants: $S_{\mathrm{low}} = 0.688$, which reflects the percentage of the total geometric area not obscured by the grid wires and $S_{\mathrm{high}} = 3/4 \times 0.617$, reflecting the smaller radial size of the high-resolution quadrant and the higher obscuration because of the finer mesh \citep{fuselier_etal:09b}. The angles $\rho$ and $\varphi$ are the angular distance from the boresight and the azimuth angle in the collimator FOV, respectively.
	
\begin{figure}
	\centering
	\resizebox{\hsize}{!}{\includegraphics{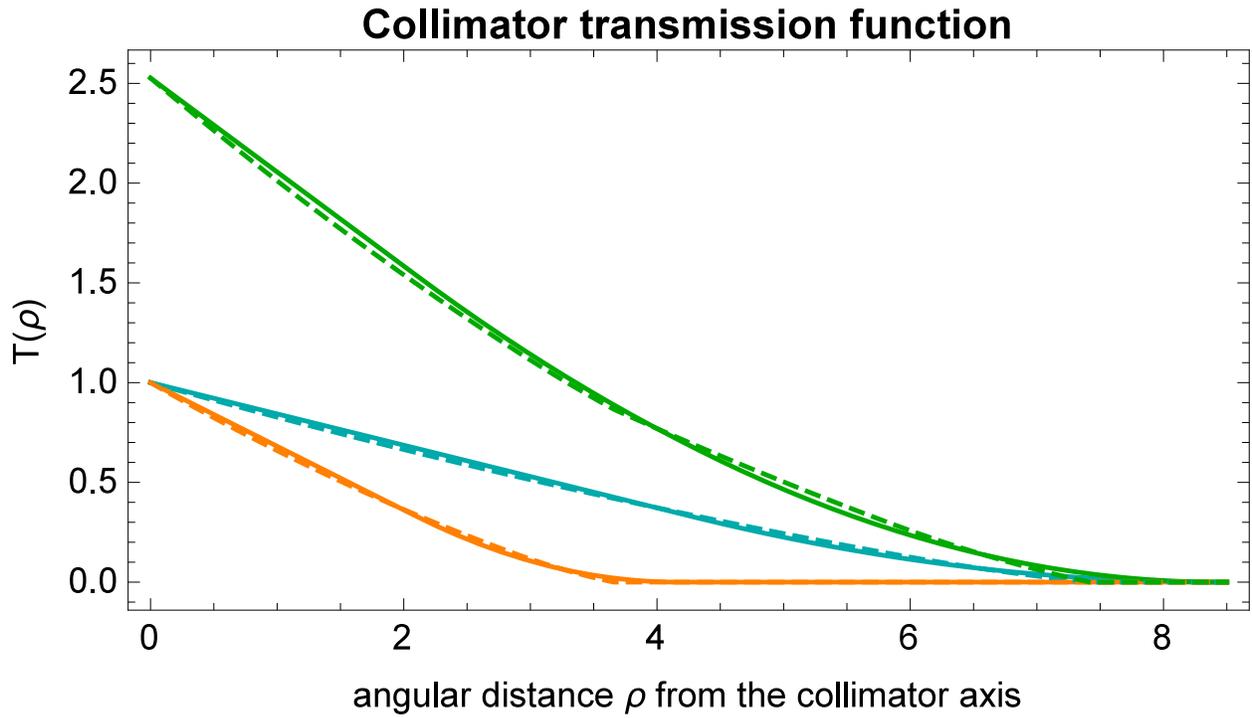}}
	\caption{Collimator transmission as a function of angular distance $\rho$ from the boresight for the high-resolution (orange) and low-resolution quadrants (blue) and the total transmission function obtained from Equation~\ref{eqCollTotal} (green). The solid lines correspond to the transmission along a line connecting the boresight with a corner of the field of view $(\overline{\varphi} = 0\degr)$ and the broken lines to the line connecting the boresight with the centers of the sides $(\overline{\varphi}= 30\degr)$.}
	\label{figCollTrans}
	\end{figure}

The collimator transmission was investigated before launch \citep[][see Figures~11 and 12]{fuselier_etal:09b} and is available at \texttt{http://ibex.swri.edu/ibexpublicdata/Data\_Release\_6/}. The numerical values for the transmission are given for both high- and low-resolution portions of the collimator for the radial lines connecting the boresight with the corner and the center of a side of the hexagonal collimator FOV. In our model, we approximated the transmission function by analytic formulae developed from simple geometric considerations based on the design of the collimator (see \citet{fuselier_etal:09b}, Figure~4): $T_{\mathrm{low,high}}\left(\rho,\varphi\right) = \tau \left(c_{\mathrm{low,high}}\tan\left(\rho\right),|\overline{\varphi}|\right)$, where $c_{\mathrm{low,high}}$ are coefficients equal to the ratio of the height of the collimator stack to the length of the edge of the hexagonal mesh. These ratios are known from the collimator calibration: $c_{\mathrm{low}}=13.47$, $c_{\mathrm{high}}=27.41$. The angle $\overline{\varphi} = \varphi - \varphi_{\mathrm{corner}}$, where $\varphi_{\mathrm{corner}}$ is the azimuth angle of the closest corner of the hexagonal mesh. The function $\tau\left(x,\overline{\varphi}\right)$ is given by the formula:
	\begin{equation}
	\tau\left( x, \overline{\varphi} \right) = \frac{1}{9} \left\{ 
	\begin{array}{ll}
	9-2\left( \sqrt{3}\sin\overline{\varphi} + 3\cos\overline{\varphi} \right)x + 2\sin\overline{\varphi}\left( \sqrt{3}\cos\overline{\varphi} - \sin\overline{\varphi} \right)x^2 & \mathrm{if} \,\, x\leq x_{b} \\
	12-12\cos\overline{\varphi}x+\left( 1+2\cos2\overline{\varphi} \right)x^2 & \mathrm{if} \,\, x_{b}<x\leq x_{e}\\
	0 & \mathrm{if} \,\, x>x_{e}	
	\end{array}
	\right.
	\label{eqColl}
	\end{equation}
where:
	\begin{displaymath}
	\begin{array}{c}
	x_{b} = \frac{3}{3\cos\overline{\varphi} - \sqrt{3}\sin\overline{\varphi}} \\
	x_{e} = \frac{6}{3\cos\overline{\varphi} + \sqrt{3}\sin\overline{\varphi}}
	\end{array}
	\end{displaymath}
A plot of the transmission function is presented in Figure~\ref{figCollTrans}, while the orientation of the FOV in the \emph{IBEX} reference system (i.e., the orientation relative to the scanning direction) is shown in Figure~3 in \citet{bzowski_etal:12a}.
	
\subsubsection{Integration over the collimator in the analytic version}
\label{sec:IntCollaWTPM}
Integration of the ISN~He flux over the collimator transmission function in the analytic version of the model is performed iteratively. The collimator FOV is divided into equal-area pixels according to the HealPix tessellation scheme with $N_{\theta} = 3$, $N_{\phi}  = 4$ \citep{gorski_etal:05a}. In this scheme, the sphere is divided into two symmetrical polar caps and an equatorial band. The division between the polar cap and equatorial band areas is such that their areas (solid angles) are identical. In our application, only the polar cap is relevant because its latitudinal range exceeds the angular radius of the collimator FOV. The polar cap is further split into four identical (and thus equal-area) lobes, which all meet at the pole. These lobes can be regarded as mega-pixels, which are further split into identical quadrants, i.e., smaller pixels. The subdivisions can further go as fine as needed. The centers of the pixels are located on rings that are parallel small circles on the sphere. Effectively, for $N_{\mathrm{side}}-1$ subdivisions, the whole sphere is covered with $N_{\mathrm{pix}} = 12 N_{\mathrm{side}}^2$ identical diamond-like pixels and $N_{\mathrm{side}}$ is referred to as the tessellation number. Necessarily, the area of a pixel in a given tessellation is equal to $\Delta\Omega_{N} = 4 \pi/\left(12 N_{\mathrm{side}}^2\right)$ and the sequence of tessellations follows the simple rule $N_{\mathrm{side}} = 2^k$, $k = 0,1,\ldots$.
	
\begin{figure}
	\centering
	\includegraphics[scale=0.6]{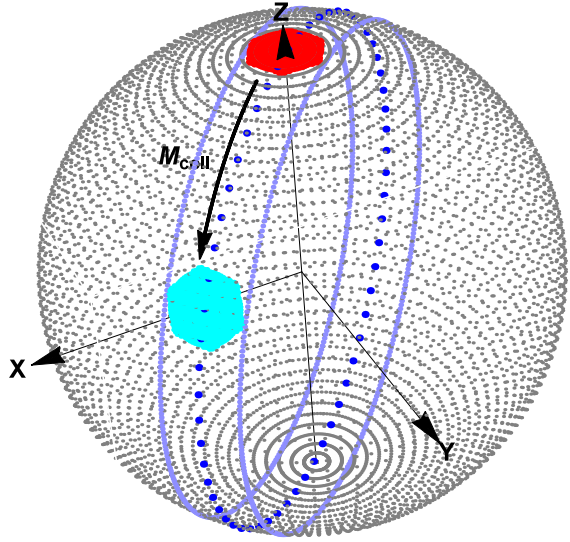}
	\caption{Illustration of positioning of the virtual collimator in the calculations done using the analytic version of WTPM. The hexagonal aperture is first mapped on the HealPix grid at the north ecliptic pole (red hexagon, actually composed of dots corresponding to the centers of individual pixels). Then the orientation of the sky strip scanned on a given orbit is selected by defining the spin axis coordinates $\left(\lambda_{\mathrm{P}},\phi_{\mathrm{P}}\right)$ in the selected celestial coordinate frame (here the ecliptic) centered at \emph{IBEX}. With this, the collimator boresight scans the great circle, sampling the sky at the points marked by the large blue dots. The blue solid circles represent the boundaries of the scanned strip. With the transmission function tabulated for the angular coordinates of the red dots, the virtual collimator is then rotated to one of its working positions, represented by spin-angle $\psi$ along the scanned strip, which corresponds to the ecliptic (longitude, latitude) $=\left(\lambda_{\psi},\phi_{\psi}\right)$. The rotation is effected by the transformation $\mathrm{\mathbf{M}}_{\mathrm{coll}}$, defined in Equation~\ref{eqMatrColl}. The collimator aperture in one of the working positions is marked by the cyan hexagon, which is composed of tessellation points actually used in the simulations.}
	\label{figCollTransMatr3D}
	\end{figure}

In the approach used in the analytic version of WTPM, we first put the collimator boresight in the north pole of the sphere and select the pixels that fill in the hexagonal FOV (see the red hexagon in Figure~\ref{figCollTransMatr3D}). Thus, for a given tessellation number, we have a fixed number $N_{\mathrm{pix}}$ of pixels that represent the collimator FOV. The transmission factors $T(\rho, \varphi)$ are pre-calculated for each pixel in all relevant tessellations and stored for a given tessellation as $T_i$, $i\in\{ 1,\ldots,N_{\mathrm{pix}}\}$. The coordinates of the pixel centers are stored as Cartesian unit vectors in a selected coordinate system. In aWTPM it is the ecliptic system, but in principle it can be any other system, e.g., heliographic or equatorial. To calculate the collimator transmission function for spin-axis pointing $\left(\lambda_{\mathrm{P}}, \phi_{\mathrm{P}}\right)$ and spin-angle $\psi$, which corresponds to the ecliptic longitude $\lambda_{\psi}$ and latitude $\phi_{\psi}$, the centers of the pixels of the collimator FOV are rotated using the following transformation:
	\begin{equation}
	\mathrm{\mathbf{M}}_{\mathrm{coll}} = \left(  
	\begin{array}{ccc}
	\sin\xi \sin\lambda_{\psi} - \cos\xi \cos\lambda_{\psi}\sin\phi_{\psi} & \cos\xi \sin\lambda_{\psi}+\cos\lambda_{\psi} \sin\xi\sin\phi_{\psi} & \cos\lambda_{\psi}\cos\phi_{\psi} \\
	-\cos\lambda_{\psi}\sin\xi - \cos\xi \sin\lambda_{\psi}\sin\phi_{\psi} & \sin\xi \sin\lambda_{\psi}\sin\phi_{\psi} - \cos\xi \cos\lambda_{\psi} & \cos\phi_{\psi}\sin\lambda_{\psi} \\
	\cos{\xi}\cos\phi_{\psi} & -\cos\phi_{\psi}\sin\xi & \sin\phi_{\psi}
	\end{array}
	\right),
	\label{eqMatrColl}
	\end{equation}	
where $\xi = 15\degr$ is the inclination angle of the hexagonal FOV to the center line of the visibility strip on the sky. This gives the coordinates of the pixels for the selected spin-axis pointing and spin-angle (see the cyan hexagon in Figure~\ref{figCollTransMatr3D}). We denote the list of these positions as $\omega_i = \left(\psi_i, \alpha_i\right)$, $i\in\left\{1,\ldots, N_{\mathrm{pix}}\right\}$. They make a list of directions for which we will calculate the differential flux $\Phi\left(\omega,t;\vec{\pi} \right)$, defined in Equation~\ref{eqDiffFlux}, to be averaged over the collimator FOV.	

With the virtual collimator appropriately positioned on the sky, we calculate an approximation to the collimator-averaged flux $F^{(N_{\mathrm{side}})}\left( \lambda_{\mathrm{P}},\phi_{\mathrm{P}},\psi,t;\vec{\pi} \right)$ based on Equation~\ref{eqCollAvDef} using the following sum:	
	\begin{equation}
	F^{\left( N_{\mathrm{side}} \right)} \left( \lambda_{\mathrm{P}}, \phi_{\mathrm{P}}, \psi, t; \vec{\pi} \right) = \frac{\sum_{i=1}^{N_{\mathrm{pix}}}{T_{i}\Phi \left( \psi, \Omega_{i},t; \vec{\pi} \right)} }{\sum_{i=1}^{N_{\mathrm{pix}}}{T_i}}.
	\label{eqJSFAppr}
	\end{equation}	
Starting from tessellation $N_{\mathrm{side}} = 2^4$, we iterate calculating $F^{(N_{\mathrm{side}})}$, increasing $k$ by one (thus effectively quadrupling the total number of pixels), until $|F^{(2 N_{\mathrm{side}})}/F^{(N_{\mathrm{side}})} -1| < 0.01$: when this condition is fulfilled, we consider the collimator-averaged flux as successfully converged and adopt the result as $F\left(\lambda_{\mathrm{P}},\phi_{\mathrm{P}},\psi,t;\vec{\pi} \right)=F^{(2N_{\mathrm{side}})}$.

Examples of the collimator transmission function $T$, the differential flux $\Phi$, and their products $\Phi T$ are shown in Figure~\ref{figCollTransmissionFluxGrid} for three example orbits: 61 (i.e., before the yearly peak of the ISN~He signal observed by \emph{IBEX}), 64 (the peak orbit), and 68 (well after the peak). 

\subsubsection{Integration over the collimator in the numerical version}
\label{sec:IntCollnWTPM}
Integration of the ISN~He flux over the collimator transmission function in the numerical version of WTPM is carried out in a totally different way. First, the differential flux $\Phi\left(\psi,\Omega,t;\vec{\pi}\right)$, given by Equation~\ref{eqDiffFlux}, is tabulated within the whole visibility strip of the sky for a given time $t$ and spin-axis orientation $\left(\lambda_{\mathrm{P}},\phi_{\mathrm{P}}\right)$. The tabulation is done on a regular mesh in the heliographic spherical coordinates, with constant pitch in each coordinate, in a two-step process. First, the differential flux $\Phi$ is calculated from Equation~\ref{eqDiffFlux} with a pitch of $0.703125\degr$ in each coordinate. Then, the mesh is further subdivided using bi-cubic interpolation so that the flux is tabulated with a constant pitch of $0.703125\degr/4 = 0.17578125\degr$, and its coordinates are converted to the spacecraft coordinates (spin-angle and elevation). Now, the virtual collimator boresight is put to a spin-angle $\psi$ and the differential flux points within the angular radius of the collimator FOV are selected. Subsequently, the coordinates of the tabulated differential flux are converted to the collimator coordinates $\left(\rho,\varphi\right)$. The collimator coordinates make a spherical reference system, with the north pole corresponding to the collimator boresight at the spacecraft coordinates $\left(\psi,0\right)$. With this, integration over the collimator FOV begins, starting from the general formula for integration in the spherical coordinates:
	\begin{equation}
	F^{\left( N_{\mathrm{side}} \right)} \left( \lambda_{\mathrm{P}}, \phi_{\mathrm{P}}, \psi, t; \vec{\pi} \right) = \frac{\int\limits_{\mathrm{FOV}}{\Phi\left( \psi, \rho, \varphi,t; \vec{\pi} \right)T\left( \rho, \varphi \right)\sin\rho \mathrm{d}\rho \mathrm{d}\varphi}}{\int\limits_{\mathrm{FOV}}{T\left( \rho, \varphi \right) \sin\rho \mathrm{d}\rho \mathrm{d}\varphi}}.
	\label{eqMQCollimInt}
	\end{equation}
The integration is done numerically. 

The collimator FOV is split into equal-area pixels defined in the collimator coordinates. Note that these pixels have nothing to do with the HealPix pixels discussed in the former section. The collimator aperture is first divided in radial distance into two parts, with division at $\rho'=\sim4.5\degr$. The inner part is then subdivided into $\left(\Delta\varphi,\Delta\rho\right)$ sectors, with $\Delta\varphi = 7.5\degr$. In the radial direction, the mesh boundaries are defined so that $\cos\rho_i = 1 -\frac{i}{n} \left(1 - \cos R\right)$, where $R = 9.0\degr$ is the maximum angular radius of the aperture. For the region at $\rho' > 4.5\degr$, $\Delta\varphi = 3.75\degr$ and $\cos \rho_i=1-\frac{2i-i_{\mathrm{end}}}{n}\left(1-\cos R \right)$, with $i_{\mathrm{end}} = 20$. The exact value for $\rho'$ is calculated from the equation $\cos\rho'=1-\frac{i'}{n} \left(1-\cos R\right)$, where $i'$ is the lowest value of $i$, for which $\cos \rho' \geq \cos 4.5\degr$. All pixels have equal areas, equal to $S = \Delta\varphi \left(\cos \rho_i  - \cos \rho_{i+1}\right) \left(\pi/180\right)$.

The contribution from one sector $n$ of the virtual collimator is calculated as 
	\begin{equation}
	F_{\mathrm{S},n} = \sum_{i=1}^{N_{i}}{\Phi\left( \rho_{i}, \varphi_{i} \right)T\left( \rho_{i},\varphi_{i} \right)}; \quad T_{{\mathrm{S},n}}=\sum_{i=1}^{N_i}{T\left( \rho_i,\varphi_i\right)},
	\label{eqCollSector}
	\end{equation}	
where $N_i$ is the number of flux points that are inside the collimator sector, $\left(\rho_i,\varphi_i\right)$ are collimator coordinates of the $i$th flux point, $T$ is the collimator transmission function defined in Equation~\ref{eqCollTotal}, and $\Phi\left( \rho_i,\varphi_i\right)$ is the differential flux of ISN~He defined in Equation~\ref{eqDiffFlux} and calculated for the coordinates corresponding to the collimator coordinates $\left(\rho_i,\varphi_i \right)$.

The full collimator-averaged flux $F$ is calculated as  
	\begin{equation}
	F = \frac{\sum_{n=1}^{N}{F_{\mathrm{S},n}}}{\sum_{n=1}^{N}{T_{\mathrm{S},n}}}.
	\label{eqMQCollimIntQuadrature}
	\end{equation}
In the case that the regular sector exceeds the hexagonal perimeter of the aperture, it enters the calculation with a weight $k/n$, where $k$ is the number of differential flux elements that belong to the portion of the sector that is inside the aperture. 	

The method of calculating the collimator-integrated flux in the numerical version of WTPM may seem much more complex than the method used in the analytic version regarding the calculation over the collimator FOV. However, this method works fine within the computation framework implemented on a computer cluster. Calculating the differential flux is the most computationally demanding portion of the entire simulation task and thus, to enable performing parameter fitting in a reasonable time, must be parallelized. To maintain balance between the development effort and the calculation time, the most practical way turned out to be organizing the calculations of the differential flux by separate instances of the program, launched in separate cluster cores. This, however, hampers cross-talk between results of calculations of individual differential flux values, so it is practical to tabulate the differential flux for a given time moment and different directions on the sky. If the tabulation is not sufficiently dense, it can be refined by interpolation, computationally much less demanding. A benefit of such an organization of calculations is that with the differential flux tabulated for the whole \emph{IBEX}-Lo visibility strip one can select the boresight of the collimator arbitrarily without too much of additional effort, which facilitates an efficient calculation of the flux averaged over spin-angle bins. This latter step is the subject of the following section. 

We have verified that the methods described in the present and preceding sections return results that agree within $1\%$ for identical parameters and ionization models.

\subsection{Integration of the flux over the spin-angle bins}
\label{sec:spinBinMethod}
As shown, e.g., by \citet[][Figures~7 and 8]{bzowski_etal:12a}, the signal from the ISN~He gas is expected to be close to a Gaussian function as a function of spin-angle. Since our simulations must reproduce the signal averaged over $\Delta\psi=6\degr$ spin angle bins, the curvature of the collimator-averaged flux $F\left(\psi\right)$, defined in Equation~\ref{eqMQCollimInt}, must be appropriately taken into account. This should be done by taking average values over the $6\degr$ bins: 
	\begin{equation}
	\langle F\left( \psi_{k} \right) \rangle_{\Delta \psi} = \int\limits_{\psi_{k} - \Delta \psi /2}^{\psi_{k} + \Delta \psi /2}{F \left( \psi \right) \mathrm{d}\psi} / \Delta \psi
	\label{eqSpinAngleAverDef}
	\end{equation}
where $\psi_k$ is the spin-angle of the center of the $k$th bin.	

For the pixels where $F\left(\psi\right)$ is almost linear, simply taking the middle value for the bin may be sufficient. However, the width of the signal is just a few $6\degr$ bins, and in practice, the curvature of the signal inside the bins does play a role, varying from orbit to orbit and from bin to bin. We analyzed the behavior of the simulated signal by comparing results of the numerical integration of the signal tabulated every 1/8 of a degree and integrated over $6\degr$ bins using the trapezoidal rule with results of integration by polynomial quadratures of various orders on much less dense mesh. We found that maintaining a $1\%$ accuracy requires tabulating the flux every $1.5\degr$ in spin-angle and approximating the signal within a bin by a polynomial of the fourth order. This polynomial is then analytically integrated within the boundaries of a given bin, which results in a quadrature.

The formula for the signal averaged over a $6\degr$ bin in spin-angle $\langle F \rangle_{\Delta \psi}$  is the well-known Boole's rule:
	\begin{equation}
	\langle F \rangle_{\Delta \psi} = \left( 7 F_{1} + 32 F_{2} + 12F_{3}+32F_{4}+7F_{5} \right)/90 
	\label{eqSpinAngleAver}
	\end{equation}
where $F_3$ is the collimator-averaged flux simulated for the center of the bin and the other $F_i$ are the flux simulated for the consecutive points inside the bin, spaced by $1.5\degr$ of spin-angle. $F_1$ and $F_5$ correspond to the boundaries of the bin and thus can be reused in the calculation of the bin-averaged flux in the neighboring bins. This formula is used in both versions of WTPM.

\subsection{Integration of the flux over good time intervals}
\label{sec:timeIntegr}
Similarly as in the case of the integration over the bins, the integration over the good time intervals is carried out using quadratures. We found that sufficiently accurate results are obtained when one tabulates the collimator- and bin-integrated flux with a 0.5 day pitch over the High Altitude Science Operations (HASO) interval and uses the fourth order polynomial quadrature. An important difference in comparison with integrating over spin-angle, however, is in the integration boundaries: good time intervals vary from season to season and orbit to orbit. Thus, one needs to calculate the coefficients of the approximating polynomials to obtain indefinite integrals and then to evaluate them in the boundaries defined by the boundaries of actual good time intervals. Thus, there is no prerequisite that the integration boundaries conform with the boundary points of the quadrature.

Denoting $F_{t_i}$ the collimator- and spin bin-integrated flux for a time $t_i$, we take five equidistant time steps $t_1,\ldots,t_5$, with $\delta t = t_{i+1} - t_i = 0.5$ day (the time for this calculation is converted into days since the beginning of a given orbit) and calculate $F_{t_1}$, $F_{t_2}$, $F_{t_3}$, $F_{t_4}$, $F_{t_5}$. With them, we define the polynomial $P(t)$ approximating the flux for the time interval $\left(t_1, t_5\right)$ as 
	\begin{equation}
	P\left( t \right) = A t^4 + B t^3 + C t^2 + D t + E
	\label{eqFluxTPoly}
	\end{equation}
and we calculate the coefficients from the following formulae:
	\begin{equation}
	\begin{array}{lll}
	A & = & F_{t_1} - 4 F_{t_2} + 6 F_{t_3} - 4 F_{t_4} + F_{t_5} \nonumber \\
	B & = & 2 \left( \delta t \left( -F_{t_1} + 2 F_{t_2} - 2 F_{t_4} + F_{t_5} \right) - 2 \left( F_{t_1} - 4 F_{t_2} + 6 F_{t_3} - 4 F_{t_4} + F_{t_5} \right) t_3 \right) \nonumber \\
	C & = & \delta t^2 \left( -F_{t_1} + 16 F_{t_2} - 30 F_{t_3} + 16 F_{t_4} - F_{t_5} \right) + t_3  \left( 6 \delta t \left( F_{t_1} - 2 F_{t_2} + 2 F_{t_4}+F_{t_5} \right) + {} \right. \nonumber \\
	& & {} \left. + \left( 6 F_{t_1} - 24 F_{t_2} + 36 F_{t_3} - 24 F_{t_4} + 6 F_{t_5} \right) t_3 \right) \nonumber \\
	D & = & \delta t^3 \left( 2 F_{t_1} - 16 F_{t_2} + 16 F_{t_4} - 2 F_{t_5} \right) + t_3 \left( \delta t^2 \left( 2 F_{t_1} - 32 F_{t_2} + 60 F_{t_3} - 32 F_{t_4} + 2 F_{t_5} \right) + {} \right. \nonumber \\
	& & {} \left. + t_3 \left( \delta t \left( -6 F_{t_1} + 12 F_{t_2} - 12 F_{t_4} + 6 F_{t_5} \right) + \left( -4 F_{t_1} + 16 F_{t_2} - 24 F_{t_3} + 16 F_{t_4} - 4 F_{t_5} \right) t_3 \right) \right) \nonumber \\
	E & = & 24 \delta t^4 F_{t_3} + t_3 \left( \delta t^3 \left( -2 F_{t_1} + 16 F_{t_2} - 16 F_{t_4} + 2 F_{t_5} \right) + t_3 \left( \delta t^2  \left( -F_{t_1} + 16 F_{t_2} - 30 F_{t_3} + 16 F_{t_4} - F_{t_5} \right) + {} \right. \right. \nonumber \\
	 & & {} \left. \left. + t_3 \left( \delta t \left( 2 F_{t_1} - 4 F_{t_2} + 4 F_{t_4} - 2 F_{t_5} \right) + \left( F_{t_1} - 4 F_{t_2} + 6 F_{t_3} - 4 F_{t_4} + F_{t_5} \right) t_3 \right) \right) \right). 
	\end{array}
	\label{eqPolyCoeff}
	\end{equation}

With the coefficients calculated, we can integrate Equation~\ref{eqFluxTPoly} over time, obtaining an indefinite integral in the form of a polynomial of the fifth order, and substitute for time $t$ the integration boundaries $t_{\mathrm{GT1},i}$, $t_{\mathrm{GT2},i}$ of the $i$th good time interval for a given orbit. These are denoted as $I_{\mathrm{GT1},i}$, $I_{\mathrm{GT2},i}$:
	\begin{equation}
	\begin{array}{lll}
	I_{\mathrm{GT1},i} & = & \left( t_{\mathrm{GT1,i}} \left( E + t_{\mathrm{GT1,i}} \left( D/2 + t_{\mathrm{GT1,i}} \left( C/3 + t_{\mathrm{GT1,i}} \left( B/4 + \left( A t_{\mathrm{GT1,i}} \right) / 5 \right) \right) \right) \right) \right) \nonumber \\
	I_{\mathrm{GT2},i} & = & \left( t_{\mathrm{GT2,i}} \left( E + t_{\mathrm{GT2,i}} \left( D/2 + t_{\mathrm{GT2,i}} \left( C/3 + t_{\mathrm{GT2,i}} \left( B/4 + \left( A t_{\mathrm{GT2,i}} \right) / 5 \right) \right) \right) \right) \right) \nonumber \\	
	\end{array}
	\label{eqPolyInt}
	\end{equation}
and finally the flux integrated over the good time interval $i$ takes the form:
	\begin{equation}
	\langle F \rangle _{GT,i} = \left( I_{\mathrm{GT2},i} - I_{\mathrm{GT1},i} \right) / \left( 24 \delta t^4 \right).
	\label{eqOneGT}
	\end{equation}
	
If the initial tabulation does not cover the whole orbit, the missing interval is covered with another set of five equidistant times, starting from the previous time $t_5$, and the procedure described by Equations~\ref{eqPolyCoeff} through \ref{eqOneGT} continues. Ultimately, we have the flux integrated over all $N_t$ intervals of good times for a given orbit and we calculate the flux averaged over spin-angle bin $k$ and all good times from the formula:
	\begin{equation}
	\langle F \left( \lambda_{\mathrm{P}}, \phi_{\mathrm{P}}, \psi_{k}; \vec{\pi} \right) \rangle _{\Delta \psi, \mathrm{GT}} = \frac{\sum_{i=1}^{N_t}{\langle F \rangle _{\mathrm{GT},i}}}{\sum_{i=1}^{N_t}{\left( t_{\mathrm{GT2},i} - t_{\mathrm{GT1},i} \right)}}
	\label{eqAvrFlux}
	\end{equation}	
Tabulating the bin-averaged flux with a 0.5 day step implies that the orbital arc is at least 2.5 days long. In a few cases when the HASO time for an orbit was shorter, we use the three-point quadrature, with approximating a polynomial of second order. 	
	
Numerical experiments showed that using this complex scheme is needed when one accounts for the spacecraft motion relative to the Earth, as is discussed in detail in Section~\ref{sec:scmotion}. The relevant effects are presented in Figure~\ref{figEffectsTimeInt}.

Equation~\ref{eqAvrFlux} gives the collimator-, spin-angle-, and good-times- averaged flux in physical units. To compare this flux with observations, we must rescale it so that it represents the collimator-, spin-angle-, and good-times-averaged count rate in individual bins for a given orbit. This procedure is presented in the following section, with no need to refer to the absolute calibration of the instrument.

\subsection{Rescaling the averaged flux from physical units to count rate}
In the absence of background, the count rate $c_k$ for a given spin-angle bin $k$, averaged over good time intervals for a given orbit, is directly proportional to the time-, spin-angle-, and collimator-averaged flux $F_k=\langle F\left( \lambda_{\mathrm{P}},\phi_{\mathrm{P}},\psi_{k};\vec{\pi} \right) \rangle_{\Delta\psi,\mathrm{GT}}$ from Equation~\ref{eqAvrFlux}, calculated for a parameter set $\vec{\pi}$. The proportionality coefficient $a$ is constant for a given observation season. It depends on details of the instrument setting and sensitivity, and on the energy of the atoms, which depends on the adopted parameter set $\vec{\pi}$. Given the simulated flux values calculated from Equation~\ref{eqAvrFlux} and observed count rates $c_k$, $k = \left\{1,\ldots,N_{\mathrm{data}} \right\}$, where $N_{\mathrm{data}}$ is the total number of $6\degr$ bins taken for the analysis from all orbits for a given observation season, we find $a$ by analytical minimization of $\chi^2$:
	\begin{equation}
	\chi^2\left( a \right) = \sum_{j=1}^{N_{\mathrm{data}}}{\sum_{i=1}^{N_{\mathrm{data}}}{\left( a F_i - c_i \right) \left( a F_j - c_j \right) w_{ij}}}.
	\label{eqChiSq}
	\end{equation}	
In this equation, $w_{ij}$ is the element of a matrix $\mathrm{\mathbf{W}}$ being the inverse covariance matrix for the data (for details see \citet{swaczyna_etal:15a}, this issue). Equation~\ref{eqChiSq} is a simple quadratic function of $a$. Thus, it takes the minimum value for $a$ equal to:	
	\begin{equation}
	a = \frac{\sum_{j=1}^{N_{\mathrm{data}}}{\sum_{i=1}^{N_{\mathrm{data}}}{ w_{ij} \left( F_i c_j + F_j c_i \right)}}}{2 \sum_{j=1}^{N_{\mathrm{data}}}{\sum_{i=1}^{N_{\mathrm{data}}}{w_{ij}F_i F_j}}},
	\label{eqScaleFact}
	\end{equation}	
which we adopt as the scaling factor to convert the simulated flux to the observed count rate. Basically, scaling the simulated flux to the observed count rate is a portion of searching for an optimum parameter set $\vec{\pi}$. We describe it here because it must be done before the simulated flux can be compared with the data and because it can be done analytically, in contrast to searching for the values of the parameters $\vec{\pi}$ of the assumed distribution function. 	

\subsection{Outlook and summary of model description}
\label{sec:Outlook}
Two potentially significant effects are currently left out of the model. One of them is the possible sensitivity of the registered count rate due to the energy of the helium atom impacting the conversion surface and the distribution of the sputtered products, as the He is not observed directly by \emph{IBEX}-Lo \citep{wurz_etal:08a}. The other is a small perturbation of the atom trajectories by the Earth's gravity. Both of them are the subject of research \citep[][this issue, respectively]{galli_etal:15a, kucharek_etal:15a}. The first one is approximated in the present version of our model by adopting a sharp threshold in the low boundary of integration over speed \citep[see the discussion by][]{sokol_etal:15a}, the other one was shown by \citet{kucharek_etal:15a} to be potentially important mostly during fall seasons of ISN observations when the atom impact energy is so low that they are not visible for \emph{IBEX}-Lo anyway \citep{galli_etal:15a}. Including them in WTPM is possible and will be done if it is proved that it is needed. 

Table~\ref{tab:compWTPM} summarizes the description of the analytic and numerical versions of the WTPM. The similarities and differences are gathered by the elements of the model to simulate the ISN gas in the heliosphere. Most of the parts are general with application to any detection/observation scheme and some have special application to \emph{IBEX} (see more in \citet{bzowski_etal:15a}; \citet{swaczyna_etal:15a}). 
	\clearpage
	\begin{table}
	\begin{centering}
	\caption{Comparison resume of aWTPM and nWTPM \label{tab:compWTPM}}
		\begin{tabularx}{\textwidth}{>{\hsize=0.5\hsize}X|>{\hsize=1\hsize}X|>{\hsize=1\hsize}X}
		\tableline
		 & aWTPM & nWTPM  \\ \tableline
		Code language & \texttt{Wolfram Research Mathematica} & \texttt{Fortran} and \texttt{C} \\ \tableline
		Adopted model of gas & classical hot model & hot model with variable ionization \\ \tableline
		Distribution function in the LIC & \multicolumn{2}{>{\hsize=2\hsize}X}{Single Maxwell--Boltzmann distribution, but any other can be easily adopted} \\ \tableline
		Ionization & photoionization + charge exchange  + electrons, at the time of detection, for the ecliptic plane, 
with instantaneous values for the calculation moment, $1/r^2$, available via Data Release~9 & photoionization + charge exchange  + electrons, for the current position at the atom's trajectory (time, distance, latitude),  
variable in time; other models can be applied  \\ \tableline
		Detector position & \multicolumn{2}{>{\hsize=2\hsize}X}{Exact \emph{IBEX} spin-axis, location in space, velocity, and position \citep{schwadron_etal:15a, swaczyna_etal:15a}; any other can be easily incorporated} \\ \tableline
		Initial conditions for atom orbit calculation set in the S/C frame & The state vector in the LIC is calculated analytically, and the result is used to obtain both the distribution function value and the survival probability & The state vector in the LIC is calculated analytically, and the result is used to obtain value of the distribution function in the LIC; survival probability is calculated from numerical atom tracking in the space- and time-variable ionization environment \\ \tableline
		Stop distance for atom tracking & Fixed, currently set to 150 AU; can use anything up to infinity & Fixed, currently set to 150 AU for the Maxwell--Boltzmann term; stop when 150 AU is slightly exceeded for the survival probability calculations; tested up to $\sim5000$~AU\\ \tableline
		\end{tabularx}
	\end{centering}
	\end{table}
	
	\begin{table*}
	\begin{centering}
	\caption{Table~\ref{tab:compWTPM}, continued.}
		\begin{tabularx}{\textwidth}{>{\hsize=0.5\hsize}X|>{\hsize=1\hsize}X|>{\hsize=1\hsize}X}
		\tableline
		 & aWTPM & nWTPM  \\ \tableline
		Differential flux calculations & \multicolumn{2}{>{\hsize=2\hsize}X}{Integrated in the SC reference frame; integration boundaries for atom speed are selected individually for each direction on the sky and calculate iteratively using the trapezoidal rule; boundaries are selected so that (1) only hyperbolic orbits are allowed and (2) $\Delta_n = 10^{-5}$ of the atoms in the LIC are potentially excluded ($\sim4.5\sigma$ included); can implement a finite energy sensitivity threshold} \\ \tableline
		Absolute scaling & \multicolumn{2}{>{\hsize=2\hsize}X}{Calculations done in physical units} \\ \tableline
		Collimator response function & \multicolumn{2}{>{\hsize=2\hsize}X}{Analytical function based on the pre-flight calibration (Equation~\ref{eqCollTotal} and Figure~\ref{figCollTrans}); other functions can be applied} \\ \tableline
		Integration over collimator & Signal integration for a given orbit, time moment, and spin-angle of the collimator boresight, using HealPix tessellation, iterated with increasingly fine resolution until convergence; differential flux for each HealPix pixel is calculated ``on the fly'' (Section~\ref{sec:IntCollaWTPM}) & Entire visibility strip for a given orbit and time moment first tabulated at a fixed grid in the heliographic spherical coordinates, subsequently interpolated to a finer mesh using a bi-quadratic interpolation; this map is subsequently integrated for each desired spin-angle pointing of the collimator, using a different scheme than in aWTPM (Section~\ref{sec:IntCollnWTPM}) \\ \tableline
		Calculation of flux for $6\degr$ bin & \multicolumn{2}{>{\hsize=2\hsize}X}{Calculation by Boole's rule with sampling with a  $1.5\degr$ step (Equations~\ref{eqSpinAngleAverDef} and \ref{eqSpinAngleAver}); any other scheme can be easily applied}  \\ \tableline
		Sampling in time & Central HASO time per orbit, but any other can be applied at a cost of an increase of computational time; any time integration scheme can be applied &	Integration over good time intervals using a polynomial method (Equations~\ref{eqFluxTPoly} through \ref{eqAvrFlux}); any time integration scheme can be applied\\ \tableline
		\end{tabularx}
	\end{centering}
	\end{table*}

	\begin{table*}
	\begin{centering}
	\caption{Table~\ref{tab:compWTPM}, continued.}
		\begin{tabularx}{\textwidth}{>{\hsize=0.5\hsize}X|>{\hsize=1\hsize}X|>{\hsize=1\hsize}X}
		\tableline
		 & aWTPM & nWTPM  \\ \tableline
		Signal assembly sequence & The collimator integrated flux is calculated individually for any selected spin-angle. & The collimator  integrated flux is calculated in series for selected spin-angles \\ 
		 & (1) Integrate over speed & (1) Integrate over speed, tabulate differential flux over visibility strip, and interpolate to a finer mesh \\ 
		 & (2) Integrate over collimator & (2) Tabulate collimator-integrated flux at a fixed spin-angle grid.  \\ 
		 & (3) Calculate spin-angle integrated flux using quadrature & (3) Calculate spin-angle integrated flux using quadrature.  \\ 
		 & Scheme used by \citet{sokol_etal:15a} & (4) Integrate (3) over good time intervals using quadratures \\
		& & Scheme used by \citet{bzowski_etal:15a}.\\ \tableline
		Main application & Tests and general studies of ISN~He. Dedicated to calculations on a personal computer. & Fit of the ISN parameters; other species like H, Ne, O, D can be easily calculated; dedicated to huge serial calculations on a cluster\\ \tableline
		Contact author & J.~M.~Sok{\'o}{\l} (\email{jsokol@cbk.waw.pl}) & M.~A.~Kubiak (\email{mkubiak@cbk.waw.pl}) \\ \tableline	
	\end{tabularx}
	\end{centering}
	\end{table*}		
	
\clearpage
\section{Cross-validation of the two versions of WTPM}
\label{sec:crossVal}
The two versions of the WTPM, presented in Section~\ref{sec:modelDescr}, are constructed based on the same main approach to atom tracking. They differ in implementation (aWTPM in \texttt{Mathematica}, nWTPM in \texttt{Fortran/C}), reproduction of the FOV of the collimator, the ability of a detailed reproduction of the ionization losses in the heliosphere, and averaging the signal over good times. Since the aWTPM is dedicated to testing and investigating various effects in the ISN~He modeling, it uses a simplified ionization model (the ionization rate is fixed in time and its value selected for the time of detection, changing with solar distance as $1/r^2$). This simplification is used to keep the time of computation reasonably short. Currently this version is not used to average the signal over time, but this function is easy to add if needed. In the numeric WTPM the ionization losses are implemented in a more sophisticated way: with the latitudinal dependence of the photoionization, charge exchange reactions, and electron impact as well as a realistic heliocentric distance-variation of the electron impact ionization taken into account. The survival probability is calculated with all variations of the ionization rate in time taken into account by numerical integration. The advantage of the numeric WTPM is that the user can code ionization in any suitable way and in further parts of the paper we show how various assumptions about ionization losses in the heliosphere affect the modeling of the ISN~He flux.

The goal for both versions of the code was to achieve an agreement to at least $1\%$ in the collimator- and spin-angle bin-averaged flux for the two codes run for an identical ionization model, i.e., with the nWTPM degraded to the simplified assumptions of aWTPM. The goal of a $1\%$ agreement, and thus cross-validation, was pursued at all levels in the calculation, starting from the state vectors of the atoms in the source region, through determination of the integration boundaries and calculating the differential flux on the sky (Equation~\ref{eqDiffFlux}), flux averaged over the collimator FOV (Equation~\ref{eqCollAvDef}), to the flux averaged over spin-angle bins (Equation~\ref{eqSpinAngleAverDef}). In the following, we show that this goal has been accomplished.

\begin{figure}
	\centering
	\resizebox{\hsize}{!}{\includegraphics{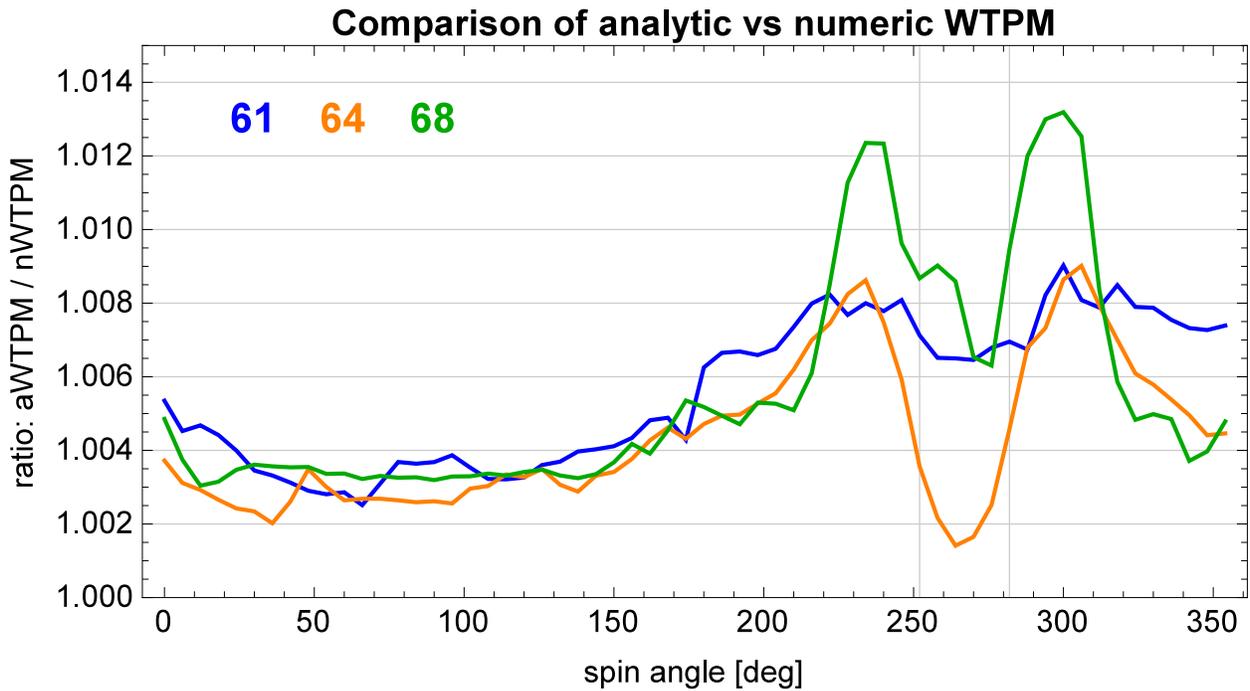}}
	\caption{Ratio of analytic to numeric WTPM simulations of the ISN~He flux, averaged over spin-angle bins and calculated with the simplified assumption on the ionization losses (ionization at the time of detection with $1/r^2$ dependence on solar distance). Different colors mark different orbits, indicated by the numbers in the plot. The vertical lines mark the spin-angle range for the data used in the analysis of ISN~He by \citet{swaczyna_etal:15a} and \citet{bzowski_etal:15a}. The ISN~He peak is close to spin-angle 264.}
	\label{figJSvsMKbase0}
	\end{figure}

Figure~\ref{figJSvsMKbase0} presents a comparison of the calculation of ISN~He flux done by the analytic and numeric versions of WTPM independently with the same assumption about ionization losses (ionization for the time of detection changing with solar distance as $1/r^2$). As it is presented in the figure, both codes yield practically identical results, with an accuracy on average of better than $1\%$ for the full range of spin-angles. In the range of the primary ISN~He, the best accuracy is for orbit 64 (up to $0.4\%$); for the orbits well before and after the peak orbit the accuracy drops to $0.8\%$. The largest discrepancies are for the so-called wings of the primary flux and they reach about $1.2\%$ for orbit 68 for the worst pixels. For the spin-angles where the flux is extremely weak, like spin-angles from $20\degr-150\degr$, the accuracy is high ($0.3\%$). 

The systematic differences between results of the two codes visible in Figure~\ref{figJSvsMKbase0} are well understood and can be eliminated if needed, but at a very high calculation cost. The small systematic underestimation of the total flux by nWTPM, manifested by an aWTPM/nWTPM ratio between $1.002$ and $1.004$ in the left-hand portion of Figure~\ref{figJSvsMKbase0} exists because the numerical atom tracking for the calculation of survival probability in nWTPM typically overshoots the tracking distance limit. Since far away from the Sun the atom tracking procedure makes large steps, in practice the actual stop distance exceeds the limit by $\sim10$~AU, which results in a small overestimate of the ionization loss compared to the losses calculated with the stop distance equal 150~AU, adopted in aWTPM. This effect can be eliminated by forcing the stop conditions in nWTPM, which would be at a calculation cost that is not justified by the accuracy enhancement. The wavy behavior in the right-hand side of Figure~\ref{figJSvsMKbase0} is due to the limit imposed on the resolution of integration over the collimator transmission function in aWTPM. We have verified that increasing the resolution limit eliminates most of these systematic features. Since increasing the resolution by one step in the HealPix system requires a four-fold increase in the number of points within the FOV to calculate, it also increases the total calculation time. We decided to not increase the accuracy of integration over the collimator FOV in aWTPM since it is not used for data fitting, and the accuracy obtained is inside the declared $1\%$ of model uncertainty. Since the small systematic differences between the two models are well understood, we decided to not strive for an extra boost in agreement, which clearly could be obtained, but at the cost of a prohibitive increase in the calculation time.

\section{Discussion of magnitude of various details affecting the ISN~He modeling}
\label{Sec:effects}
In this section we present cross-validation of the two strains of WTPM, show substantiation for the algorithms and numerical solutions used in WTPM and discuss the significance of some effects and the related uncertainties taken into account in the modeling of ISN~He gas. We illustrate results for three orbits for the 2010 observation season: 61 (the first orbit taken into account in the ISN~He gas analysis by \citet{bzowski_etal:12a}), 64 (the orbit in which the maximum flux was observed), and 68 (an orbit that is challenging for modeling because the collimator is just skimming the ISN~He beam and a significant contribution from ISN~H is expected). When appropriate, we show results for selected individual $6\degr$ bins centered at spin-angle of $246\degr$, which typically is located at a far wing of the signal, $264\degr$, which is at the peak of the signal, and $276\degr$, which is approximately in the middle of the slope of the signal at the opposite side of the maximum (see the purple dots in Figure~\ref{figEffectsSpinAngIntFluxTab6}). In doing so, we cover most of the typical beam versus collimator FOV boresight geometries and the full range of energies of the atoms relative to the spacecraft, common for the modeling of the primary ISN~He population. This is intended to show that WTPM is able to cope with all those situations while maintaining a numerical precision of $\sim1\%$, which is better than the uncertainties in the data \citep[see][this issue]{swaczyna_etal:15a}.  

In the following subsections, we show the results from the analytic version of WTPM except for the subsections where we present effects of time and heliolatitude dependence of the ionization rate on the simulated flux (Section~\ref{sec:ionEffects}) and high-resolution sampling of data for investigation of spin-angle averaging (Section~\ref{sec:spinBins}), for which the results from the numeric version of WTPM are presented. 

\subsection{Effect of spin-axis pointing in or out of ecliptic plane}
\label{sec:spinAxTilt}
Expected modification of the ISN~He signal due to various tilts of the spin-axis with respect to the ecliptic plane is important in the context of apparent differences in the fitted ISN~He parameters obtained from the portions of the observations carried out with different tilts, as during the 2013/2014 season \citep{leonard_etal:15a, mccomas_etal:15a}, when the spin-axis was alternated between $\sim0\degr$ and $-4.9\degr$ tilts. For the 2014/2015 season, a different tilt change scheme was planned, with the axis tilt alternating between $0\degr$ and $+5\degr$. The effect of various tilts of the spin-axis on analysis of the ISN He is also studied by \citet{mobius_etal:15b}.

\begin{figure}
	\centering
	\resizebox{\hsize}{!}{\includegraphics{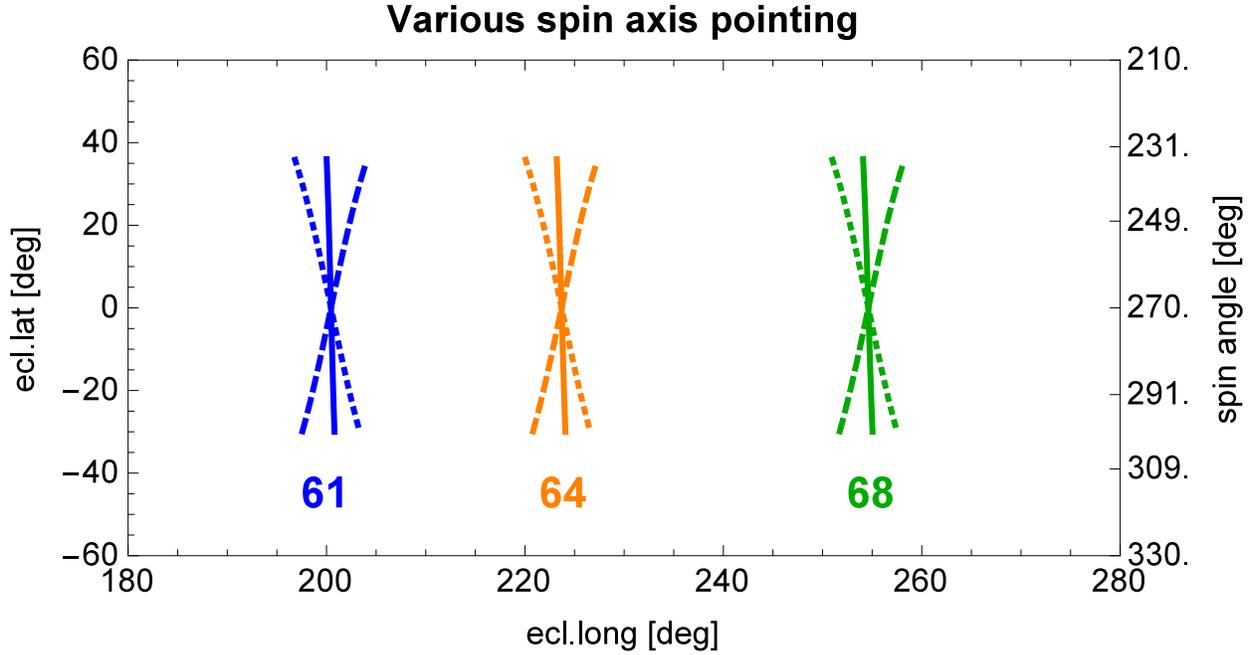}}
	\caption{Lines of sight of the collimator boresight for orbits 61, 64, and 68 for the cases of various spin-axis tilt. The solid line is the true pointing with the spin axis close to the ecliptic plane ($\epsilon = \sim0.7\degr$), the dashed line is spin-axis tilted $-5\degr$ below the ecliptic plane, and the dotted line is the spin-axis pointed $+5\degr$ above the ecliptic plane. The right-hand vertical axis is scaled in the spin-angles for orbit 64 to provide reference.}
	\label{figSpinAxisPointingSky}
	\end{figure}

\begin{figure}
	\centering
	\resizebox{\hsize}{!}{\includegraphics{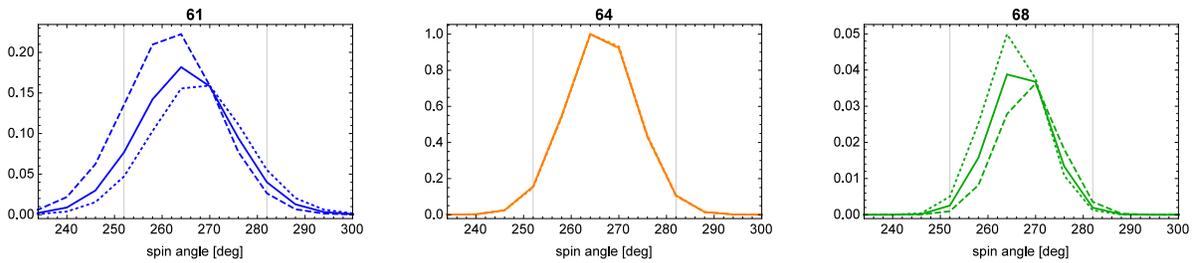}}
	\caption{Simulated bin-averaged flux (Equation~\ref{eqSpinAngleAverDef}) normalized to the maximum value for the season (orbit 64, spin-angle bin 264), calculated for different spin-axis tilts. The solid lines show the simulations with the true spin-axis pointing, i.e.,  close to the ecliptic ($\epsilon\simeq0.7\degr$), the dashed lines show the simulations with the spin-axis tilted to $\epsilon=-5\degr$ with respect to the ecliptic and dotted lines show the simulations with $\epsilon=+5\degr$ above the ecliptic. Note that the right-hand (southern) branches change relatively little with the change in the spin-axis tilt, while the left-hand (northern) branches vary substantially in orbits 61 and 68, while the change in the spin-axis tilt has a vanishing effect on the flux in orbit 64.}
	\label{figSpinAxisPointing}
	\end{figure}

Tilting the spin axis by a few degrees above or below the ecliptic plane results in a small change in the orientation of the FOV in the sky (as shown in Figure~\ref{figSpinAxisPointingSky}), which translates into sampling different portions of the ISN~He beam. This results in markedly different signals for orbits before and after the peak orbit, but practically no change is seen in the peak orbit, as illustrated in Figure~\ref{figSpinAxisPointing}. 
	
Figure~\ref{figSpinAxisPointingSky} presents the spin-angle-averaged flux for orbits 61, 64, and 68, normalized by the maximum value for the season (specifically: by the value calculated for spin-angle 264, orbit 64), simulated for three different spin-axis tilts: the true one, which was close to the ecliptic plane ($\epsilon\simeq0.7\degr$), and the two opposite settings with $\epsilon=-5\degr$ and $\epsilon=+5\degr$ below/above the ecliptic plane. The tilt of the spin axis shifts the position of the local peak for each orbit, with the largest shift for the orbits most distant from the peak orbit. For the orbits with maximum flux observed, the modification of the peak position is very small. The change due to different spin-axis tilt is mostly seen in the branch of the flux before the peak for the given orbit, i.e., for spin-angles less than 264, when $\epsilon < 0$, which means the northern hemisphere of the sky. 

If this effect is properly addressed in the simulations, tilting the spin axis in the observations should not affect the inferred parameters of ISN~He gas. If, however, some phenomenon left out from the current model modifies the gas either in front of or inside the heliosphere, results of fitting for data from orbits with one tilt of the axis may systematically vary from results obtained for orbits with a different tilt. The modification of the interstellar gas distribution at the source region either should break the symmetry of the gas distribution outside the last collision distance (see discussion in Section~\ref{sec:rFin}), or systematically modify the gas entering the heliosphere, effectively causing a north--south asymmetry in the flow. An example of the latter effect could be differential filtration in a non-axially symmetric outer heliosheath. Thus it is important to have available observations for different tilt angles of spin-axis because they may bring important insight into possible departures of the ISN~He flow near or inside the heliosphere from the assumptions typically made in the analysis, i.e., an axial symmetry of the flow around the inflow axis and the spatial uniformity of the parent distribution. Such departures may possibly be modified by differential charge-exchange ionization in the outer heliosheath, where the secondary ISN~He population is expected to be produced at the expense of atoms from the primary population. 

\subsection{Effect of stop distance for atom tracking}
\label{sec:rFin}
Using a finite heliocentric distance for tracking atoms in WTPM has physical grounds. The theory used in the classical hot model of neutral interstellar gas in the heliosphere is constructed under the assumption that the gas is collisionless and that ionization falls off with the square of the solar distance, down to $0$ at infinity. Neither is true in reality. The main factors that seem to disturb this assumptions are collisions of ISN~He atoms with each other and with ambient interstellar matter.

At $\sim7500$~K, a typical collision energy for He atoms is $\sim10$~eV. At collision energies of $\sim10$~eV, the main collision reaction affecting neutral He atoms is elastic collisions with protons and H atoms. For a total density of $\sim0.2$~cm$^{-3}$ in the LIC the mean free path (mfp) for this reaction is $\sim120$~AU. The cross section for resonant charge exchange between He atoms and He~$^+$ ions is similar to the cross section for the H--H$^+$ collisions, and since He is approximately ten-fold less abundant than H, the mfp for charge-exchange collisions for He in the LIC is on the order of 1000 AU. Thus the effective mfp against collisions in the unperturbed LIC will be $\sim100$~AU. The collision rate in the outer heliosheath will be even larger (thus, the mfp shorter) because of the increase in density and temperature of the matter expected in this region. Inside the heliopause, where no charged population of interstellar matter exists, and the neutral component (both H and He) dominates, the density of the ambient matter is reduced approximately by a factor of two (because the ionized component does not penetrate the heliopause), which still leaves a non-negligible collision rate. Thus the region of interest can be treated neither as collision-dominated, nor as collision free.

Inside the termination shock, this collision rate becomes practically negligible in comparison with the travel time to the Sun. Hence, a useful image of this problem is the following: there exists a finite distance inside which no collisions happen, but outside of which the gas is collisionally mixed. We refer to this distance as the distance of last collision. We estimate the value of this parameter to be $\sim150$~AU from the Sun and set the tracking distance $r_{\mathrm{fin}}$ to this value. 

In addition to collisions, the gas in front of the heliosphere is subjected to solar gravitation. Gravitation attracts the atoms toward the Sun and increases their speeds, i.e., their kinetic energies with respect to the Sun. Collisions tend to destroy the flow ordering that is building up due to the Sun's gravity and may at least partially annihilate the speedup effect by transferring the increasing momentum to the degrees of freedom perpendicular to the direction toward the Sun (an isotropization effect). If the gas is dominated by collisions, then a MHD model of accretion should be used to describe its physical state. The other extreme is the approach due to \citet{danby_camm:57}, who describe the behavior of the fully collisionless accretion. The true behavior of the gas must be somewhere in between, but to our knowledge, this topic has not been thoroughly investigated. Therefore we adopt a scenario of a homogeneous and uniform distribution of interstellar gas outside the last collision distance and a fully collisionless gas inside it. 

The effect of gravity practically does not affect the gas temperature even for $r_{\mathrm{fin}}= 150$~AU. Let us assume with some exaggeration that the collisions are very effective in randomizing the atom motion and that consequently, the entire increase in kinetic energy of the atoms due to the action of solar gravity between infinity and $r_{\mathrm{fin}}$ goes into heating of the gas, with the bulk speed unchanged due to the conservation of energy. For an atom that in infinity had energy corresponding to a speed of 25.5~\kms, as obtained by \citet{bzowski_etal:15a}, the increase in its kinetic energy between infinity and $r_{\mathrm{fin}} = 150$~AU will be by 1.8\%. Thus the thermal energy of the gas, and consequently its temperature, will be increased by this percentage, and for $T_{\mathrm{ISN}} = 7440$~K, the temperature at $r_{\mathrm{fin}}$ will be equal to 7570~K, i.e., larger by just $\sim 130$~K. Such a small increase is much less than the uncertainty in the temperature determination using all of the methods presented in this special issue \citep{bzowski_etal:15a,mobius_etal:15b,schwadron_etal:15a}. Hence we conclude that it is reasonable to adopt the limiting distance for atom tracking approximately equal to the distance of last collision for the atoms approaching the Sun, i.e., at $\sim150$~AU and to maintain that the flow speed and temperature of the gas found from the model fitting to data will yield representative values for the gas much farther away from the heliosphere.
	\begin{figure}
	\centering
	\begin{tabular}{cc}
	\includegraphics[scale=0.5]{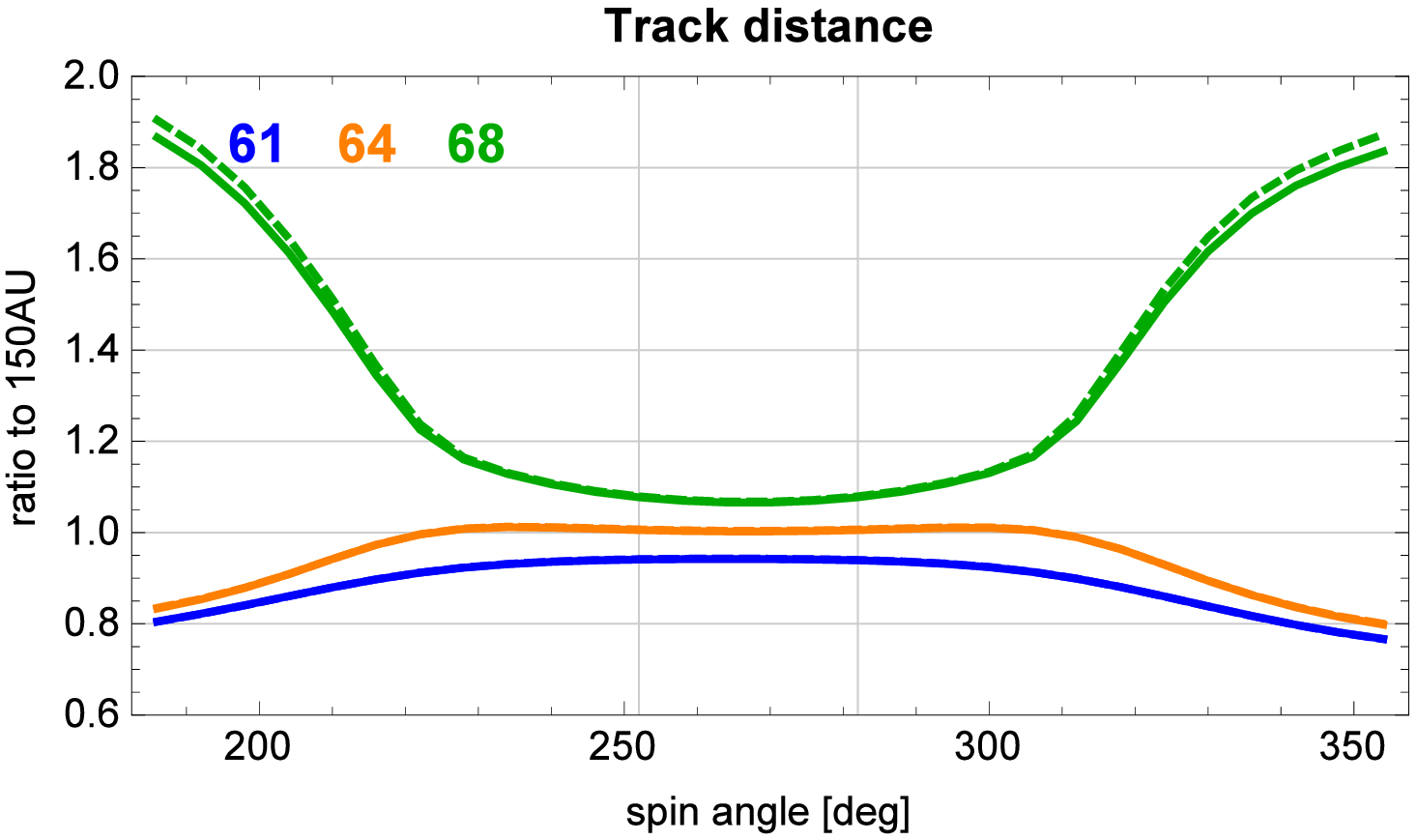}&\includegraphics[scale=0.5]{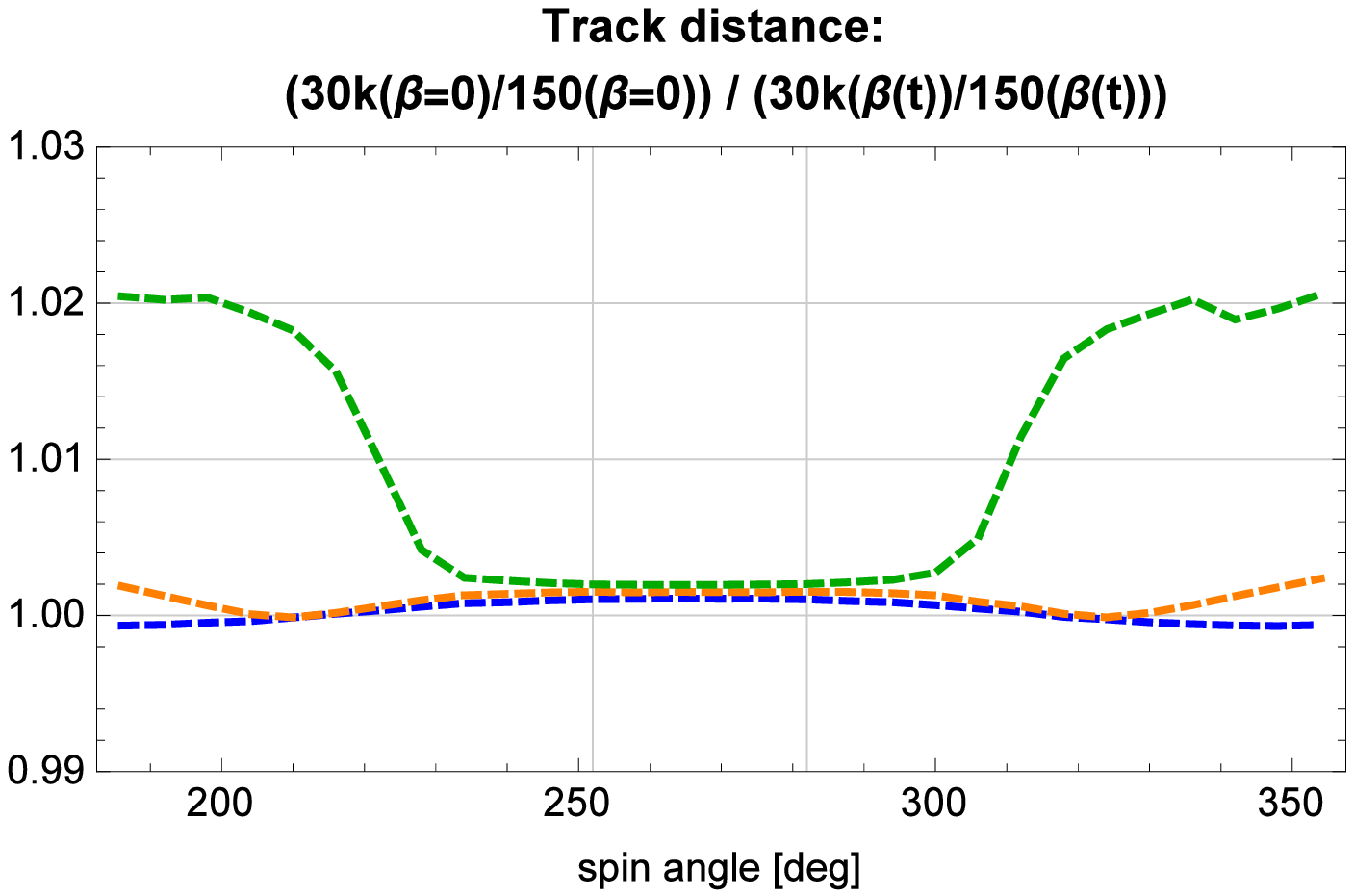} \\
	\end{tabular}
	\caption{Left-hand panel: ratio of the signal modeled with stop distance equal $30,000$~AU to $150$~AU. The vertical lines indicate the spin-angle range of primary ISN~He observed by \emph{IBEX}. Solid lines present the calculation with the total ionization given for the times of detection with a $1/r^2$ dependence with solar distance, and dashed lines represent the calculation with ionization equal zero. Right-hand panel: ratio of the solid to dashed lines from the above figure.}
	\label{figEffectStopDist}
	\end{figure}
	\begin{figure}
	\centering
	\resizebox{\hsize}{!}{\includegraphics{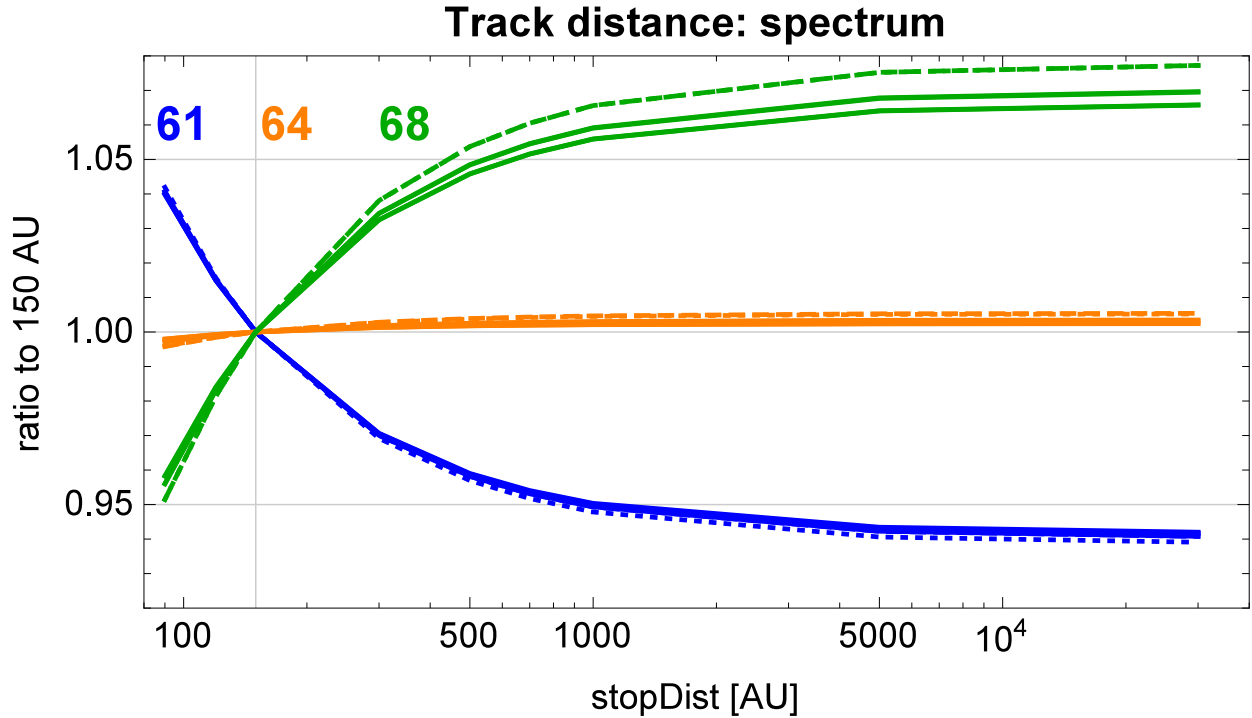}}
	\caption{Ratio of the signals modeled with various stop distances to the signal tracked to 150~AU, shown as a function of adopted stop distance for six $6\degr$ spin-angle bins from $252\degr$ (dashed line) to $282\degr$ (dotted line, the intermediate are solid). Lines of the same color show the $6\degr$ spin-angles from the range where the primary ISN~He is typically observed (spin-angles 252--282) marked with vertical lines in Figure~\ref{figEffectStopDist}.}
	\label{figEffectStopDistSpectrum}
	\end{figure}

To assess the influence of the finite tracking distance on the modeled signal in comparison with the typically adopted tracking distance at infinity, we calculated the expected flux for orbits 61, 64, and 68 tracking to $150$~AU and to $30,000$~AU and either for the true ionization rates, coming out from the adopted model, or for null ionization. In addition, we repeated the same simulations for a number of intermediate tracking distances between 150 and $30,000$~AU. Results are shown in Figures~\ref{figEffectStopDist} and \ref{figEffectStopDistSpectrum}. In the first of these figures, we show the ratios of the signals with tracking to $30,000$~AU to the signal with tracking to $150$~AU for the full range of spin-angles in the ram hemisphere. In the range of spin-angles occupied by the ISN~He signal, systematic differences in the simulated signal of $\sim6\%$ were obtained (see the left-hand panel of Figure~\ref{figEffectStopDist}). The change has a systematic character and is directed downward for pre-peak orbits and upward for the post-peak orbits. The reason for this was the action of solar gravity: the differences for the cases with and without ionization are on the order of the thickness of the lines in the figure. The differences in the signal shape due to neglecting the ionization between 150 and $30000$~AU are on a level of $0.2\%$ for the ISN He spin-angle range (see the right-hand panel of Figure~\ref{figEffectStopDist}), below the numerical accuracy of the model. On the other hand, the differences due to the action of solar gravity are not small and certainly finding an optimum tracking distance, with the effects of collisions and solar gravity, deserves a more in-depth study. Figure~\ref{figEffectStopDistSpectrum} suggests that for a tracking distances between $\sim1000$ and $5000$~AU from the Sun, the modification of the signal by solar gravity with collisionless assumption becomes less than $\sim1\%$. 

\subsection{Integration of the flux over the spin-angle bins}
\label{sec:spinBins}
The \emph{IBEX}-Lo data used for ISN~He gas analysis are integrated over $6\degr$ bins in spin-angle and over good time intervals for individual orbits. In this section, we discuss the efficient method adopted to approximate the flux within each $6\degr$ spin-angle bin, given as the average over the characteristic spin-angle range for the given bin (see Equation~\ref{eqSpinAngleAverDef}). The method should provide the desired accuracy with the smallest calculation load.  

\begin{figure}
	\centering
	\includegraphics[scale=0.7]{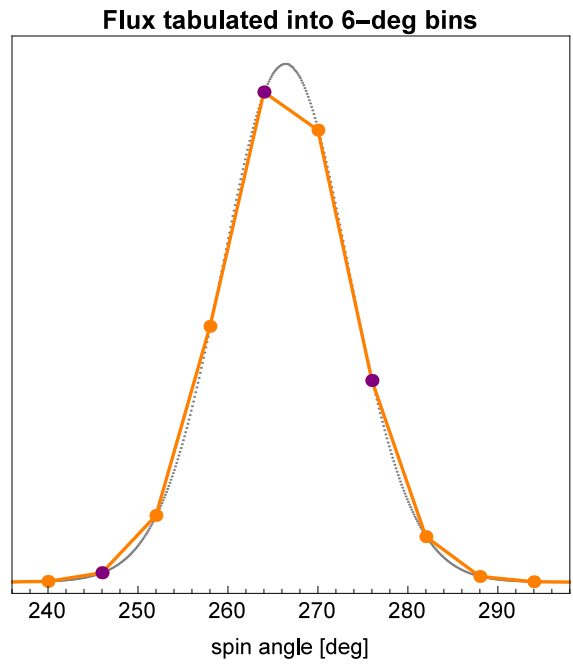}
	\caption{Collimator-integrated flux as a function of spin-angle sampled with $0.125\degr$ step (tiny gray points) and at the centers of the $6\degr$ bins (thick dots). Purple dots mark the selected spin-angle bins used, e.g. to show the change of the flux with time in Figure~\ref{figEffectsTimeInt}.}
	\label{figEffectsSpinAngIntFluxTab6}
	\end{figure}

We adopted as accurate the results of averaging over the flux sampled at a uniform mesh with $0.125\degr$ step and integrated over $6\degr$ bins using the trapezoidal rule. Taking this simulation as baseline, we compared results of three methods, simple and easy to implement, to obtain the simulations averaged over $6\degr$ bins: (1) tabulating the flux with a $6\degr$ step at the center of the bin (thick dots in Figure~\ref{figEffectsSpinAngIntFluxTab6}), (2) arithmetic averaging of the flux sampled every $1\degr$ (the method used by \citet{bzowski_etal:12a} and \citet{kubiak_etal:14a}), and (3) integrating a polynomial representation of the flux, sampled every $1.5\degr$, according to the formula from Equation~\ref{eqSpinAngleAver}.  

\begin{figure}
	\centering
	\resizebox{\hsize}{!}{\includegraphics{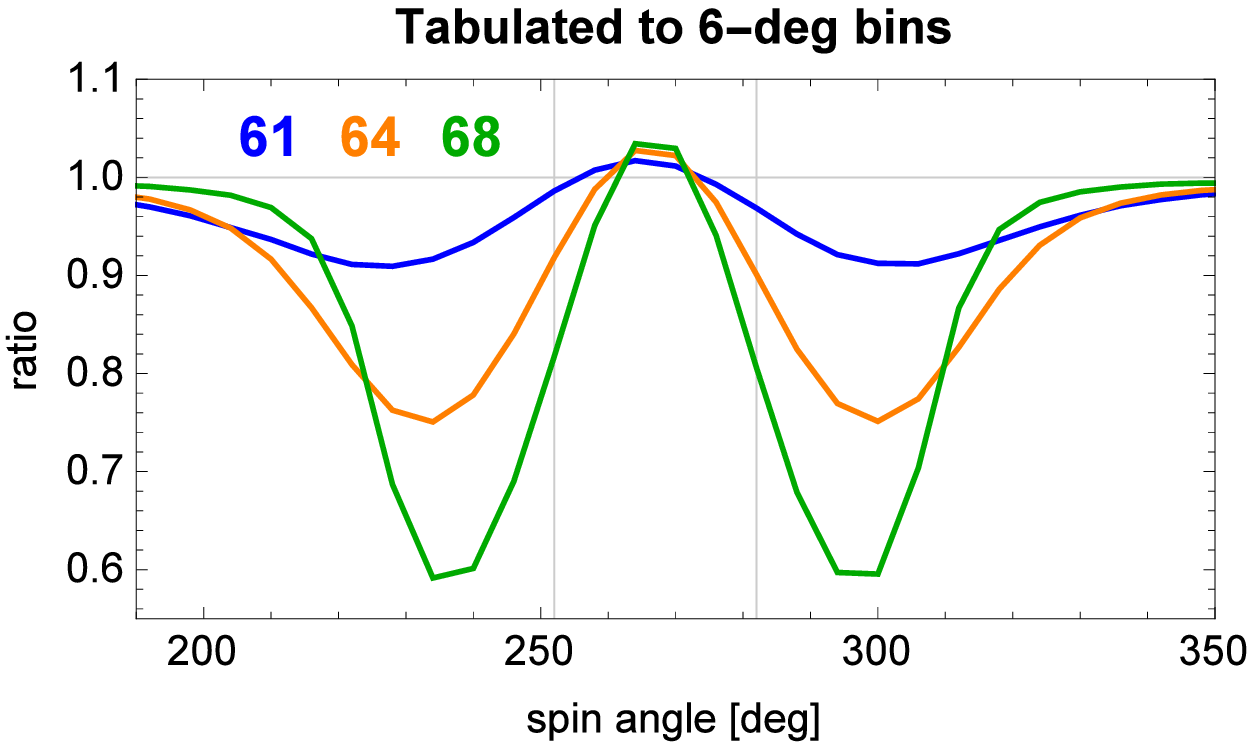}}
	\caption{Ratio of the flux tabulated at the center of each $6\degr$ (orange dots in Figure~\ref{figEffectsSpinAngIntFluxTab6}) to the flux sampled with a fixed step of $0.125\degr$ (gray points in Figure~\ref{figEffectsSpinAngIntFluxTab6}), integrated using the trapezoidal rule. The vertical lines present the typical range of spin-angles where the primary ISN~He is observed. The bias of the results due to the non-optimal sampling of the flux in spin-angle is presented for orbits 61 (blue), 64 (orange), and 68 (green). The deviations increase with the increase of the detector's ecliptic longitude and exceed the statistical accuracy of the data.}
	\label{figEffectsSpinAngIntTab6}
	\end{figure}

Solution (1) is the worst. Generally, it gives just $\sim1.5\%$ accuracy within the ISN signal range, but for orbit 61 the accuracy is reduced to $10\%$. The accuracy drops with the increasing Earth's longitude down to about $40\%$ for spin-angles corresponding to far wings of the flux for orbit 68, as illustrated in Figure~\ref{figEffectsSpinAngIntTab6}. The estimates for the accuracy of the central (maximum) bins are $\sim3\%$, but the statistical accuracy of the data in these pixels is largest and thus the flux estimate must be very good too. A comparison of the orange line connecting the thick dots with the tiny gray points in Figure~\ref{figEffectsSpinAngIntFluxTab6} illustrates the amount of information ignored when the true flux is approximated by simple tabulation for the center of each bin. The strongest differences occur in the portion of the signal where the curvature as a function of spin-angle is the largest, i.e., at the peak and in the bottom of the wings. In all, approximating the bin averages by the center value for the $6\degr$ bins is not accurate enough for fitting the ISN inflow parameters.

\begin{figure}
	\centering
	\resizebox{\hsize}{!}{\includegraphics{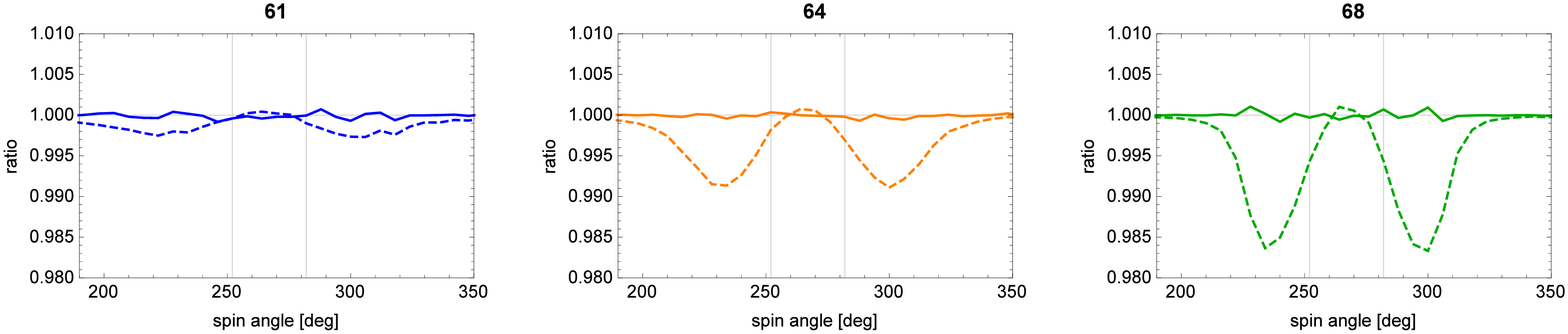}}
	\caption{Ratio of the flux averaged over $6\degr$ bins calculated using various averaging methods to the bin-averaged flux sampled with a step of $0.125\degr$, integrated using the trapezoidal rule, shown as a function of spin-angle for orbits 61, 64, and 68. Dashed lines: the ratio for the flux calculated as arithmetic averages over $6\degr$ bins with sampling every $1\degr$; solid lines: the ratio for the flux sampled with a step of $1.5\degr$, averaged over $6\degr$ bin using a fourth order polynomial formula (Equation~\ref{eqSpinAngleAver}).}
	\label{figEffectsSpinAngInt}
	\end{figure}

Arithmetic averaging over simulations sampled with a $1\degr$ step (method (2)) gives much better results; the uncertainty is not lower than $2\%$ for the worst orbit 68, i.e., only a little worse than the difference in the simulation of $F\left(\psi\right)$ between both versions of WTPM. But this method still features some systematic deviations as a function of spin-angle (see Figure~\ref{figEffectsSpinAngInt}). The latter effect almost vanishes for method (3), which gives the best approximation of the signal over spin-angle from the three methods investigated. When tabulating the flux every $1.5\degr$ we need to calculate fewer points and the boundary values for a given spin-angle bin can be used twice to calculate the bin-averaged flux for the neighboring bins. The accuracy of the reproduction of the accurate result of the simulation  is better than $0.1\%$, i.e., much better than the precision of simulated $F\left(\psi\right)$. Thus, averaging over spin-angle bins does not introduce any significant additional error. In all, the calculation load in this aspect is reduced by $\sim30\%$ in comparison method (2), the approach used by \citet{bzowski_etal:12a} and \citet{kubiak_etal:14a} and, additionally, the accuracy is higher. We have verified that using lower-order polynomials does not always provide a sufficient accuracy, while using a higher order method would not necessarily bring better results, but certainly would increase the calculation load in comparison with method (2). Therefore we recommend method (3) for use in fitting the ISN~He flow parameters.

\section{Integration of the flux over good time intervals and the importance of  the spacecraft orbital velocity}
\label{sec:scmotion}
	\begin{figure}
	\centering
	\resizebox{\hsize}{!}{\includegraphics{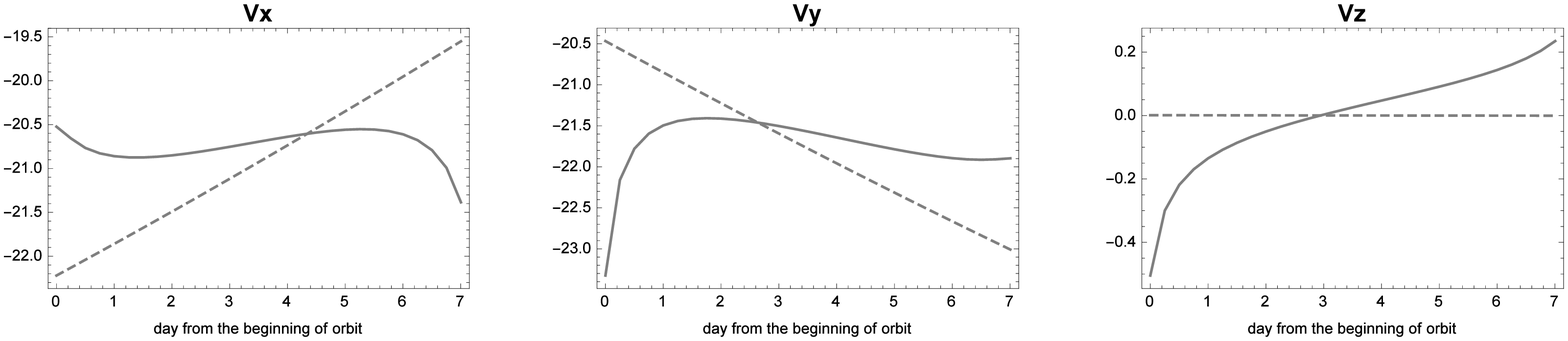}}
	\caption{Components of the Cartesian ecliptic coordinates of the velocity vector for \emph{IBEX} (solid line) and Earth (dashed line) relative to the Sun as a function of days during one orbit, here 64. The magnitude of the variation of the \emph{IBEX} velocity is approximately $2$~\kms, but the correlation of speed variations with the simulated flux changes shown in Figure~\ref{figEffectsTimeInt} is evident. The time intervals shown correspond to the HASO intervals, i.e., the intervals when science data are taken by \emph{IBEX} instruments.}
	\label{figVelocityVectors}
	\end{figure}
	\begin{figure}
	\centering
	\resizebox{\hsize}{!}{\includegraphics{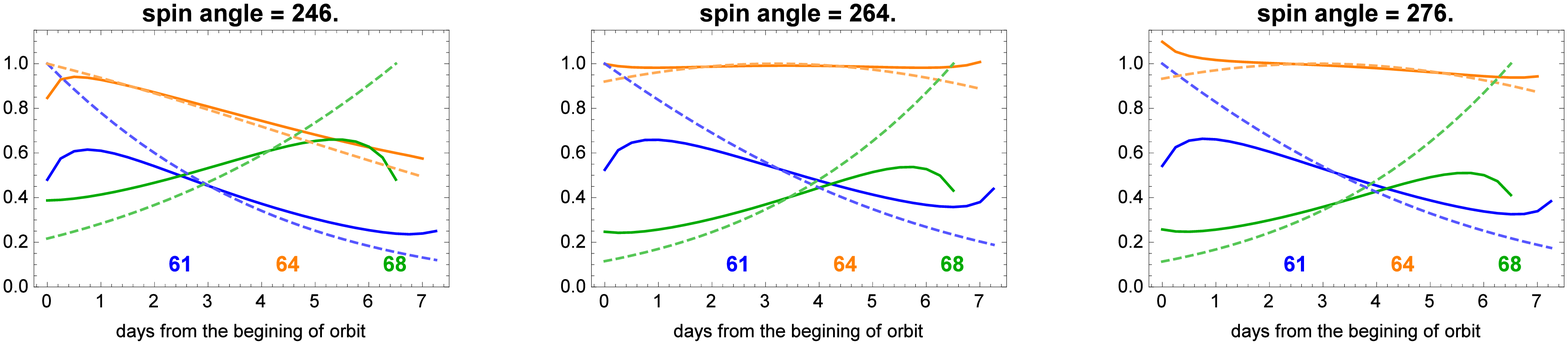}}
	\caption{Relative time variation of the flux for selected spin-angles (246, 264, 276: the points marked in purple in Figure~\ref{figEffectsSpinAngIntFluxTab6}) for orbits 61, 64, 68, sampled for the entire HASO times with a timestep of 0.25~day. The solid lines show the flux simulated with the real \emph{IBEX} velocity vectors, and the dashed lines represent the flux simulated for the case when only the Earth's velocity is used in the computations. Lines of a given color are normalized by dividing the corresponding flux $F\left(\psi, t\right)$ by $F_{\mathrm{max}}(\psi, t_{\mathrm{max}})$ for the case with only Earth's velocity. The drop or increase in the flux at the beginning and end of the HASO times, shown by the solid lines, is due to the rapid increase in the velocity of the spacecraft relative to the Earth at the beginning and end of the HASO intervals (see Figure~\ref{figVelocityVectors}).}
	\label{figEffectsTimeInt}
	\end{figure}
Once the topic of averaging the flux over $6\degr$ bins is addressed, one faces the question of how to calculate the flux averaged over good time intervals for a given orbit. The flux observed in a given spin-angle bin on a given orbit varies with time. The variation with time of the potentially observed signal is on one hand due to the motion of the ISN~He beam through the FOV because of the motion of the Earth with the \emph{IBEX} spacecraft across the beam and on the other hand due to the motion of \emph{IBEX} relative to the Earth. This latter motion is illustrated in Figure~\ref{figVelocityVectors}, which shows the Cartesian coordinates of velocity vectors of the Earth and the spacecraft relative to Sun. If the motion of the spacecraft is neglected, the flux is calculated with the use of the vectors shown with broken lines. This latter motion is almost linear with constant speed during an orbit, with the change in direction by $\sim1\degr$ day$^{-1}$, so the observed flux would be changing almost linearly, with a relatively low second derivative over time, as illustrated with broken lines in Figure~\ref{figEffectsTimeInt}. But the proper velocity of the spacecraft cannot be neglected, especially at the beginning and toward the end of the HASO intervals: in these portions of the spacecraft orbit around the Earth, the spacecraft accelerates since it is far from its apogee and thus its velocity vector relative to the Sun importantly differs from the velocity of the Earth relative to the Sun. The flux variation during the orbit due to the geometric reasons is practically the only important source of signal changes with time; the variation in the ionization rate on the timescales of days modifies the ISN~He flux negligibly \citep{rucinski_etal:03}. 

\begin{figure}
	\centering
	\resizebox{\hsize}{!}{\includegraphics{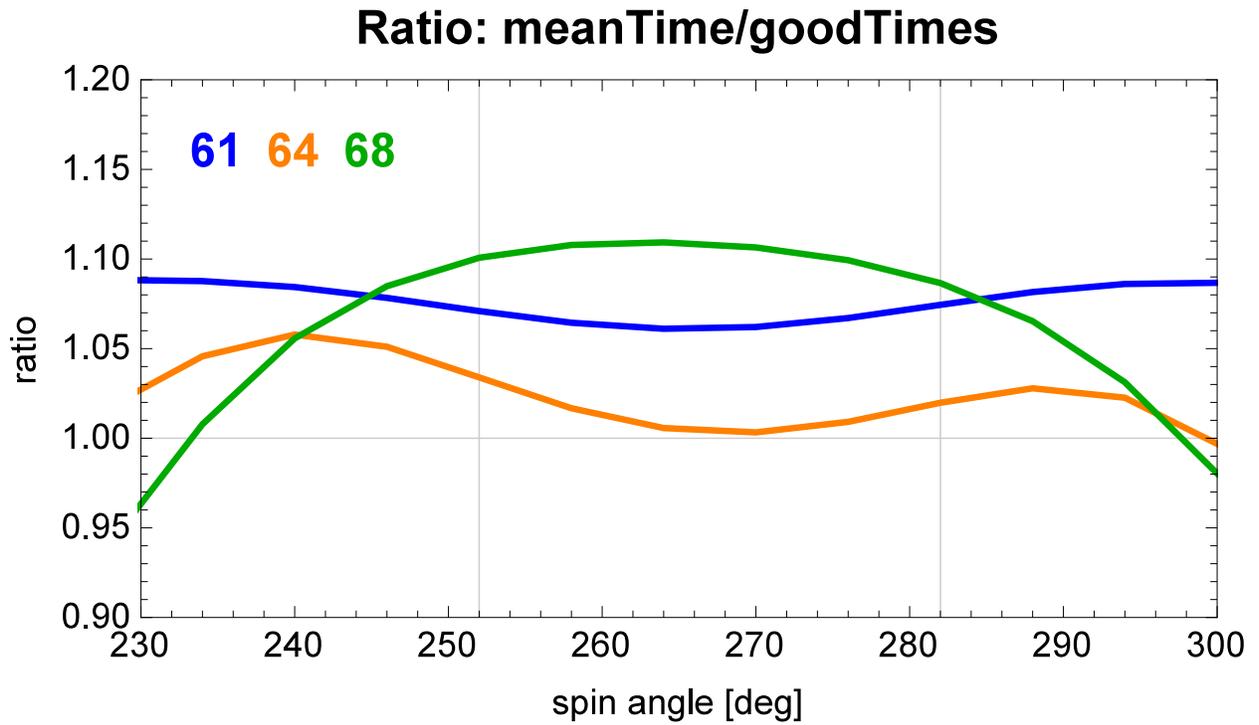}}
	\caption{Ratio of the flux calculated for the middle of HASO times to the flux averaged over good times for orbits 61 (blue), 64 (orange), and 68 (green), shown as a function of spin-angle. See \citet{bzowski_etal:15a} for the actually adopted good time intervals.}
	\label{figTimeIntRatio}
	\end{figure}

Neglecting the time variation of the flux during the orbit and representing the good-time-averaged flux by the flux calculated for the middle of the HASO interval may lead to inaccuracies exemplified in Figure~\ref{figTimeIntRatio}. The effect increases away from the peak orbits and is on the order of $10\%$. The influence of proper velocity of the spacecraft is the weakest in the peak orbits (here: orbit 64) and markedly increases for orbits before and after the peak orbit. Therefore precise reconstruction of the observation time should be implemented in the simulation program.

The prerequisite for the time-integration method is that it must be sufficiently accurate, robust for various sets of parameters of the model, efficient computationally, and easy to implement, in that order. Figure~\ref{figEffectsTimeInt} illustrates the problem that the time-averaging algorithm must address. The time variation at the beginning and end of the HASO times is strong and the flux differs considerably from the approximation of detector stationary relative to the Earth (compare the solid and broken lines of the corresponding colors). On the other hand, the variation in the flux is almost linear in the middle section of the orbit. If the good time intervals are located in the central portion of the orbit, the problem seemingly simplifies because the integration routine must integrate an almost linear function. But if one of the good time intervals is close to the beginning or the end of HASO, the integration routine must cope with a rapidly varying function with large higher-order time derivatives. 

This problem is easily solvable if one has the flux tabulated at a fine time resolution. Regrettably, adding more simulation points in time is the most costly operation from the computation viewpoint, so implementing an adjustable-step routine is computationally prohibitive. Hand-picking the best time coverage from the viewpoint of all pixels in a given orbit is, on the other hand, too labor-intensive. Therefore we decided to develop and implement the procedure described in Section~\ref{sec:timeIntegr} and we verified in a few test cases that the flux tabulated at a resolution of 0.25~day is adequately reproduced (i.e., with an accuracy of $\sim1\%$) by the polynomial model defined in Section~\ref{sec:timeIntegr}. Thus, from the mean value theorem, the integral over a subinterval is also that accurate. As non-standard as it may seem from the viewpoint of numerical art, we have verified that the proposed system works reliably for the problem at hand.

\subsection{Modification of the flux by the collimator}
\label{sec:collimModif}
\begin{figure}
	\centering
	\begin{tabular}{ccc}
	\includegraphics[scale=0.4]{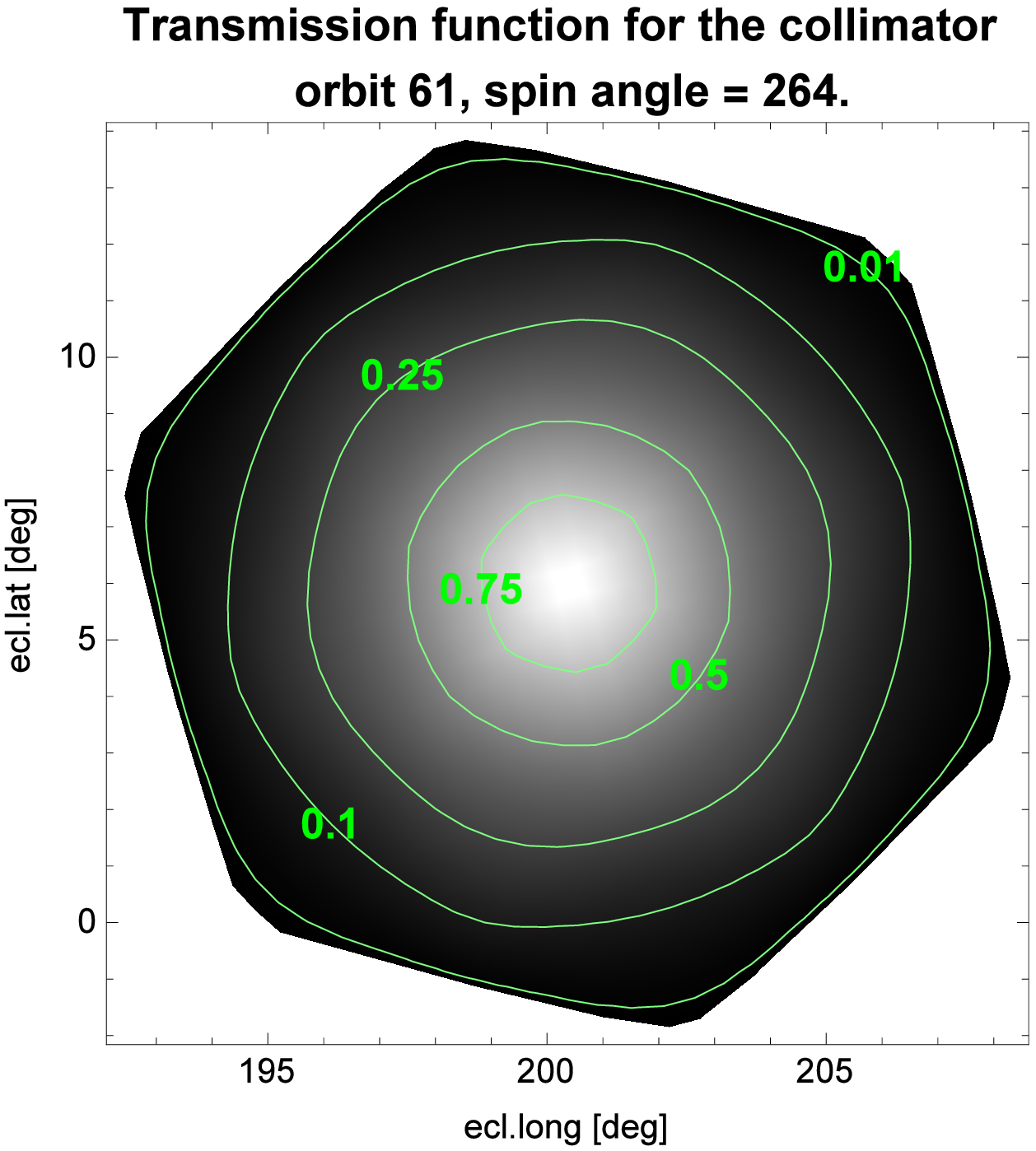} & \includegraphics[scale=0.4]{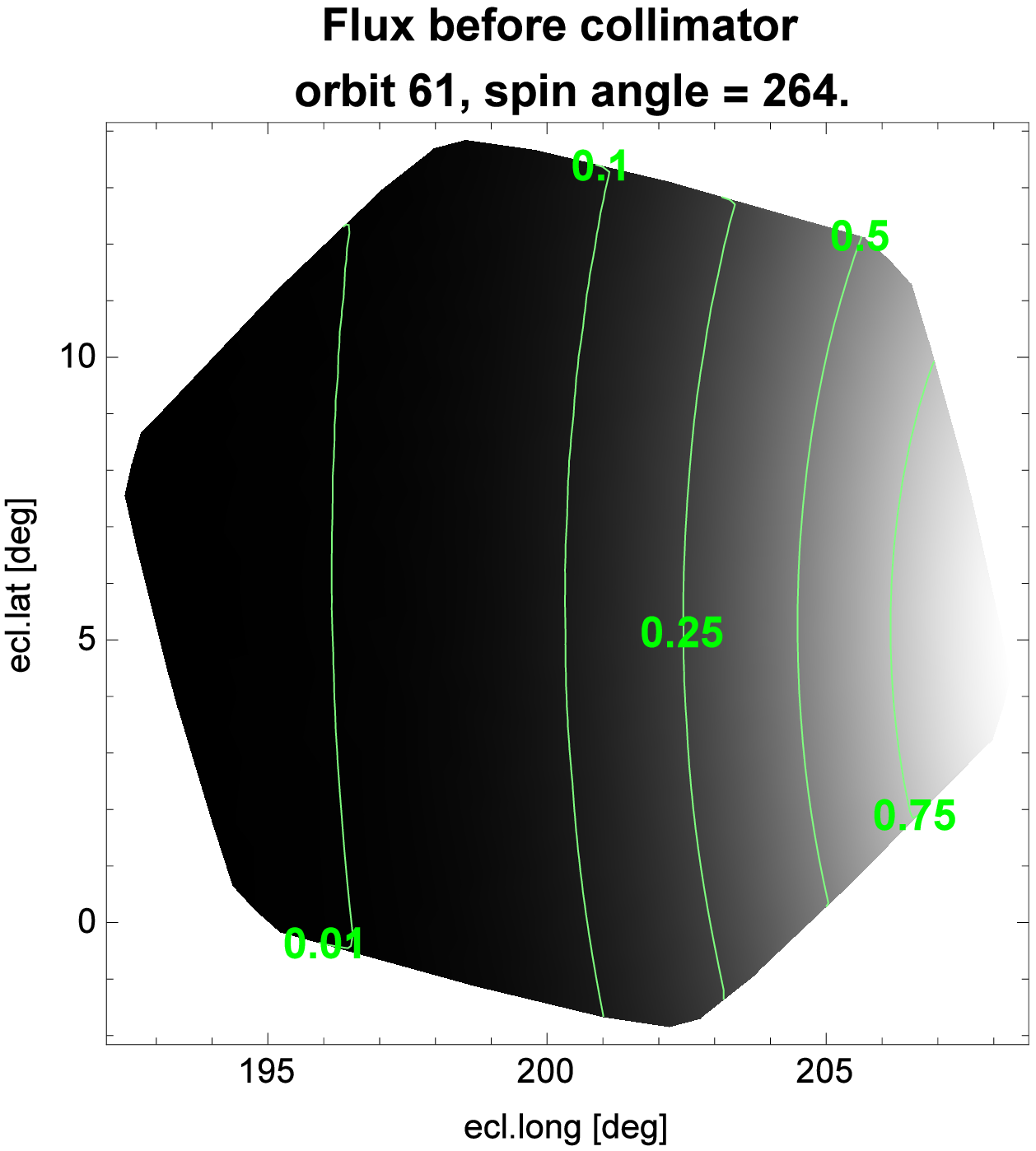} & \includegraphics[scale=0.4]{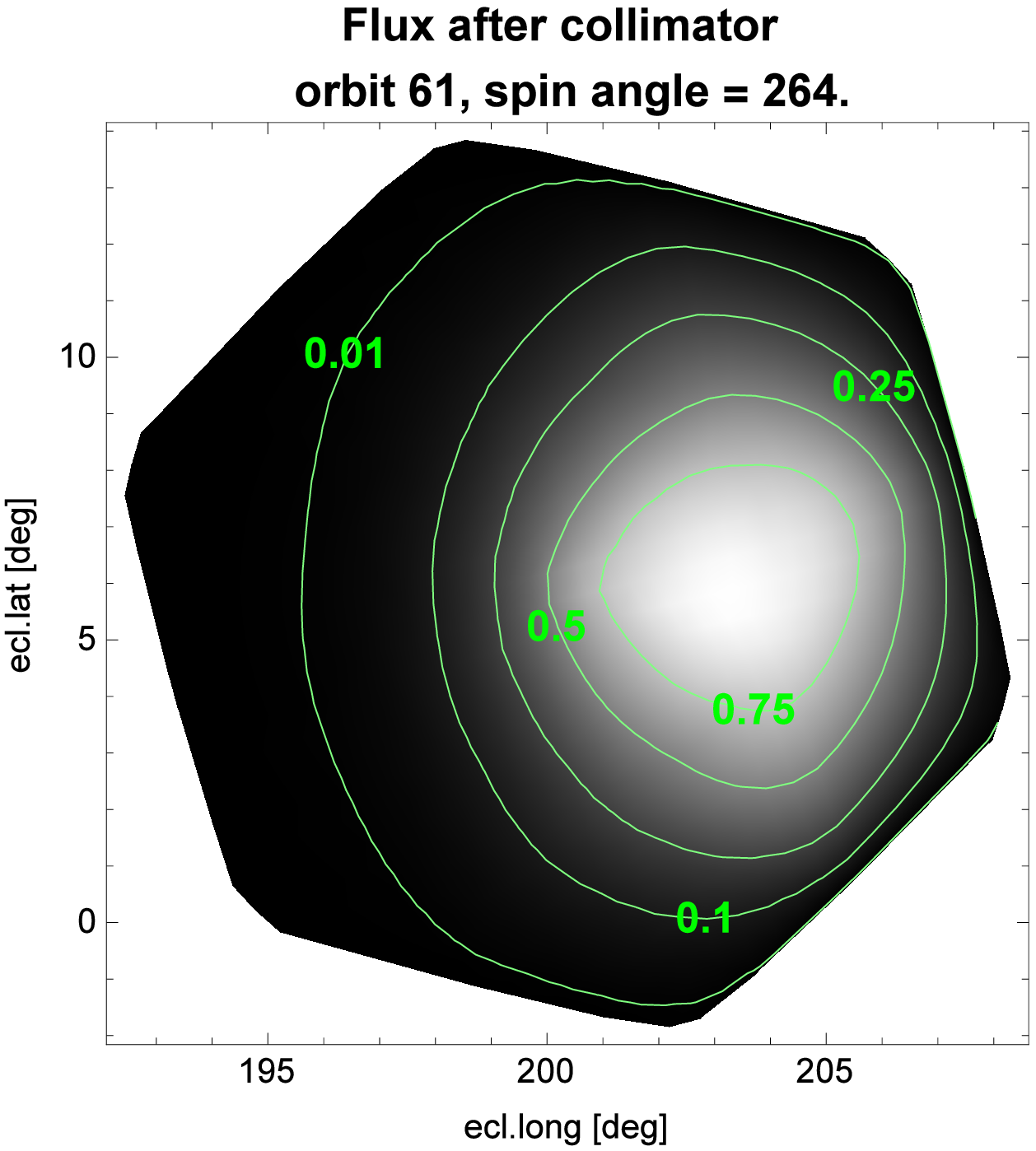}\\
	\includegraphics[scale=0.4]{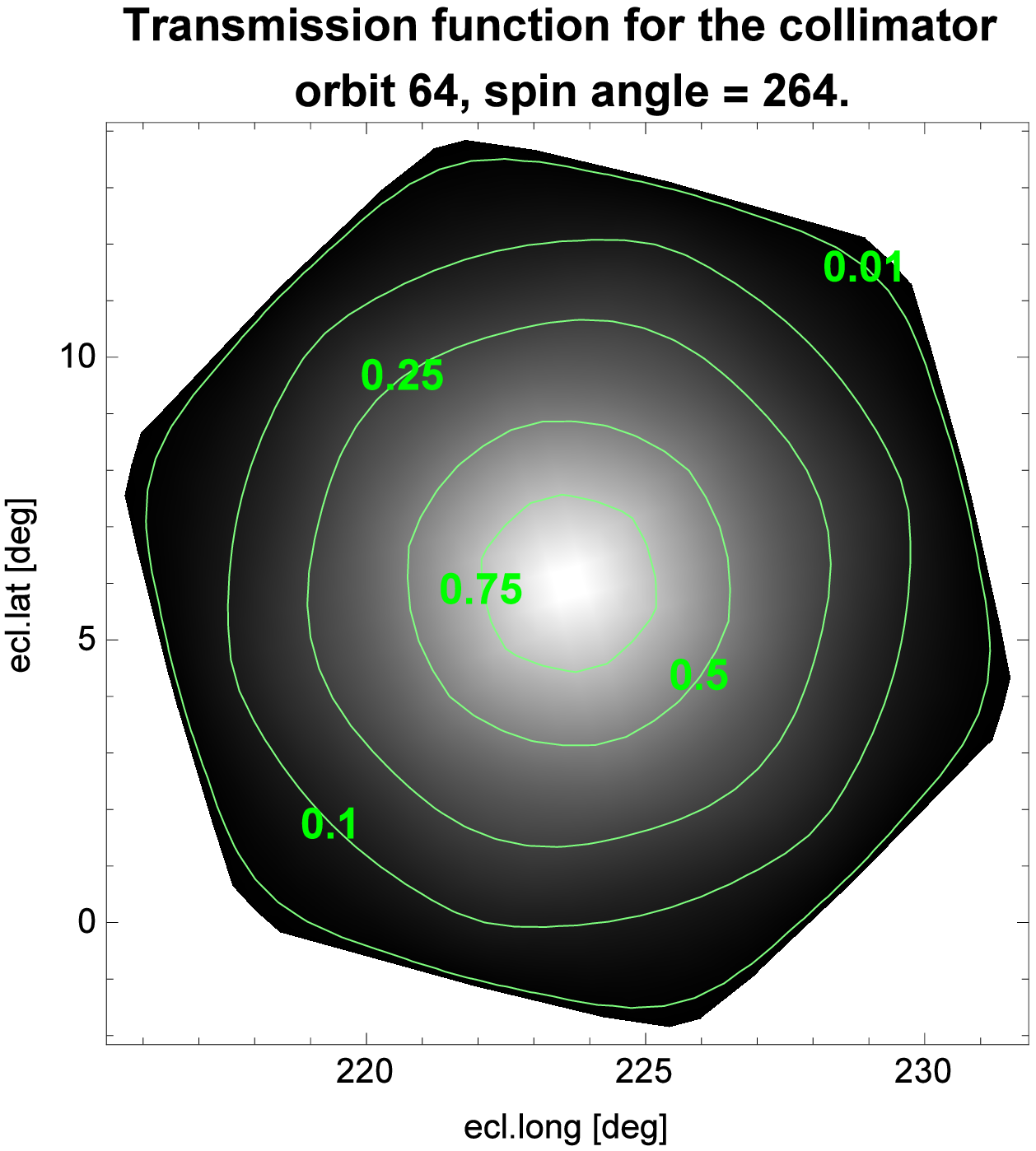} & \includegraphics[scale=0.4]{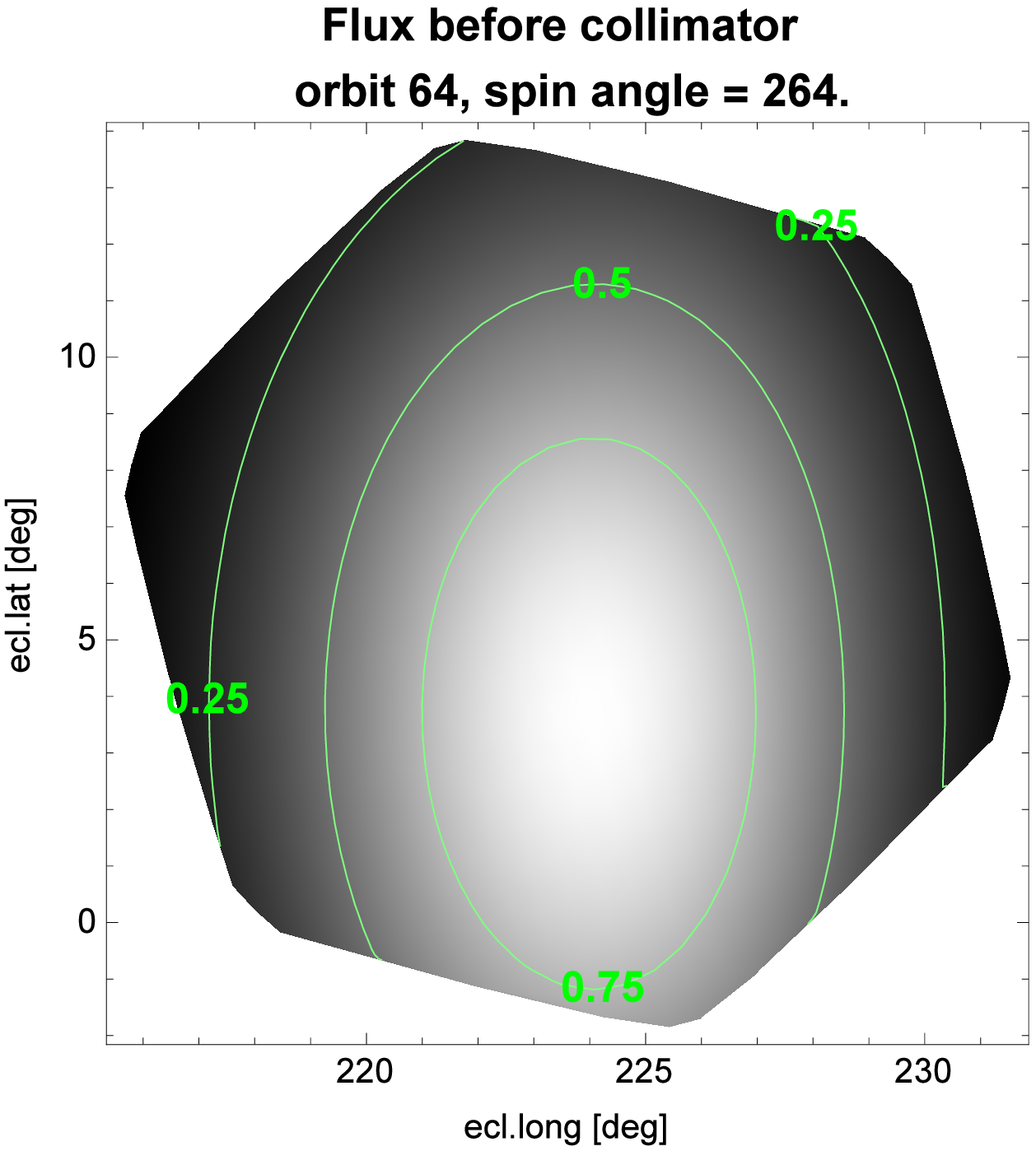} & \includegraphics[scale=0.4]{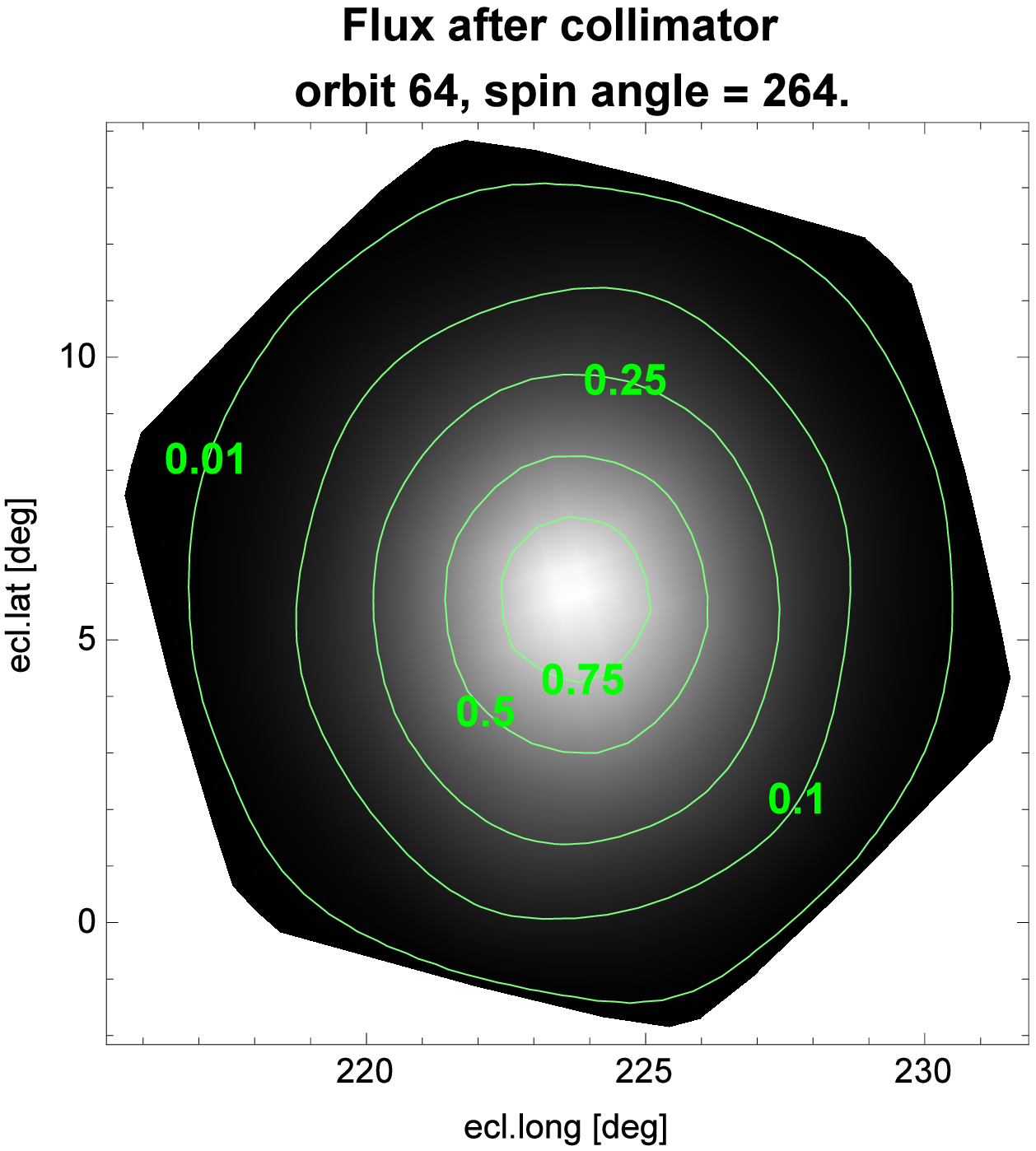}\\
	\includegraphics[scale=0.4]{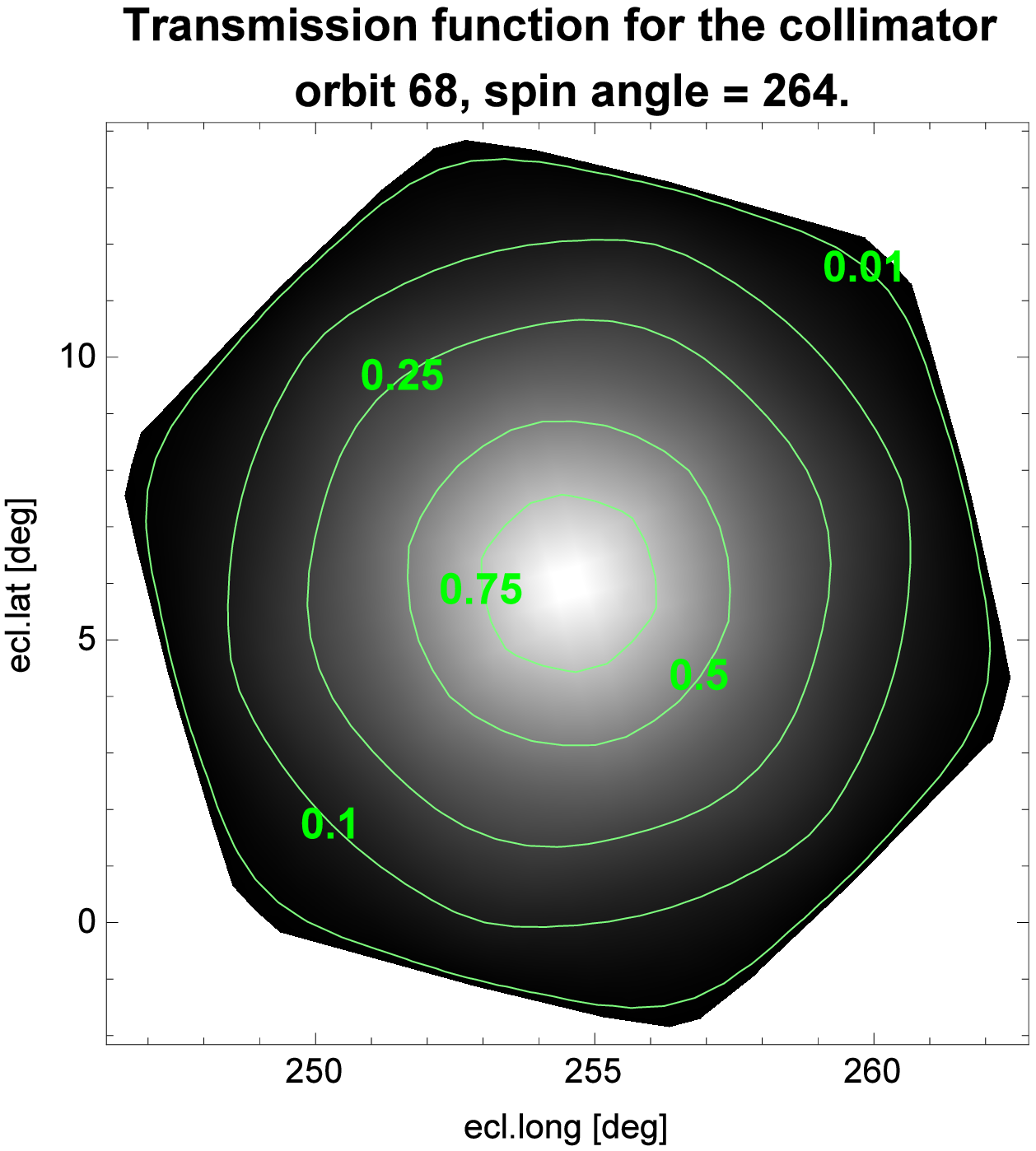} & \includegraphics[scale=0.4]{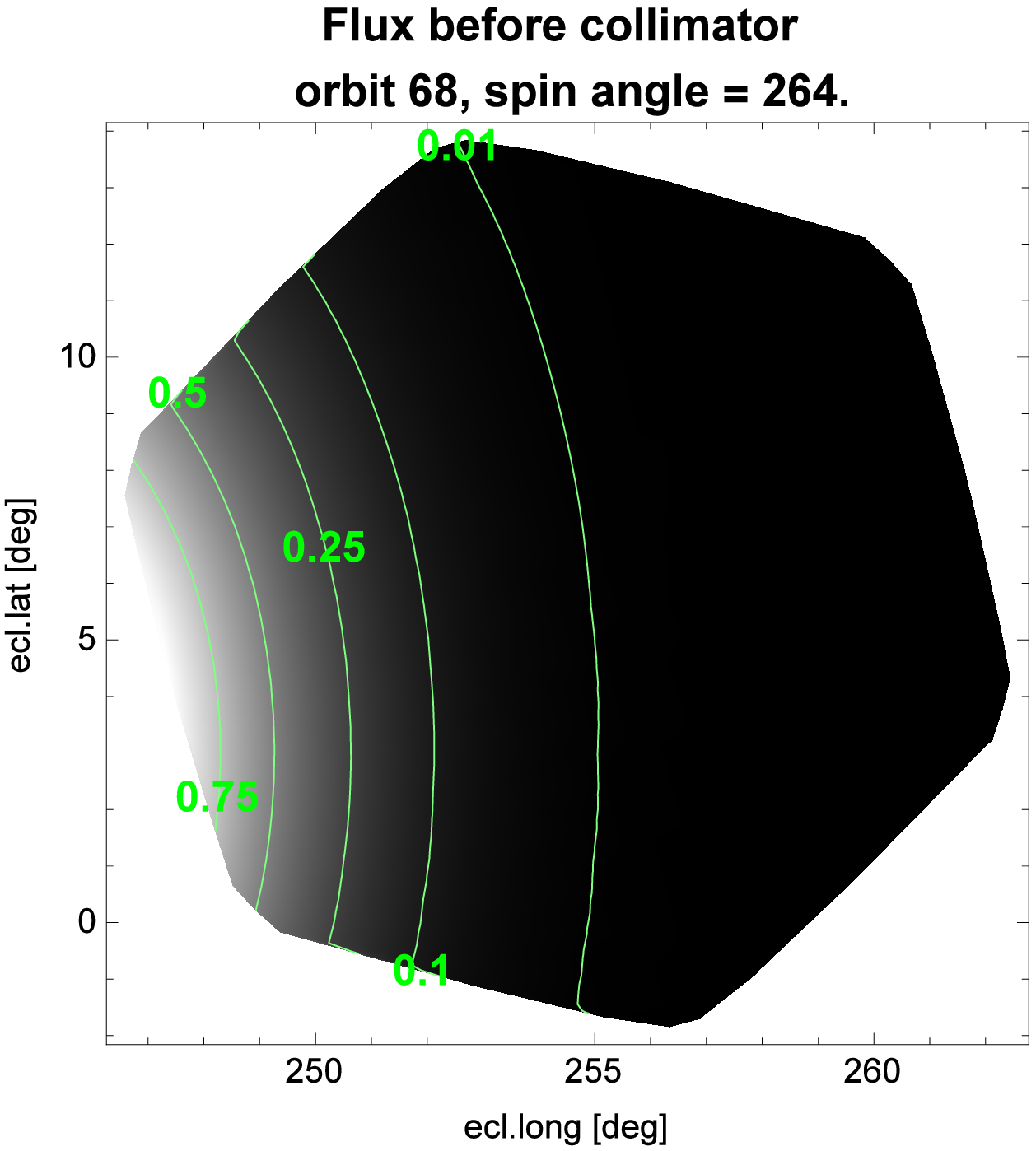} & \includegraphics[scale=0.4]{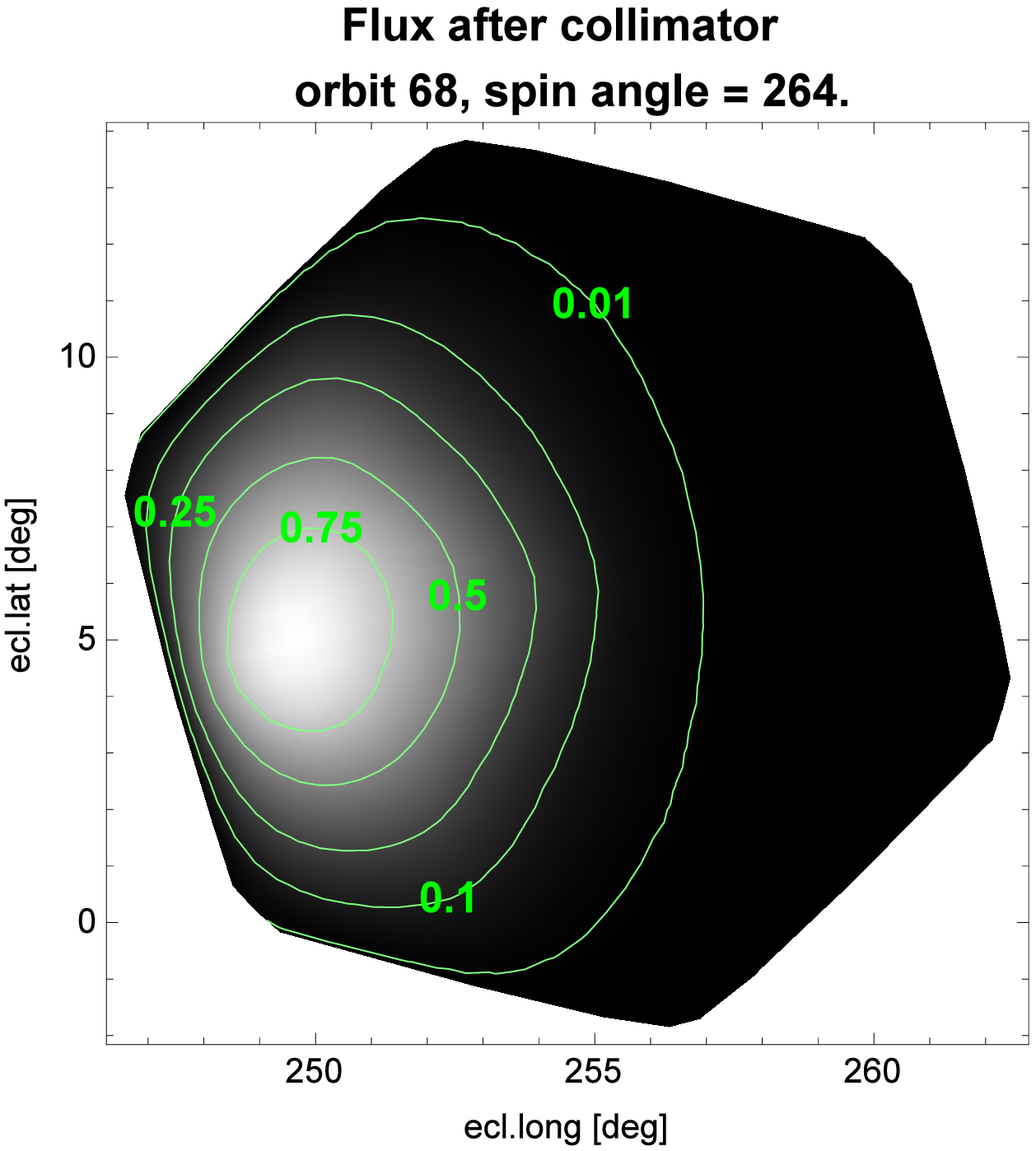}\\
	\end{tabular}
	\caption{Modification of the flux due to the collimator field of view. Three orbits of the primary ISN~He are presented, 61 for the beginning of helium ISN season, 64 for the peak of the ISN gas, and 68 for the end of the helium ISN season. For all three orbits the spin-angle 264 for the peak of the observed flux is presented. The left column shows the collimator response function for the selected orbits; these plots are almost identical with respect to the spin-axis direction in each orbit. The central column shows the flux of ISN~He as it is seen by \emph{IBEX} before transmission through the collimator, and the right columns present the flux after the transmission through the collimator.}
	\label{figCollTransmissionFluxGrid}
	\end{figure}

In this section, we present an investigation of averaging the flux over the collimator transmission function and some important aspects that must be addressed in the simulations. Depending on the orientation of the ISN~He beam relative to the collimator's FOV, different portions of the aperture play a dominant role in forming the observed signal. The maximum of the observed flux does not necessarily coincide with the collimator boresight. This is illustrated in Figure~\ref{figCollTransmissionFluxGrid}, which presents an example flux simulated for three orbits from the helium ISN season 2010 for the spin-angle of the maximum flux of each orbit (it is spin-angle 264).

Two snapshots of the flux are presented for each orbit, one before the transmission through the collimator and one just after modification by the collimator's response function. In the orbit with maximal flux per season (e.g., orbit 64 in 2010 and equivalent orbits during other seasons) the maximum of the differential flux occurs close to the collimator boresight and the flux fills the entire FOV. Consequently, the maximum of the post-collimator flux coincides almost exactly with the collimator boresight and it contributes the dominant portion of the entire signal. On the other hand, for the off-peak orbits, the maximum of the flux in the aperture occurs just at the edge of the FOV and the maximum of the collimator-processed signal occurs at the side of the collimator transmission function. Thus details of the response function and the shape of collimator must be taken into account during modeling with special attention and sufficient precision to avoid possible bias. 

\subsubsection{How important are details of the collimator shape and its response function?}
Details of the collimator response function and implementation of integration over the FOV were presented in Section~\ref{sec:collimTransmiss}. Here we discuss the significance of adopted shape and response functions of the collimator on the simulated ISN~He flux.

\begin{figure}
	\centering
	\includegraphics[scale=0.55]{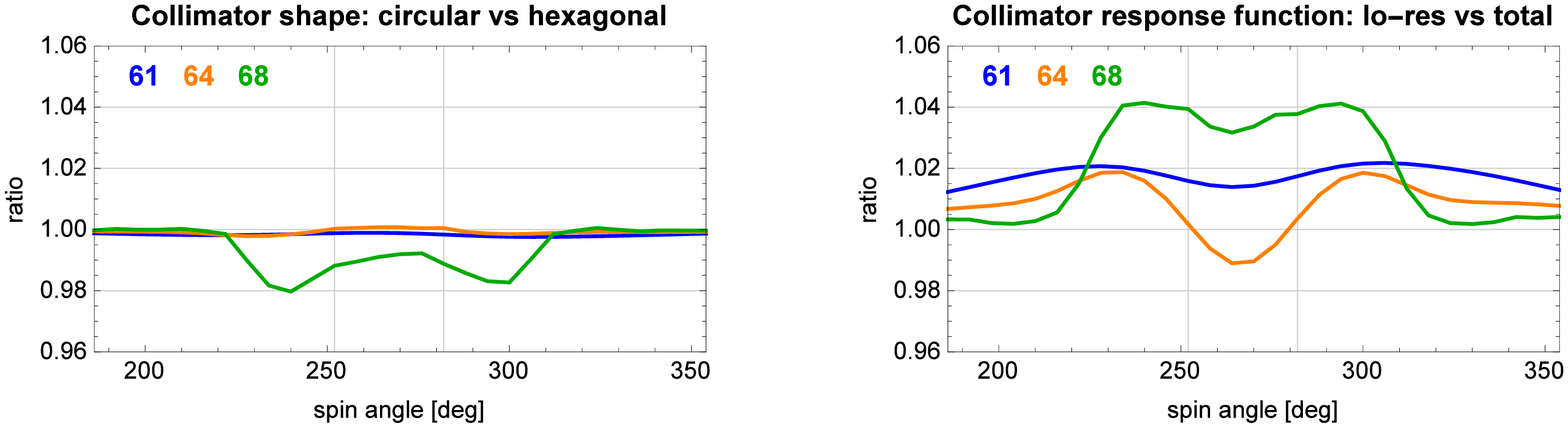} 
	\caption{Influence of different assumptions on the aperture shape and response function of the collimator on the simulated flux, shown for the observation geometry for orbits 61, 64, and 68. The color code is shown in the panels. The left panel shows the ratio of the fluxes calculated with the circular and hexagonal apertures for the same response function (according to Equation~\ref{eqCollTotal}). The right panel shows the ratio of the fluxes calculated with the response function corresponding to four low-resolution sections ($T_{\mathrm{low}}$ in Equation~\ref{eqCollTotal}) and the full model, including both the low- and high-resolution sections, for hexagonal aperture. The two vertical lines indicate the range in spin-angle where the primary ISN~He is observed.}
	\label{figEffectCollShapeResponseFun}
	\end{figure}

To assess the importance of the shape of the boundaries of the collimator, we simulated the signal with the same response function (following Equation~\ref{eqCollTotal}), but with the different shapes of the aperture boundary: circular and hexagonal. The ratios of the collimator-averaged fluxes for these two are presented in the left-hand panel in Figure~\ref{figEffectCollShapeResponseFun}. We found that there is almost no difference in the flux for orbits 61 and 64, but for orbit 68, adopting a circular boundary introduces an error of $\sim1\%$ within the spin-angle range of the ISN~He signal, and up to $2\%$ outside. It is because the signal in orbit 68 is sampled only by the edge of the collimator's FOV (see the lower row of Figure~\ref{figCollTransmissionFluxGrid}). Thus, if one does not require an accuracy better than $\sim1\%$, approximating the aperture shape by a circle is acceptable. Since implementation of the required hexagonal shape of the aperture in the simulations does not induce an additional computational burden, we recommend keeping the collimator hexagonal in shape.

We also investigated the importance of precise reproduction of the profile of the transmission function. Specifically, we checked the differences in the collimator transmission function simulated either for all four collimator quadrants of the low-resolution type, as used by \citet{bzowski_etal:12a} and \citet{kubiak_etal:14a} ($T_{\mathrm{low}}$ in Equation~\ref{eqCollTotal}), and the more realistic function, including both low- and high-resolution sections, presented in this paper (Equation~\ref{eqCollTotal}). We found that the flux is modified up to $4\%$ in the region of the main signal of the primary ISN~He. The correct flux can be either increased or decreased, depending on the orbit. This is because the placement of the ISN~He beam in the aperture changes from one orbit to another, as illustrated in Figure~\ref{figCollTransmissionFluxGrid}. Again, the largest effect is observed for the far off-peak orbit 68. The replacement of the high-resolution with the low-resolution quadrant in the simulations very likely caused the model used by \citet{bzowski_etal:12a} to be imprecise from about $1\%$ to $4\%$, depending on the simulated orbit and spin-angle. 

\subsection{The role of ionization}
\label{sec:ion}
\subsubsection{Ionization processes and their variation with time and heliolatitude}
\label{sec:ionModel}
\begin{figure}
	\centering
	\resizebox{\hsize}{!}{\includegraphics{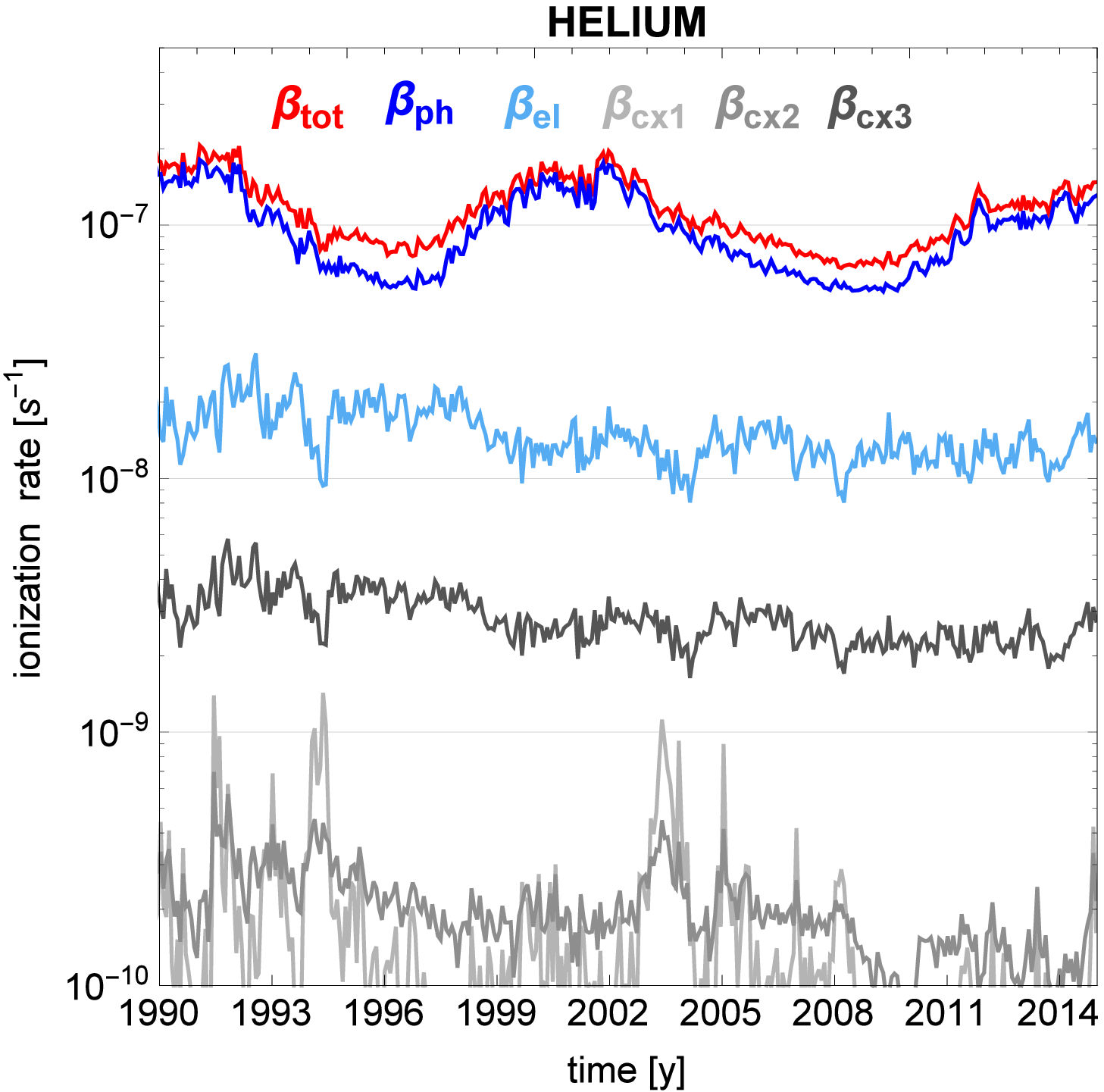}} 
	\caption{Time series of rates of the relevant ionization processes of neutral interstellar He at 1~AU from the Sun. Shown are rates for: photoionization ($\beta_{\mathrm{ph}}$), from the updated model proposed by \citet{sokol_bzowski:14a}, electron-impact for the slow solar wind ($\beta_{\mathrm{el}}$, following the model by \citet{rucinski_fahr:89, rucinski_fahr:91} and \citet{bzowski_etal:13b}), charge exchange ($\beta_{\mathrm{cx}}$) rates for all relevant reactions ($\beta_{\mathrm{cx1}}$: He + H$^+$ $\to$ H$_{\mathrm{ENA}}$ + He$^{+}_{\mathrm{PUI}}$, $\beta_{\mathrm{cx2}}$: He + $\alpha \to $H$^+_{\mathrm{sw}}$ + He$^{+}_{\mathrm{PUI}}$, $\beta_{\mathrm{cx3}}$: He + $\alpha \to$ He$_{\mathrm{ENA}}$ + He$^{++}_{\mathrm{PUI}}$) \citep{bzowski_etal:13b}, and the sum of them, the total ionization rates ($\beta_{\mathrm{tot}}$) as it is used in the analytic WTPM. In the numerical version of WTPM, $\beta_{\mathrm{tot}}$ is adopted as the baseline rate for the solar equator, but additionally, the latitudinal variations of the contributing rates are taken into account. The time series of $\beta_{\mathrm{tot}}$ are available in the Data Release~9.}
	\label{figIonRatesHe}
	\end{figure}

The ionization rate of neutral He in the heliosphere is a sum of rates of photoionization, electron-impact, and charge exchange. The latter one is practically negligible (see Figure~\ref{figIonRatesHe}), and the electron rate is important mostly inside $\sim2$~AU from the Sun because it drops with the solar distance more rapidly than $1/r^2$ (see, e.g., Figure~2 in \citet{bzowski_etal:13b}). The electron rate features a strong latitudinal anisotropy that approximately follows the latitudinal structure of the solar wind, which, together with the departures from the $1/r^2$ fall off with distance, makes it challenging to be precisely account for in an analytic expression for the total ionization losses of ISN~He. The photoionization rate in the ecliptic plane was calculated by \citet{sokol_bzowski:14a} from spectral irradiances measured by TIMED \citep{woods_etal:05a}. Charge exchange is calculated for the relative speed of the products with the latitudinal and time variation of the solar wind taken into account following the solar wind structure from \citet{sokol_etal:13a}.

The aspect of latitudinal dependence of the photoionization rate is the poorest investigated. As discussed by \citet[][pp. 67-138]{bzowski_etal:13a}, some theoretical expectations by \citet{cook_etal:80a, cook_etal:81a} and remote-sensing measurements of the coronal flux by \citet{auchere_etal:05b, auchere_etal:05c} suggest that such an anisotropy should exist and vary relatively little with solar cycle even though instantaneous fluctuations may be quite substantial (see Figure~7 in \citet{katushkina_etal:14c}). On the other hand, based on analysis of ISN~He flux on GAS/\emph{Ulysses}, \citet{witte:04} suggested that the anisotropy may be as high as $50\%$, while \citet{kiselman_etal:11a} pointed out that the solar spectrum does not vary with heliolatitude, which may imply that there is no heliolatitude dependence of the photoionization rate. The numerical version of WTPM adopts an analytic ellipsoidal model of the photoionization rate as a function of heliolatitude, described by Equation~3.4 in \citet[][pp. 67-138]{bzowski_etal:13a},  with the polar rates equal to $0.8$ of the equatorial ones. 

Recent studies \citep{snow_etal:14a, wieman_etal:14a} showed that the rate of the dominant ionization process for helium, i.e., photoionization, may be biased by systematic instrumental effects. This topic is still a subject of research, but for now we cannot rule out that the ionization model we use is systematically biased upward or downward. Discrepancies between photoionization rates calculated using different assumptions on this bias are up to $\sim20\%$ (see discussion in \citet{sokol_bzowski:14a}). 

The history of ionization at 1~AU in the ecliptic plane adopted as the baseline ionization model in this paper and the accompanying papers \citep{bzowski_etal:15a, galli_etal:15a, sokol_etal:15a, swaczyna_etal:15a} is shown in Figure~\ref{figIonRatesHe}, where in addition to the total rate, we also present the rates of individual reactions. The time series of the total ionization rate in the ecliptic plane at 1~AU used in this study is available in the Data~Release~9. The main effect of the variation in the ionization rate on the ISN~He gas at 1~AU from the Sun is a modulation of the local helium density. The scale of this effect was studied by \citet{rucinski_etal:03} for a model variation of the ionization rate, and by \citet{bzowski_etal:13b} and Sok{\'o}{\l}~et~al. (in preparation) for the realistic ionization. Variations of the ionization rate during the solar cycle cause variations in the density of ISN~He at 1~AU, and thus in the ISN~He flux, with an amplitude of $\sim2$. Detailed analysis of the effects of ionization losses on the flux measured by \emph{IBEX} is presented in the next section. 

\subsubsection{Effects of ionization losses on the absolute flux measured by \emph{IBEX}}
Attenuation of the ISN~He flux observed by \emph{IBEX}-Lo by ionization losses is approximately by a factor of $\sim1.7$ for 2010, when the ionization rate was low due to low solar activity. During higher activity times, this attenuation will be approximately two-fold larger.  Therefore, effects of ionization on the absolute flux observed by \emph{IBEX} must be taken into account when one wants to analyze data from a number of observation seasons covering an interval of changing solar activity. In fact, the first ISN~He gas observations were made in 2009/2010 during the extended solar minimum, while the most recent ones, from 2012/2013 and 2013/2014, were carried out during the maximum of solar activity. On the other hand, when data from a relatively short interval of a few months are analyzed, details of the ionization rate changes become less important, as we show in the following subsections.

\subsubsection{Importance of ionization in the analysis of ISN~He gas observed by \emph{IBEX}}
\label{sec:ionEffects}
Analysis of \emph{IBEX}-Lo observations of ISN~He gas is usually carried out for data subsets covering individual seasons \citep{bzowski_etal:12a, bzowski_etal:15a, mobius_etal:12a, leonard_etal:15a, mccomas_etal:15a}. The analysis based on the analytic interpretation model by \citet{lee_etal:12a} assumes stationary spherically symmetric ionization and is focused on moments of the observed ISN~He beam: spin-angle of the peaks and the beam widths for individual orbits. It is sometimes assumed that the ionization losses are negligible for the modeling because they do not introduce any important bias into the results. To verify this we simulated the ISN~He beam for orbits 61 through 68 either assuming zero ionization or adopting the ionization rate as it comes out from the ionization model presented in Section~\ref{sec:ionModel}. The calculations were performed using the analytic version of WTPM. With the ISN~He beam calculated for each orbit, we fitted a Gaussian function $F\left(\psi\right) = f_0 \exp \left[-\left(\psi - \psi_0\right)^2/\sigma^2\right]$ to both sets of simulations with free parameters $f_0$ (peak height), $\psi_0$ (spin-angle of the peak), and $\sigma$ (width of the peak). 	

\begin{figure}
	\centering
	\resizebox{\hsize}{!}{\includegraphics{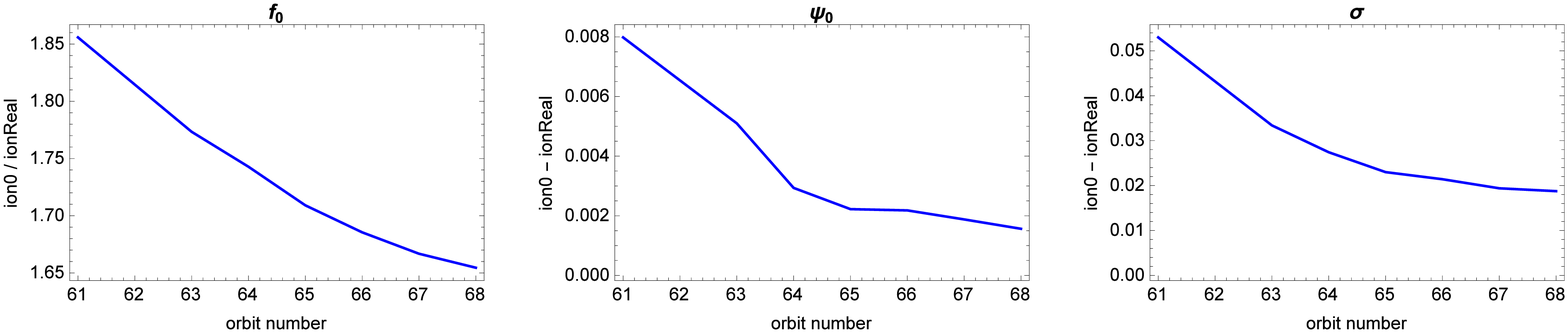}} 
	\caption{Ratio of the peak heights (left-hand panel) and differences between peak positions (middle panel) and widths of the peaks (right-hand panel) obtained for a model of ISN~He flux observed in orbits 61 through 68 for an ionization rate of 0 and an ionization realistic for the epoch of observations, given by $\beta_{\mathrm{tot}}$ shown in Figure~\ref{figIonRatesHe}. The beam parameters were obtained from Gaussian fits to the flux as a function of spin-angle.}
	\label{figIonZeroParams}
	\end{figure}

Results are shown in Figure~\ref{figIonZeroParams}. Neglecting the ionization rate virtually does not move the positions of the peak of the observed beams: the difference is on the order of $0.005\degr$. Also the width of the beams is little affected: neglecting the ionization increases the beam width by $\sim0.03\degr$, which translates into a difference in fitted temperature of $\sim20$~K. Of course, the peak heights are affected quite strongly --- the early orbits in the season by a factor of $1.8$ and the latest orbits by a factor of $\sim1.6$ --- but neglecting the ionization reduces the ecliptic longitude of the maximum flux by only $\sim0.25\degr$. 

In the analysis using the method developed by \citet{swaczyna_etal:15a}, one calculates a normalization factor to scale the model values to measured count rates and performs $\chi^2$ fitting of the ISN~He flow parameters, looking for the scaling factor separately for each test parameter set. The drivers for the fitted parameters are relations between the values of simulated data points for individual orbits and between the orbits during one observation season. Important are relations between individual data points. Ionization losses make a strongly correlated effect on all simulated data points: the prime effect is the reduction in intensity and changes of relations between the points (higher losses for some pixels, lower for others) are a secondary effect. To assess potential influence of the hypothetical bias in the ionization rate on the results of modeling the ISN~He flux observed by \emph{IBEX}, we simulated the extreme cases, i.e., one with the currently used ionization model and the other assuming an ionization rate of 0. This latter case is important as the limiting case for the systematic uncertainties of the ionization rate, mentioned in Section~\ref{sec:ionModel}. 
	\begin{figure}
	\centering
	\resizebox{\hsize}{!}{\includegraphics{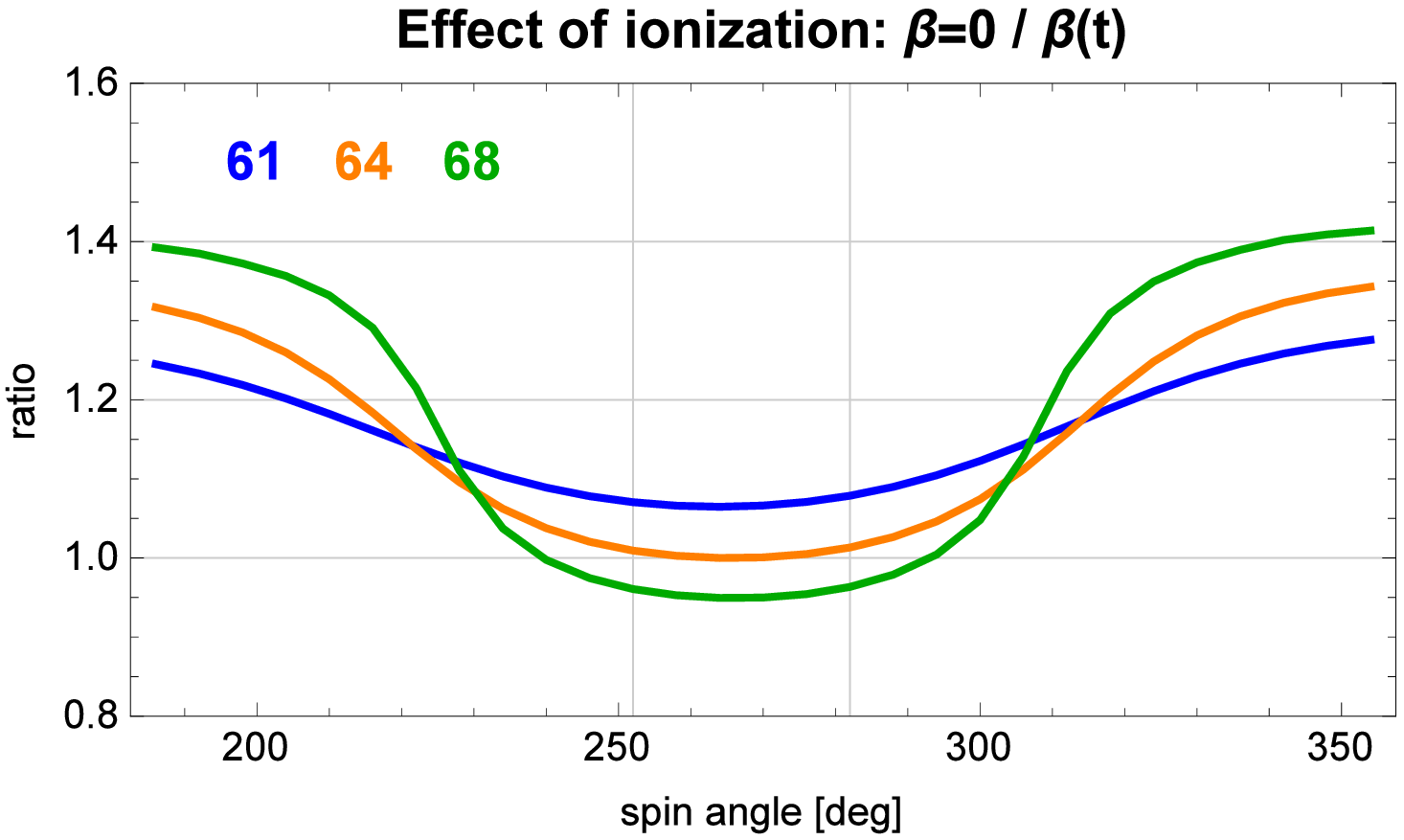}} 
	\caption{Ratio of the normalized to maximal value of the flux simulated with an ionization of zero to an ionization given for the time of detection ($\beta_{\mathrm{tot}}$ in Figure~\ref{figIonRatesHe}) for orbits 61, 64, and 68. Two vertical grids illustrate the range in spin-angle where the primary ISN~He is mainly observed. The normalization factor for the absolute fluxes is $1.74$ for orbit 64, spin-angle 264.}
	\label{figIonZero}
	\end{figure}

Consequences of neglecting of the ionization in the ISN~He modeling for the signal shape are presented in Figure~\ref{figIonZero}, which shows the ratio $q(\psi)$, defined as follows: 
	\begin{equation}
	q(\psi) = \frac{F(\psi, \beta=0)/ F(\psi_{\mathrm{max}},\beta=0)}{F(\psi,\beta (t))/F(\psi_{\mathrm{max}},\beta (t))},
	\label{eqFluxRatioIon0IonTrue}
	\end{equation}
where $\beta (t)$ and $\beta=0$ denote the cases with and without ionization, respectively, and $\psi_{\mathrm{max}}$ represents the spin-angle bin with maximal flux for a given case. 

The modification of the normalized ISN~He flux increases from the peak orbit 64 toward the side orbits (upward for pre-peak and downward for post-peak orbits for the ISN~He spin-angle range) and extend from about $5\%$ in the peak position to $10\%$ at the slopes of the signal. The discrepancies grow further with the spin-angle values and can reach $40\%$  in the most extreme case, which, however, is for spin-angles less interesting for the studies on the ISN~He primary population. Hence, it is not appropriate to neglect the ionization altogether if one wants to model a detailed distribution of the signal in the $6\degr$ bins, as is needed in the analysis method presented by \citet{swaczyna_etal:15a}. The deviations strongly exceed the measurement uncertainties, except for the pixels at the far wings of the measured signal. 

\begin{figure}
	\centering
	\resizebox{\hsize}{!}{\includegraphics{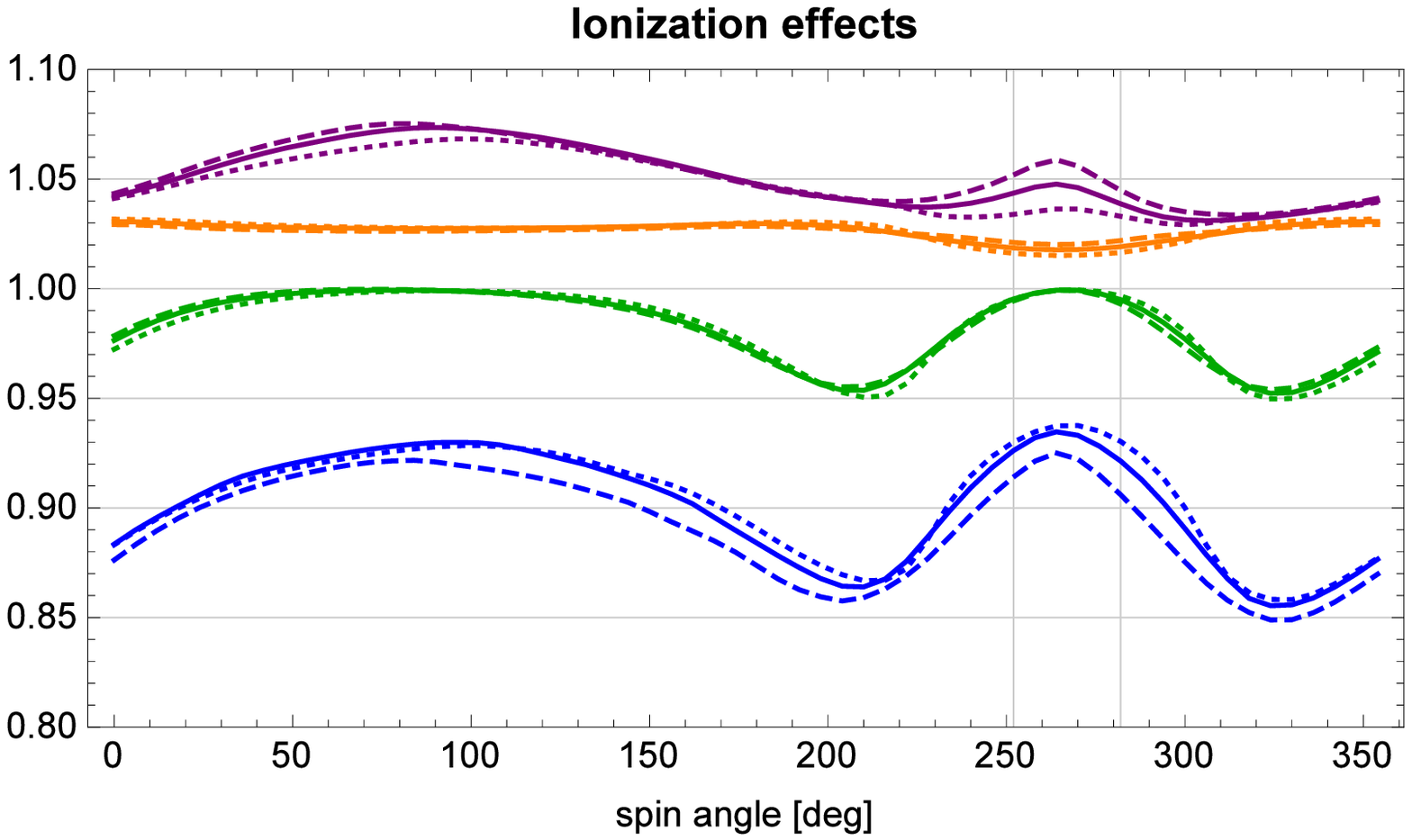}} 
	\caption{Effects of various components of the total ionization rate on the absolute level of the signal, simulated for the primary ISN~He population using the numeric version of WTPM. Shown are results for three orbits: 61 (dashed), 64 (solid), 68 (dotted). Green lines present the ratio of simulations for spherically symmetric photoionization to simulation with photoionization modulated with heliolatitude (effect of latitudinal anisotropy of photoionization). Orange lines show the ratio of calculations with the 3D photoionization to the ionization being a sum of the 3D photoionization and charge exchange reactions with solar wind protons and $\alpha$-particles (effect of charge exchange). Purple lines illustrate the ratio of the total ionization without accounting for the electron impact-ionization to ionization with electron impact-ionization for slow solar wind included (role of electrons). Blue lines present the ratio of simulations with the total ionization ($\beta_{\mathrm{tot}}$ in Figure~\ref{figIonRatesHe}) for the time of detection given only by in-ecliptic values (similar as Figure~\ref{figJSvsMKbase0}) to ionization with the history, latitudinal anisotropy, and correct electron-impact distance-relation taken into account. The vertical lines mark the spin-angle range of observations of the primary ISN~He population.}
	\label{figMKionEffects}
	\end{figure}

In the following subsections, we will investigate results of various effects in the ionization rate model used for analysis of ISN~He gas. Results of this analysis are collected in Figure~\ref{figMKionEffects}.
	
\subsubsubsection{Effect of latitudinal anisotropy of photoionization}
The effect of latitudinal anisotropy of photoionization on simulation of ISN~He flux is illustrated by the green lines in Figure~\ref{figMKionEffects}. From the viewpoint of ISN~He gas analysis it is negligible for all orbits, the difference between the spherically symmetric and anisotropic ionization rate are on the order of $1\%$ at the boundary of the signal region used in the analysis, and nearly null for the spin-angle bins at the peak. Potentially, it might be of some importance for the Warm Breeze orbits, which feature a much wider distribution of the signal: not surprisingly, the signatures of the hypothetical latitudinal anisotropy of the photoionization rate are largest for the spin-angle ranges corresponding to the solar poles.
	
\subsubsubsection{Effect of charge exchange}
Th effect of charge exchange with solar wind particles is illustrated by the orange lines in Figure~\ref{figMKionEffects}. We compare the flux calculated with photoionization only with the flux calculated assuming ionization rate as a sum of the photoionization and the charge exchange rate, taking latitudinal anisotropy into account in both cases. The effect for the absolute flux level is $\sim1.5\%$ for the peak of the signal, much less for the shape of the signal. Thus charge exchange ionization is negligible for the ISN~He observed by \emph{IBEX}.

\subsubsubsection{Effect of electron ionization}
The effect of electron-impact ionization is illustrated by the purple lines in Figure~\ref{figMKionEffects}. Electron ionization modifies the absolute flux by a few percent (from $3\%$ at the peak of orbit 68 to $\sim6\%$ at the peak of orbit 61, with a $5\%$ modification for orbit 64). Thus, the effect on the orbit-to-orbit ratios of the peak bins is comparable to the uncertainty due to the Poisson statistics for the peak pixels and practically negligible as much less than this uncertainty in all other pixels. 

\subsection{All departures from the standard model together}
In this section we show a comparison of the flux simulated assuming only spherically symmetric ionization given by the sum of all relevant processes with the values taken for the moment of the calculation for a given orbit, but otherwise invariable (i.e., no time dependence of the ionization rate along the trajectory) with the full model of the ionization rate, i.e., for the time-dependent ionization, with heliolatitude anisotropy and not $1/r^2$ dependence of electron impact rate. This is illustrated with the blue lines in Figure~\ref{figMKionEffects}. All details of the ionization rate together reduce the total ISN~He flux from $5\%$ to $15\%$, depending on the orbit and spin-angle. The effect as a function of spin-angle within individual orbits is small (on a level of $1\%$ between the peak and the wings), and from orbit to orbit it is approximately $\pm2\%$, with pre-peak orbits systematically reduced and post-peak orbits enhanced. The $2\%$ effect is on the order of Poisson uncertainty of the peak pixels and is much less in the other pixels.

In summary, details of the ionization rate are of minor importance for analysis of individual seasons of ISN~He measurements. However, they may become important when one analyzes several seasons together using the method discussed by \citet{swaczyna_etal:15a}, especially if they are from the times of markedly different solar activity. The main factor will be the change in the solar photoionization rate which is the most effective ionization for ISN He, which may modify the absolute level of the flux by a factor of two from solar minimum to maximum. Thus a lack of credible ionization model may in this case hamper finding a statistically satisfactory solution.

\section{Summary and conclusions}
We developed a new version of the WTPM, specially tailored for analysis of interstellar neutral atom flux observed by \emph{IBEX}. The model now has two strains, aWTPM and nWTPM, which are complementary to each other. We present them in detail, in the terms of both the physical assumptions and the implementation aspects, and show that they give results that agree to at least $1\%$ when run under identical assumptions (Figure~\ref{figJSvsMKbase0}). aWTPM uses a simplified approach to the calculation of ionization losses, but due to implementation details it is well suited for investigating effects of various physical and measurement aspects, like, e.g., non-Maxwellian distribution function of ISN~He in the LIC \citep[][this volume]{sokol_etal:15a}, or various approximations to the collimator transmission function (Figure~\ref{figEffectCollShapeResponseFun}). nWTPM is a heavy-duty version for mass-scale calculations, needed to fit the model parameters to the data, and includes fully time- and latitude-dependent ionization losses. nWTPM is a strongly optimized and refined version of the WTPM model used by \citet{bzowski_etal:12a, bzowski_etal:13b, bzowski_etal:14a}, \citet{kubiak_etal:13a, kubiak_etal:14a}, \citet{rodriguez_etal:13a, rodriguez_etal:14a}, \citet{park_etal:14a}, and \citet{mccomas_etal:15a} in their analyses of various species of interstellar gas in the heliosphere, observed by \emph{IBEX} or \emph{Ulysses}. aWTPM was used by \citet{sokol_etal:15a} and \citet{galli_etal:15a} in the search for the fall peak in ISN~He and discussion of the expected low-level ``haze'' in the sky due to extended wings of the Warm Breeze and ISN~He populations. A brief comparison of aWTPM and nWTPM is provided in Table~\ref{tab:compWTPM} at the end of Section~\ref{sec:Outlook}.

We analyzed the influence of a number of effects that may be tempting to neglect in the simulation and show how they affect the results of simulations needed to fit the data using the method developed by \citet{swaczyna_etal:15a}. These effects are listed in Table~\ref{tab:effects} with commentaries on their significance. The significance of these effects in the analysis method developed by \citet{lee_etal:12a} is presented by \citet{mobius_etal:15a}; an exception is the influence of the ionization rate for the determination of the flux maximum longitude along the Earth's orbit, which we present in Section~\ref{sec:ionEffects} (Figure~\ref{figIonZeroParams}). 

Generally, most of the effects we have considered modify the signal by a few percent in the spin-angle range  characteristic for the primary ISN~He population, but much stronger just outside it, where the Warm Breeze discovered by \citet{kubiak_etal:14a} is visible. We conclude that in order to maintain a homogeneous accuracy for all simulated data points, one needs to take almost all the listed effects into account in the calculation because they are of comparable strength. We point out that for the purpose of fitting a model to the data, one must consider the precision needed in the simulations of individual data points, which is directly related to the measurement uncertainties and correlations between various data points. This aspect is discussed in an accompanying paper by \citet{swaczyna_etal:15a}. 

WTPM in its present version seems to be a tool very well suited to analysis of \emph{IBEX}-Lo measurements of ISN neutrals, which feature an unprecedentedly high signal-to-noise ratio of $\sim1000$. We were able to streamline and refine the algorithm so that the code now runs faster and is more accurate than it was previously. Results of this analysis are presented in the accompanying papers by \citet{bzowski_etal:15a}, \citet{sokol_etal:15a}, and \citet{galli_etal:15a}.

	\clearpage
	\begin{table}
	\begin{center}
	\caption{Resume of effects included in WTPM and their significance in the modeling of ISN~He flux observed by \emph{IBEX}-Lo \label{tab:effects}}
		\begin{tabularx}{\textwidth}{>{\hsize=.5\hsize}X|>{\hsize=.5\hsize}X|>{\hsize=1.5\hsize}X}
		\tableline
		Effect & Section, Equation, Figure & Commentary and Recommendation \\ \tableline
		Non-zero tilt of spin-axis relative to the ecliptic plane & Sections:~\ref{sec:diffFlux}, \ref{sec:spinAxTilt}, Figure:~\ref{figSpinAxisPointing} & Important, must be included; see \citet{mobius_etal:15b}. \\ \tableline
		Orbital motion of the spacecraft & Sections: \ref{sec:diffFlux}, \ref{sec:scmotion}; Figures: \ref{figVelocityVectors}, \ref{figEffectsTimeInt} & Adopting the Earth's velocity relative to the Sun instead of the vector sum of the Earth's velocity and the \emph{IBEX} velocity relative to Earth affects the result depending on the time distance of the modeled good time interval from the beginning and end of HASO times; strongly recommended at least for the orbits where good times are short and near the HASO boundaries. \\ \tableline
		Finite versus infinite distance to the source region of ISN~He atoms & Sections:~\ref{sec:ballistics}, \ref{sec:rFin}; Figure:~\ref{figEffectStopDistSpectrum} & Physical sense: the distance of  last collisions for atoms before entering the heliosphere; changing this distance from $\sim150$~AU to infinity modifies the simulated signal up to  $\pm5\%$. The effect is correlated for different orbits, but affects ISN parameter results only weakly; the main difference is in the fitted inflow speed (by $\sim0.25$~\kms), with resulting uncertainty in the other parameters due to parameter correlation. \\ \tableline
		\end{tabularx}
	\end{center}
	\end{table}
		
	\begin{table*}
	\begin{center}
	\caption{Table~\ref{tab:effects}, continued.}
		\begin{tabularx}{\textwidth}{>{\hsize=.5\hsize}X|>{\hsize=.5\hsize}X|>{\hsize=1.5\hsize}X}
		\tableline
		Effect & Section, Equation, Figure & Commentary and Recommendation \\ \tableline
		Details of collimator transmission function and shape of the aperture & Sections: \ref{sec:collimTransmiss}, \ref{sec:collimModif}; Equations: \ref{eqCollAvDef} through \ref{eqMQCollimIntQuadrature}; Figures: \ref{figCollTrans}, \ref{figCollTransMatr3D}, \ref{figCollTransmissionFluxGrid}, \ref{figEffectCollShapeResponseFun} & The broadening of the beam by the collimator must be taken into account. Approximating the collimator as fully low-resolution versus true introduces a $\sim4\%$ error in the flux, different for different orbits and pixels. The aperture shape can be approximated by a circle (deviations on the order of $1\%$ visible only when the ISN beam is skimming the FOV, e.g., orbit 61). Recommendation: approximate the hexagonal FOV by circular. \\ \tableline
		Averaging over $6\degr$ bins versus adopting center value for the bin & Sections:~\ref{sec:spinBinMethod}, \ref{sec:spinBins}; Figures:~\ref{figEffectsSpinAngIntFluxTab6}, \ref{figEffectsSpinAngIntTab6}, \ref{figEffectsSpinAngInt} & Tabulating the flux at the centers of the $6\degr$ bins instead of averaging is potentially inaccurate up to $20\%$ in some pixels. Arithmetic average for a tabulation every $1\degr$ is acceptable (errors of $\sim1\%$), much better results obtained with sampling every $1.5\degr$ and using the formula from Equation~\ref{eqSpinAngleAver}. \\ \tableline
		\end{tabularx}
	\end{center}
	\end{table*}

	\begin{table*}
	\begin{center}
	\caption{Table~\ref{tab:effects}, continued.}
		\begin{tabularx}{\textwidth}{>{\hsize=.5\hsize}X|>{\hsize=.5\hsize}X|>{\hsize=1.5\hsize}X}
		\tableline
		Effect & Section, Equation, Figure & Commentary and Recommendation \\ \tableline
		Averaging over good time intervals versus adopting middle HASO time & Sections: \ref{sec:diffFlux}, \ref{sec:scmotion}, \ref{sec:timeIntegr}; Equations: \ref{eqFluxTPoly} through \ref{eqAvrFlux}; Figures: \ref{figVelocityVectors}, \ref{figEffectsTimeInt} & Signal varies during the orbit because the beam moves through the field of view due to the spacecraft's motion with Earth. The orbit-integrated signal is affected by the uneven distribution of good time intervals during the orbit. Actual magnitude depends on details of good times, especially the distance from the HASO boundaries; recommended to average over good time intervals.\\ \tableline
		Ionization losses & Sections: \ref{sec:calcSurPro}, \ref{sec:ion}; Equations: \ref{eqWdef}, \ref{eqW}; Figures: \ref{figIonRatesHe}, \ref{figIonZeroParams}, \ref{figIonZero}, \ref{figMKionEffects} & Important for the evaluation of the absolute values, e.g., for simultaneous analysis of seasons with significantly different solar activity. Photoionization is responsible for $\sim85\%$ of the losses, electron impact for $\sim10\%$, and charge exchange for $\sim5\%$. The latitudinal anisotropy effect is negligible. When modeling one ISN season and scaling the simulations to the data, ionization effects are of secondary importance. \\ \tableline
		\end{tabularx}
	\end{center}
	\end{table*}

\acknowledgments
The authors are indebted to Eberhard M{\"o}bius and David McComas for careful reading of the manuscript and valuable suggestions.
This research was supported by Polish National Science Centre grant 2012-06-M-ST9-00455.

\bibliographystyle{apj}
\bibliography{iplbib}{}

\end{document}